\documentclass{article}

\usepackage{ssArxiv}

\usepackage[utf8]{inputenc}

\usepackage[title]{appendix}

\usepackage{amsmath,amsbsy,amsgen,amscd,amssymb,amsthm,amsfonts}
\usepackage{mathtools}
\usepackage{booktabs}
\usepackage{bm}
\usepackage{enumitem}
\usepackage{subcaption, geometry, graphicx, lipsum}
\usepackage{microtype}
\usepackage{multirow}
\usepackage[
  pagebackref=true,
  colorlinks=true,
  citecolor=MidnightBlue,
  linkcolor=Fuchsia,
]{hyperref}
\usepackage{pifont}
\usepackage[nameinlink]{cleveref}
\usepackage{url} 
\usepackage{graphicx}
\usepackage{algorithm,algorithmic}
\usepackage[round]{natbib}
\usepackage{xspace}
\usepackage{caption}
\usepackage{listings}
\definecolor{codegreen}{rgb}{0,0.6,0}
\lstdefinestyle{mystyle}{  
    commentstyle=\color{codegreen},
    keywordstyle=\color{blue},
    basicstyle=\ttfamily\footnotesize,
    breakatwhitespace=false,         
    breaklines=true,                 
    captionpos=b,                    
    keepspaces=true,                 
    showspaces=false,                
    showstringspaces=false,
    showtabs=false,                  
    tabsize=2
}

\lstdefinestyle{boxed}{  
    frame=single,
    breaklines=true,
    postbreak=\mbox{$\hookrightarrow$\space},
    commentstyle=\color{codegreen},
    keywordstyle=\color{blue},
    basicstyle=\ttfamily\footnotesize,
    breakatwhitespace=false,         
    breaklines=true,                 
    captionpos=t,                    
    keepspaces=true,                 
    showspaces=false,                
    showstringspaces=false,
    showtabs=false,                  
    tabsize=2
}

\lstdefinestyle{reasoning}{  
    breaklines=false,
    commentstyle=\color{codegreen},
    keywordstyle=\color{blue},
    breakautoindent=0pt,
    breakindent=0pt,
    basicstyle=\itshape\footnotesize,
    breaklines=true,
    breakatwhitespace=true,         
    captionpos=t,                    
    keepspaces=true,                 
    showspaces=false,                
    showstringspaces=false,
    showtabs=false,
}

\usepackage[dvipsnames]{xcolor}
\usepackage[most]{tcolorbox}

\definecolor{lightgreen}{HTML}{C4FFCF}

\crefname{assumption}{assumption}{assumptions}

\makeatletter
\newcommand\nnfootnote[1]{%
  \begin{NoHyper}
  \renewcommand\thefootnote{}\footnote{#1}%
  \addtocounter{footnote}{-1}%
  \end{NoHyper}
}

\makeatother
  
\renewcommand*\backref[1]{\ifx#1\relax \else (Cited on pg. #1) \fi}
\title{The Counterfeit Conundrum: Can Code Language Models Grasp the Nuances of Their Incorrect Generations? \\ \
}

\author{\name Alex Gu
\email{gua@mit.edu} \\
\addr{MIT CSAIL} \\
\name Wen-Ding Li$^{\star}$
\email{wl678@cornell.edu} \\
\addr{Cornell University} \\
\name Naman Jain$^{\star}$
\email{naman\_jain@berkeley.edu} \\
\addr{University of California, Berkeley} \\
\name Theo X. Olausson$^{\star}$
\email{theoxo@csail.mit.edu} \\
\addr{MIT CSAIL} \\
\name Celine Lee$^{\star}$
\email{cl923@cornell.edu} \\
\addr{Cornell University} \\
\name Koushik Sen
\email{ksen@berkeley.edu} \\
\addr{University of California, Berkeley} \\
\name Armando Solar-Lezama
\email{asolar@csail.mit.edu} \\
\addr{MIT CSAIL} \\
}

\begin{document}

\maketitle
\raggedbottom

\nnfootnote{$^{\star}$ Equal contribution, author order determined via \texttt{random.shuffle}}

\begin{abstract}
While language models are increasingly more proficient at code generation, they still frequently generate incorrect programs. Many of these programs are obviously wrong, but others are more subtle and pass weaker correctness checks such as being able to compile. In this work, we focus on these \textit{counterfeit samples}: programs sampled from a language model that 1) have a high enough log-probability to be generated at a moderate temperature and 2) pass weak correctness checks. Overall, we discover that most models have a very shallow understanding of counterfeits through three clear failure modes. First, models mistakenly classify them as correct. Second, models are worse at reasoning about the execution behaviour of counterfeits and often predict their execution results as if they were correct. Third, when asking models to fix counterfeits, the likelihood of a model successfully repairing a counterfeit is often even lower than that of sampling a correct program from scratch. Counterfeits also have very unexpected properties: first, counterfeit programs for problems that are easier for a model to solve are \textit{not necessarily} easier to detect and only slightly easier to execute and repair. Second, counterfeits from a given model are just as confusing to the model itself as they are to other models. Finally, both strong and weak models are able to generate counterfeit samples that equally challenge all models. In light of our findings, we recommend that care and caution be taken when relying on models to understand their own samples, especially when no external feedback is incorporated.
\end{abstract}

\section{Introduction}
In the past year, language models such as CodeLlama \citep{roziere2023code}, DeepSeek-Coder \citep{guo2024deepseek}, and GPT-4 \citep{openai2023gpt} have demonstrated great advances in code generation. Their success has primarily been due to their strong code generation abilities, as measured by benchmarks such as HumanEval \citep{chen2021evaluating} and MBPP \citep{austin2021program} as well as their usefulness in general-purpose code writing. While these models are able to produce correct code for impressively complex specifications, they just as often produce incorrect code.

\begin{figure*}[ht!]
    \centering
    \includegraphics[width=0.95\textwidth]{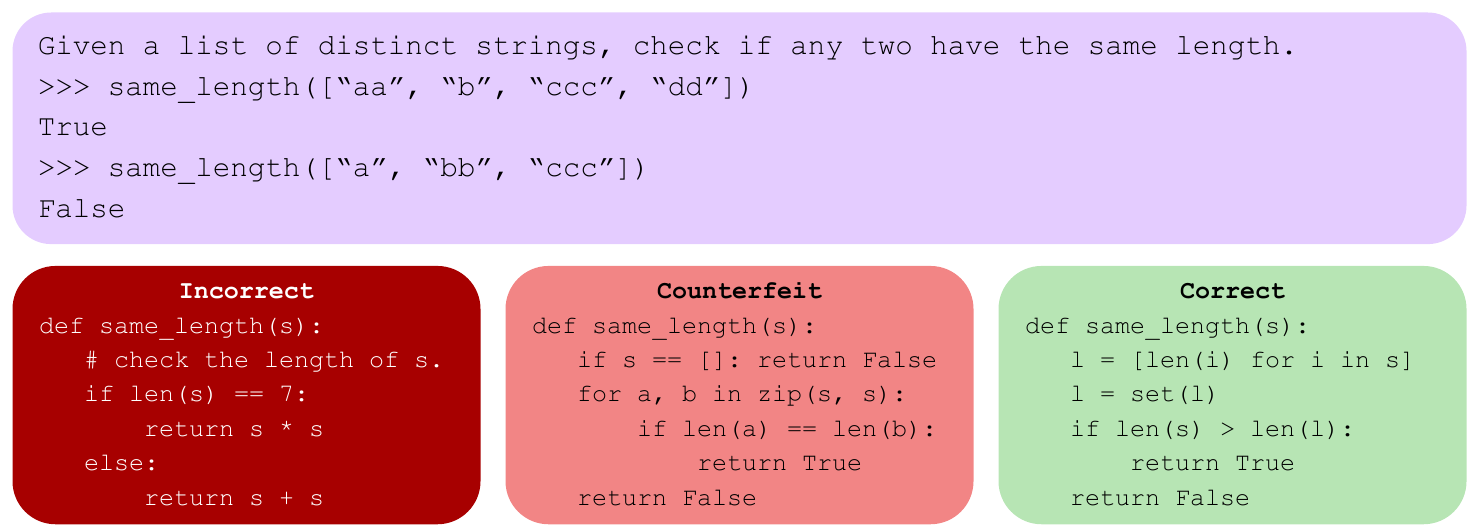}
    \caption{Example of a problem specification with incorrect, counterfeit, and correct programs.}
    \label{fig:main}
\end{figure*}

Some of these incorrect programs contain egregious mistakes, but others fail in more subtle ways. We focus our attention towards the second group, which we call \textit{\textbf{counterfeit samples}}. We define a counterfeit sample to be a program sampled from a code language model which is 1) good enough to be generated by the language model at a moderate temperature, 2) are incorrect, and 3) pass weak but nontrivial correctness checks. In this work, we study the extent to which models can understand these counterfeit programs. This final criterion of passing nontrivial correctness checks distinguishes programs with more subtle errors from those that trivially fail and are likely uninteresting. In Fig. \ref{fig:main}, we show an example of an incorrect, counterfeit, and correct program. Because we use relatively weak correctness checks, many counterfeit programs can still be easily detected as wrong by a human.

We provide empirical evidence that code language models have a shallow understanding of these counterfeit samples (Sec. \ref{sec:understanding-counterfeit}) via three evaluations: correctness checking, execution prediction, and program repair. For correctness checking, the model is asked to assess whether a short piece of code correctly implements a natural language specification (sometimes with test cases). For execution prediction, the model is given a program-input pair and asked to predict the output of executing the program on the given input. For fairness, we ensure the programs are generally short and that execution does not require complex calculations. For repair, the model is given the counterfeit program alongside its original specification and is asked to correct it. First, we find that models frequently misjudge counterfeit samples as correct. Second, models are much worse at reasoning about the execution of counterfeits than their correct companions, often executing counterfeits as if their semantics matched those of a correct program. Third, models falter at repair: the likelihood of a model successfully repairing a counterfeit example is often even lower than that of generating a correct program when sampling from scratch. As a caveat, this paper focuses on open-source models, primarily CodeLlama 34B and DeepSeek Instruct 33B. We also present limited results on GPT-3.5 and GPT-4 which suggest that GPT-3.5 behaves similarly to the open-source models while GPT-4 has a much better understanding of counterfeits. Nevertheless, we still find that GPT-4 still exhibits some of these misunderstandings.

Through further analysis, we find that counterfeit samples have other unexpected properties
(Sec. \ref{sec:analyzing-counterfeit}). We find, for example, that counterfeit samples from problems that are easier for the model to solve are \textit{not} easier to assess and only slightly easier to execute and repair, highlighting an inconsistency between generation and understanding capabilities. We also observe that models don't perceive their own counterfeit samples differently from other models' counterfeits and that models of all capability levels are able to generate equally difficult counterfeit samples.

Overall, we find that these counterfeit samples are, in a sense, adversarial to the model: models often struggle to assess their correctness, reason about their execution as if they were correct programs, and repair them at a low rate. Understanding counterfeit samples is a prerequisite to many downstream applications in which models use their own feedback to improve themselves. Therefore, in light of our findings, we recommend exercising caution in these schemes such as self-repair and model-based reranking of outputs, especially when no external feedback is incorporated.
\section{Experimental Setup}
\subsection{Generating Counterfeit Examples}
\begin{figure}[ht!]
    \centering
    \begin{subfigure}[b]{0.45\columnwidth}
        \centering
        \includegraphics[width=\linewidth]{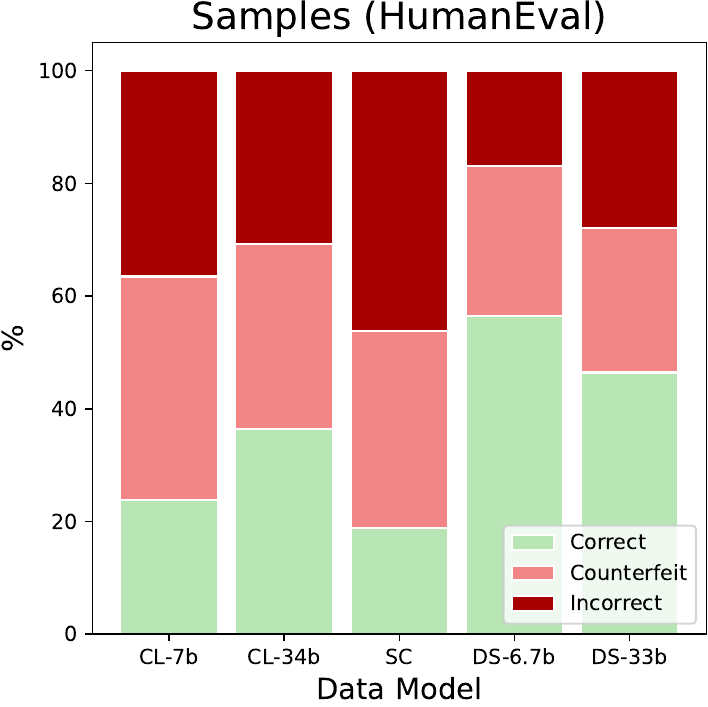}
    \end{subfigure}
    \hfill
    \begin{subfigure}[b]{0.26\columnwidth}
        \centering
        \includegraphics[width=\linewidth]{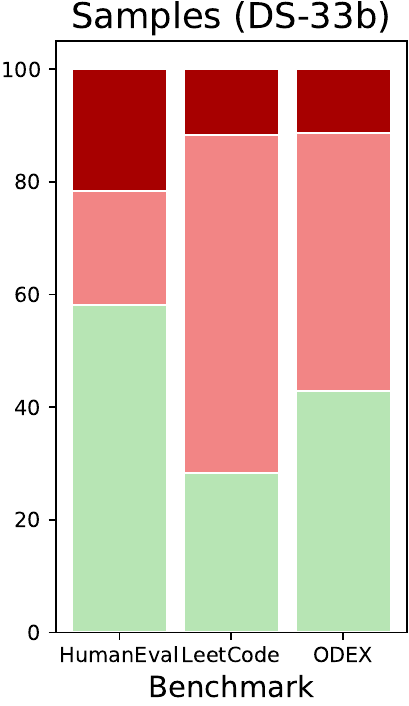}
    \end{subfigure}
    \hfill
    \begin{subfigure}[b]{0.26\columnwidth}
        \centering        \includegraphics[width=\linewidth]{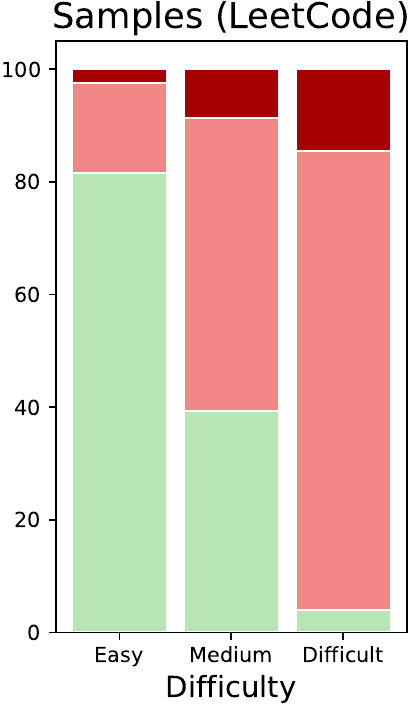}
    \end{subfigure}
    \caption{Distribution of correct, counterfeit, and incorrect samples by model, benchmark, and problem difficulty, showing their prevalence across all settings.}
    \label{fig:sample-distribution}
\end{figure}

We use three datasets: HumanEval, LeetCode, and ODEX \citep{wang2022execution}. HumanEval evaluates code generation of simple natural language descriptions, LeetCode is a harder dataset of the same flavor using programming interview practice problems, and ODEX tests knowledge of diverse Python libraries. To generate counterfeit examples, we first sample programs from CodeLlama (CL), DeepSeek-Coder Instruct (DS-I), and StarCoder (SC) at temperature $T=0.6$. Of the incorrect programs, we design a dataset-specific filter to remove incorrect programs that do not pass mild correctness criteria. For HumanEval, counterfeits are programs passing at least 10\% of EvalPlus \citep{liu2023your} tests. For LeetCode, counterfeits are programs that received a ``Wrong Answer'' verdict, which filters out programs that crashed during runtime or took too long to finish. For ODEX, counterfeits are programs that can be successfully parsed by \texttt{ast.parse} and are under 500 characters. In Fig. \ref{fig:sample-distribution}, we show the number of counterfeits generated by different models (left), benchmarks (middle), and problem difficulty levels (right), showing that counterfeits are widespread and occur in each setting.

\subsection{Code Understanding Tasks}
We evaluate counterfeits on the following tasks:

\textit{Correctness Checking}: The goal is to check whether a model-generated Python program (either correct or counterfeit) correctly implements a natural language (NL) specification. These specifications often include input-output examples. We use CoT with majority voting \citep{wei2022chain, wang2022self} and report accuracy.

\textit{Execution Prediction}: The goal is to predict the execution output of a given model-generated Python program on a specific input. We use an execution-based metric for correctness and report pass@1, the fraction of samples that are correct.

\textit{Repair}: The goal is to repair a given incorrect model-generated counterfeit program to correctly implement a given natural language specification. The model is not given any execution feedback other than the fact that the program is incorrect.

All three tasks are given to the language model (LM) in a few-shot setting. For GPT-3.5 and GPT-4, all tasks are prompted with chain-of-thought (CoT) \citep{wei2022chain}. For other models, we use CoT with majority voting ($N=10$) for correctness checking but not execution prediction (we found it did not help). For more details and full prompts, see Appendix \ref{appendix:task-evaluation-prompts}.


\subsection{Dataset Creation}
Each set of samples is curated using a single dataset (such as HumanEval) and model (such as CodeLlama 34B). Each set is balanced and consists of 5 correct and 5 counterfeit programs for each problem (problems that do not have enough programs are discarded). Overall, across HumanEval, LeetCode, and ODEX, we generate $12$ different sets of samples, each consisting of $360$ to $1190$ programs. For correctness prediction, these datasets are used directly. For execution prediction, we randomly selected input-output examples, removing pairs that require complex arithmetic or execution. We also remove the problem statement and example input-output pairs so the model focuses on executing the code. For repair, we use the counterfeit samples in each set and discard the correct samples. More details and examples can be found in Appendix \ref{appendix:detailed-experimental-setup}.

\section{Can code language models understand counterfeit samples?} \label{sec:understanding-counterfeit}
In this section, we argue that models struggle to understand counterfeit samples. Due to space limitations, we only highlight a subset of datasets and models in this plot, deferring the complete set of results to Appendix \ref{appendix:all-accuracy-results}. 
\subsection{Correctness Checking}
\begin{figure}[ht!]
    \centering
    \begin{subfigure}[b]{0.49\columnwidth}
        \centering
        \includegraphics[width=\textwidth]{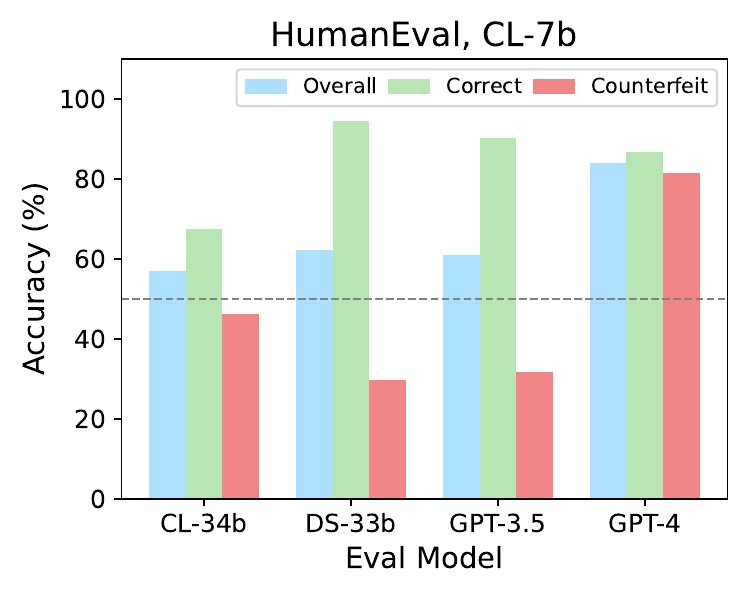}
        \subcaption{HumanEval (CL-7B)}
    \end{subfigure}
    \hfill
    \begin{subfigure}[b]{0.49\columnwidth}
        \centering
        \includegraphics[width=\textwidth]{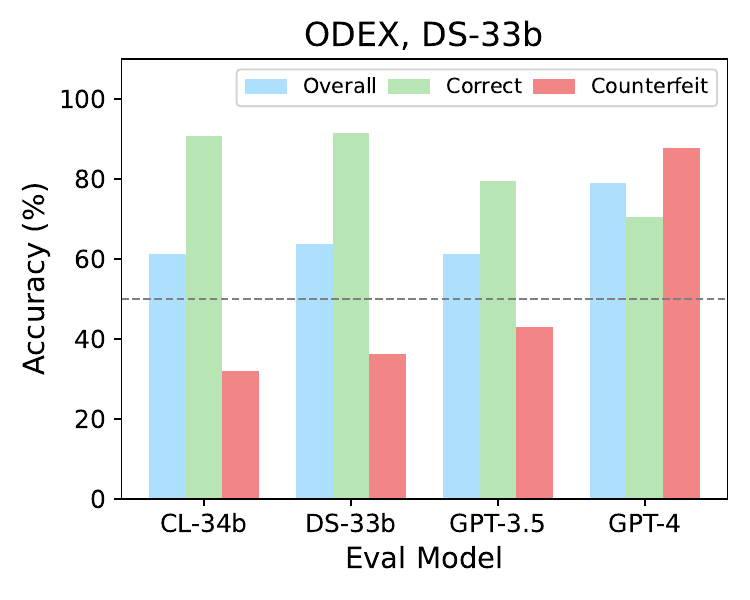}
        \subcaption{ODEX (DS-33B)}
    \end{subfigure}
    \caption{Models other than GPT-4 struggle to classify samples as correct or counterfeit and are much better at assessing the correctness of correct samples than counterfeit samples.}
    \label{fig:execution-accuracy}
\end{figure}

We begin by examining whether language models can correctly identify whether a program is correct or counterfeit given the natural language specification. In Fig. \ref{fig:execution-accuracy}, we plot the accuracy of CodeLlama 34B, DeepSeek-Coder 33B, GPT-3.5, and GPT-4 on balanced datasets of correct and counterfeit programs for HumanEval and ODEX. For the first three models, the blue bars indicate that correctness checking accuracy is at about 60\% for both of these datasets, which is only slightly better than the 50\% random-guessing baseline. This indicates that models generally fail to distinguish between correct and counterfeit samples. In addition, by comparing the green and red bars, we observe that the performance of these three models on correct samples is much higher than their performance on counterfeit samples, showing that models are biased towards thinking that counterfeit samples are actually correct. On the other hand, GPT-4 is much better (but not perfect) at this task with an accuracy at around 80\% for both datasets. We also observe that in contrast with the rest of the models (including those not shown here, see Fig. \ref{fig:correctness-checking-counterfeit}), GPT-4 is \textit{not} biased towards predicting that these samples are correct. However, GPT-4 still falters around 20\% of the time, and we qualitatively analyze some of these remaining GPT-4 failures in Sec. \ref{sec:qual-analysis}.

\subsection{Execution Prediction}
\begin{figure}[ht!]
    \centering
    \begin{subfigure}[b]{0.49\columnwidth}
        \centering
        \includegraphics[width=\textwidth]{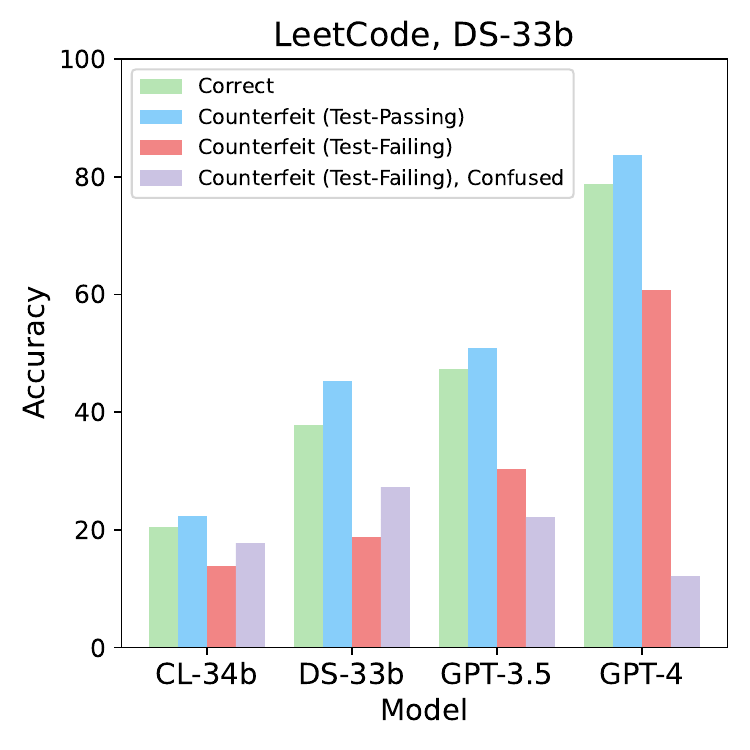}
        \subcaption{LeetCode (DS-33B)}
    \end{subfigure}
    \hfill
    \begin{subfigure}[b]{0.49\columnwidth}
        \centering
        \includegraphics[width=\textwidth]{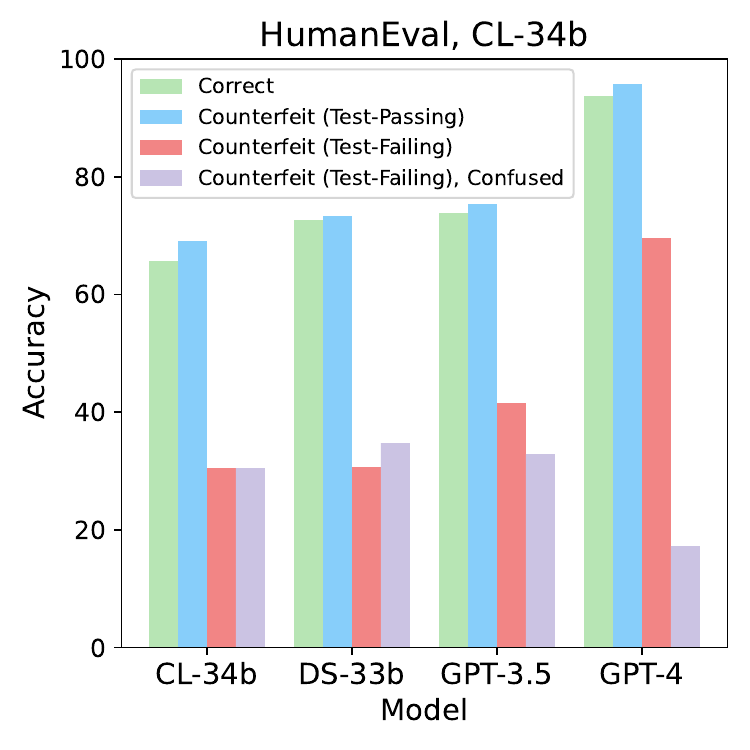}
        \subcaption{HumanEval (CL-34B)}
    \end{subfigure}
    \caption{Models are much better at executing correct samples than counterfeit samples, and even often execute counterfeit samples as if they were correct.}
    \label{fig:execution-prediction-accuracy}
\end{figure}

Next, we assess the ability of models to predict the execution behavior of counterfeit samples. In Fig. \ref{fig:execution-prediction-accuracy}, we plot the execution accuracy of the previous four models on two datasets, LeetCode generated by DS-33B and HumanEval generated by CL-34B. 

In this task, each sample includes a program (correct or counterfeit) $P$ and an input $I$. The accuracy of the correct samples are shown in the green bars. Because counterfeit programs still pass a subset of tests, we distinguish their execution samples into two groups. We call samples where $P$ passes test $I$ \textit{test-passing} counterfeit samples and the rest as \textit{test-failing} counterfeit samples. The execution prediction accuracies of these samples are shown in blue and red, respectively. In purple, we show the proportion of test-failing counterfeit samples where the model actually predicted the output of the correct program. Note that samples counting towards the red accuracy are disjoint from those counting towards the purple accuracy.

Overall, we observe that models have a difficult time distinguishing the semantics of a counterfeit program from their correct counterparts, suggesting they may have a shallow understanding of program semantics. By comparing the green and blue bars with the red bar, we see that models fail much more  at executing counterfeit programs when the semantics are incorrect. The purple bars provide further evidence of this: models other than GPT-4 frequently execute counterfeit programs as if they had the semantics of a correct program, sometimes even more often than their true semantics (red). For GPT-4, the effect is much less pronounced but still present, as GPT-4 still performs much better on correct and test-passing counterfeits than test-failing counterfeits. Despite having such a high performance, it was still confused for a sizable number of test-failing counterfeit samples, predicted the output of the correct program rather than the correct execution result. Overall, as models only see the programs and not the problem statements, this suggests that they may be hallucinating the semantics of incorrect programs. This provides further evidence that models are poor at distinguishing correct programs from counterfeit programs.

\subsection{Repair}
\label{sec:repair}
Finally, we probe the model's ability to repair the counterfeit samples it has generated.
Although this task may appear to simply boil down to code generation,
prior work has highlighted that code understanding forms an integral part of the repair pipeline since achieving good performance hinges on the
model's (in)ability to generate accurate textual explanations of \emph{what} is wrong with the code \cite{olausson2024repair}; as such, self-repair may give us further insight into the model's capabilities.

Prior work has shown that when given information about \emph{which} unit test failed, many models are capable of repairing incorrect Python programs at rates that exceed their baseline pass rates \cite{chen2024teaching,olausson2024repair}.
In this section, we press the model even harder by not giving any execution signal whatsoever, instead simply informing it that the program did not pass; thus, successful repair depends entirely on the model's own ability to understand the program and its relationship to the specification.
Importantly, the success rate of repair must be compared to the baseline pass@1 rate,
since a sample can also be ``repaired'' simply by drawing another unconditional sample from the model.
Details of the experimental setting, and the prompt used for this task, are given in Sec.~\ref{sec:task-eval}-\ref{sec:prompts}.

\begin{figure}[ht!]
    \centering
    \begin{subfigure}[b]{0.49\columnwidth}
        \centering
        \includegraphics[width=\textwidth]{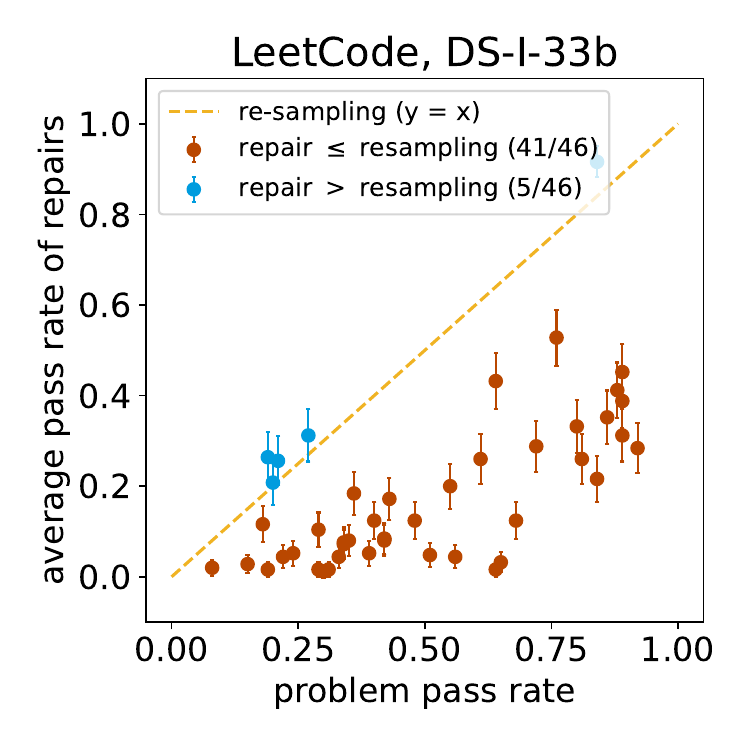}
        \subcaption{LeetCode (DS-I-33B)}
    \end{subfigure}
    \hfill
    \begin{subfigure}[b]{0.49\columnwidth}
        \centering
        \includegraphics[width=\textwidth]{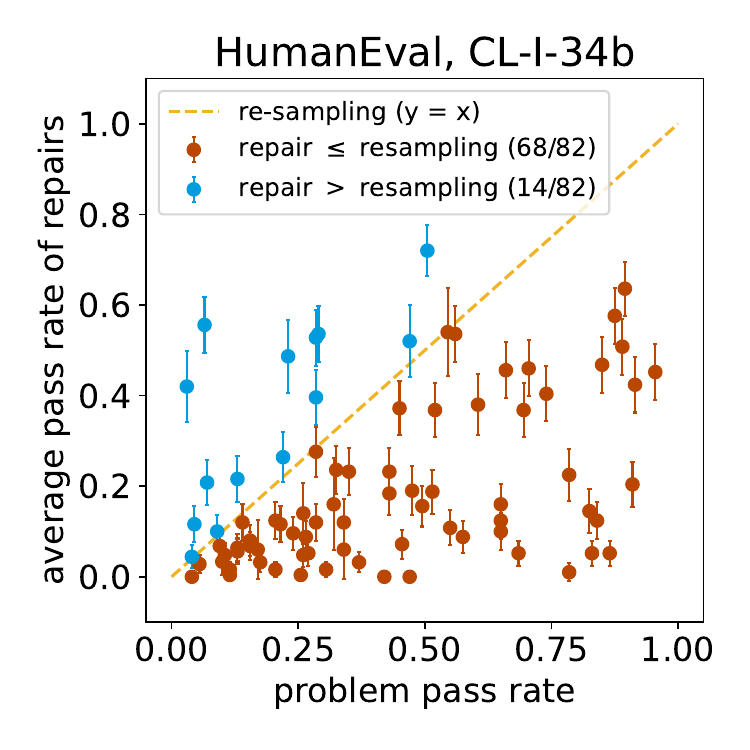}
        \subcaption{HumanEval (CL-I-34B)}
    \end{subfigure}
    \caption{In the absence of execution information, we find that repair underperforms resampling in almost all settings. Samples above the $y=x$ resampling baseline have been coloured in blue for clarity. See Appendix~\ref{appendix:repair} for full results. Vertical lines indicate 95\% confidence intervals over repair samples.}
    \label{fig:repair-accuracy}
\end{figure}

Fig.~\ref{fig:repair-accuracy} shows the results for CodeLlama 34B\footnote{Since repair is a task that depends heavily on the model adhering to instructions such as actually repairing the programs, rather than re-generating them from scratch, we use the instruction-tuned version CodeLlama 34B-Instruct for these experiments.} and DeepSeek 33B when repairing their own programs on HumanEval and LeetCode (respectively). The full set of results are in Appendix~\ref{appendix:repair}.
In these figures, each point is the mean success rate of repair for a particular problem; points above the line $y=x$ (which corresponds to a pass rate equal to that of the simple resampling strategy) thus indicate successful repair, while points below it indicate that the model could not reliably debug and repair the programs.
We note that although repair appears somewhat successful with DeepSeek-33B on HumanEval (Fig.), beating out the baseline for 35/81 problems, in all other settings a strong majority of the points lie below the line $y=x$. In other words, the success rate of repair is---for most tasks---significantly below what one would achieve with the simple resampling strategy.
This evidence shows that models cannot reliably repair counterfeit samples, which suggests that they could not understand why these programs were deemed incorrect.

\section{Do counterfeit samples from different models or problems differ significantly?} \label{sec:analyzing-counterfeit}
\subsection{Is it easier for models to understand counterfeit samples from problems it finds easier?}
\begin{figure*}[ht!]
    \centering
    \begin{subfigure}[b]{0.31\textwidth}
        \centering
        \includegraphics[width=\textwidth]{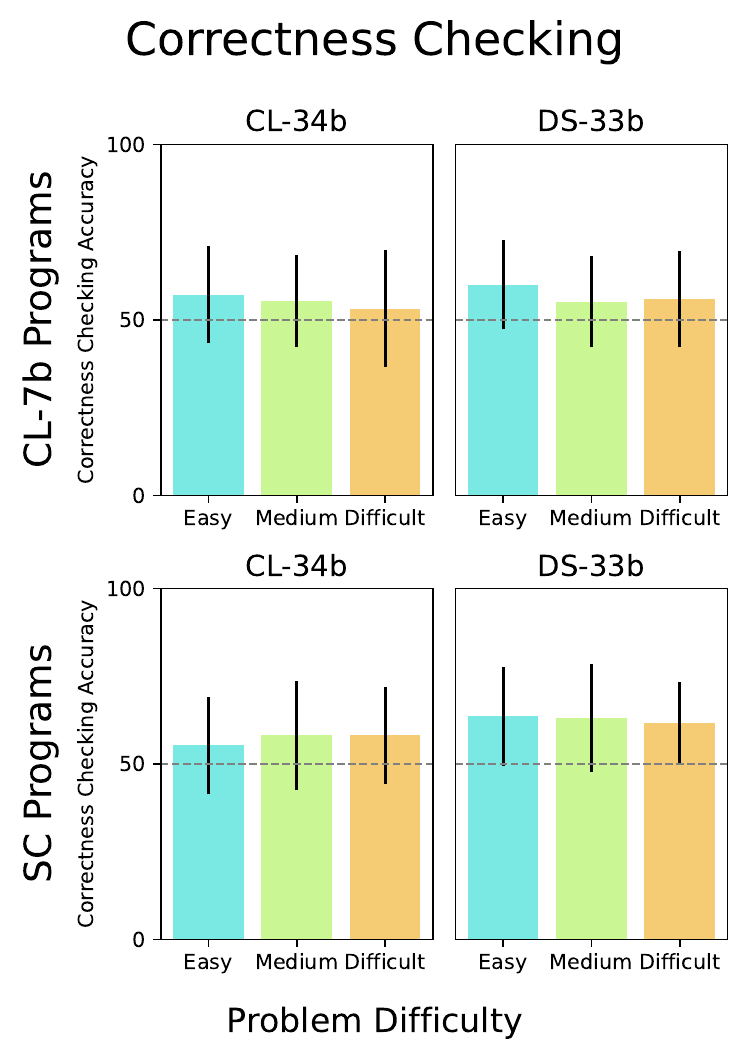}
        \caption{Correctness Checking}
    \end{subfigure}
    \hfill
    \begin{subfigure}[b]{0.31\textwidth}
        \centering
        \includegraphics[width=\textwidth]{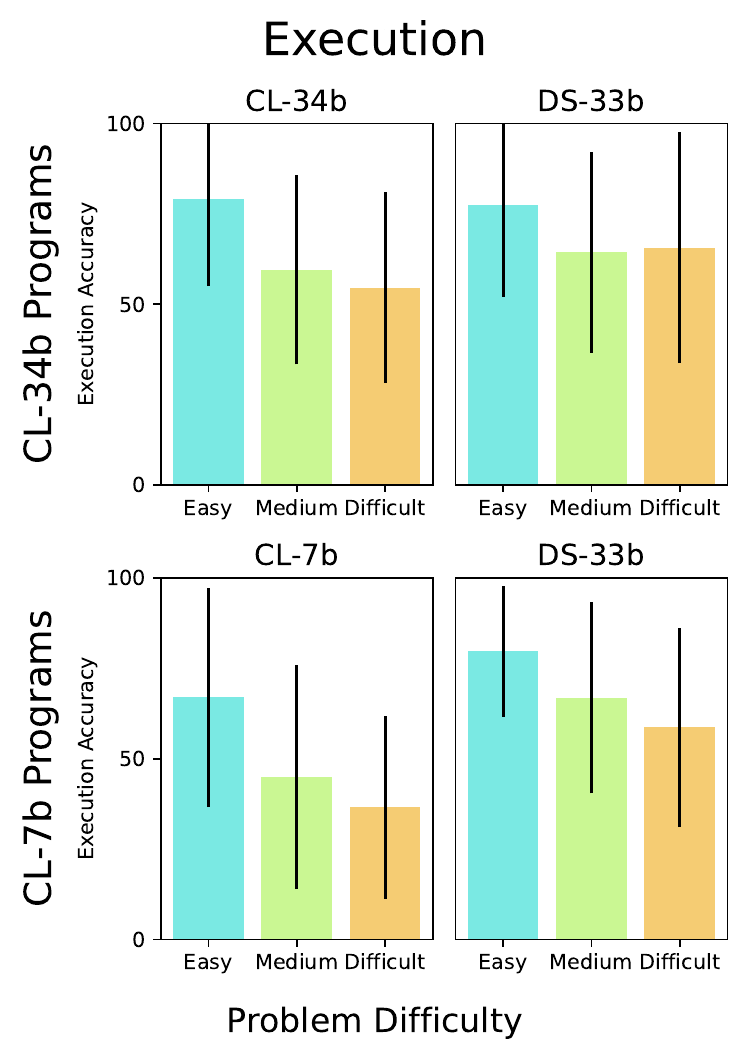}
        \caption{Execution}
    \end{subfigure}
    \hfill
    \begin{subfigure}[b]{0.31\textwidth}
        \centering
        \includegraphics[width=\textwidth]{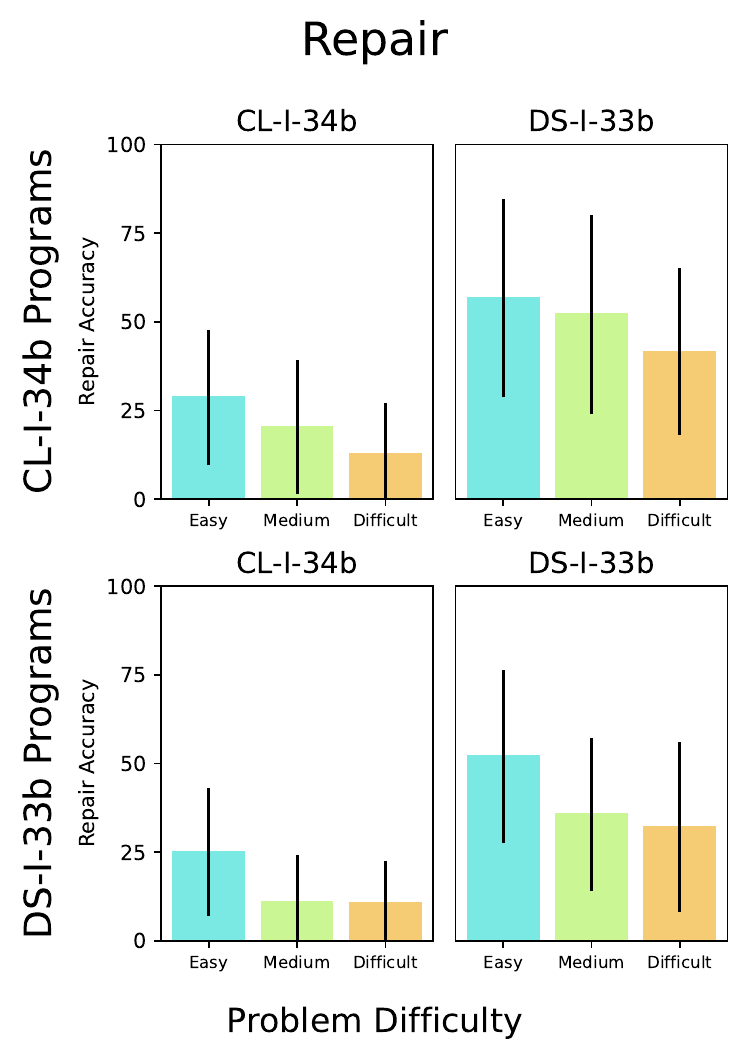}
        \caption{Repair}
    \end{subfigure}
    \caption{Mean accuracy of correctness checking, execution, and repair on HumanEval (error bars are 1 std). A problem's difficulty is determined by the pass@1 of the evaluation model. In general, correctness checking accuracy is uncorrelated with problem difficulty, while both execution and repair accuracies are only very weakly correlated.}
    \label{fig:difficulty-sample}
\end{figure*}

Intuitively, if a given programming problem is easy for a model to solve (meaning it has a high pass@1 rate), we might believe models understand how to solve that problem. If that is the case, then models should be able to better understand both correct and counterfeit samples for that problem. To test if this is true, we bucket problem difficulties into easy, medium, and hard by a model's pass@1 on that problem. We then calculate the average correctness checking, verification, and repair accuracy for each of the tasks for each problem difficulty bucket. In Fig. \ref{fig:difficulty-sample}, we show a subset of these results on HumanEval; 
full results for verification and execution are given in Sec~\ref{appendix:difficulty-correctness-full} and \ref{appendix:difficulty-execution-full}.

Surprisingly, we find that 1) correctness checking accuracies are relatively uncorrelated with problem difficulty, while 2) execution ability and the success rate of repair exhibit a modest amount of correlation with problem difficulty.
We find that these trends are generally robust across HumanEval, LeetCode, and ODEX, although the amount of correlation exhibited in the repair task varies (but is, at best, modest).

\begin{figure}[H]
\centering
\includegraphics[width=0.75\columnwidth]{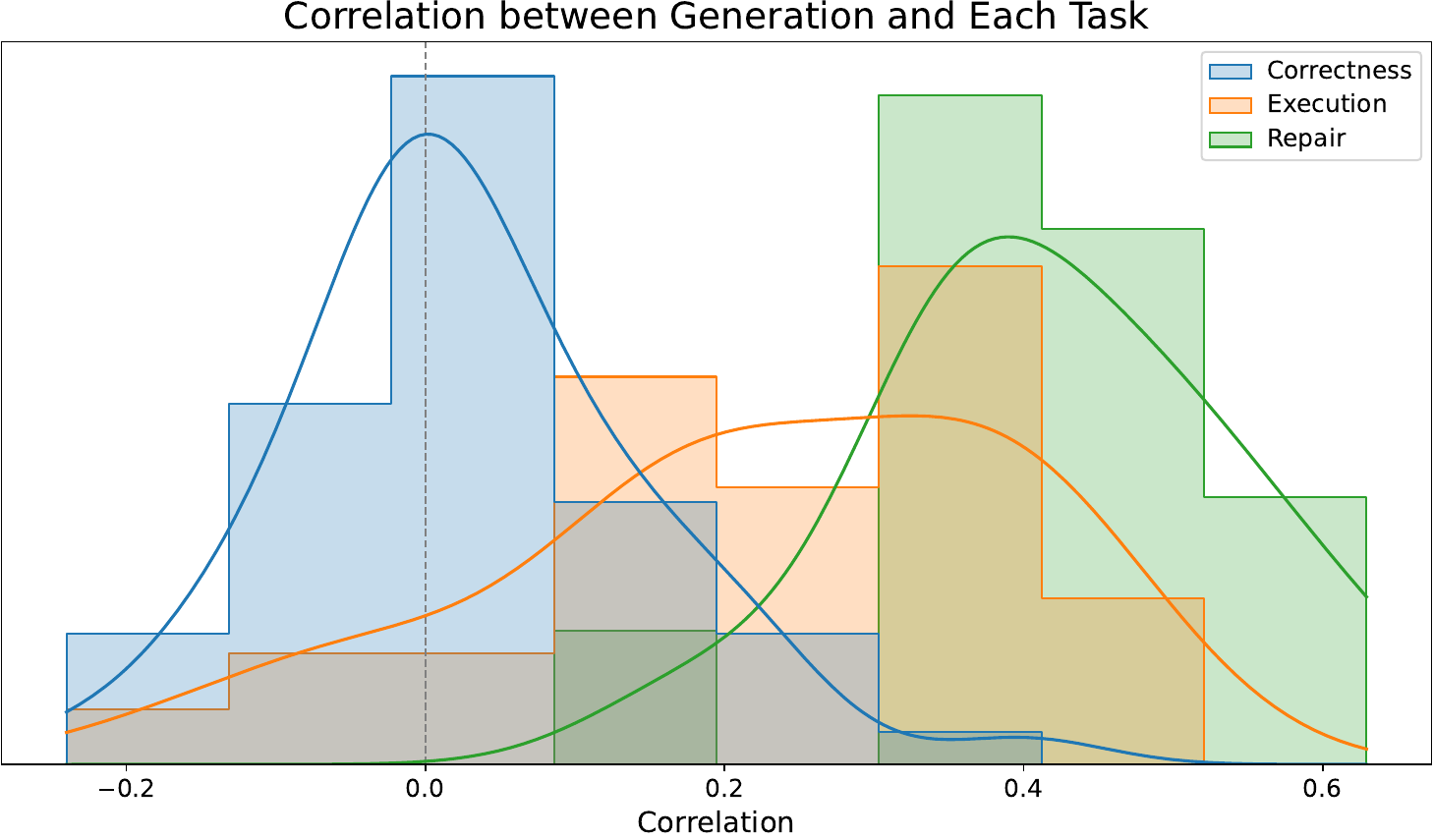}
\caption{Correlation between generation and each of our three tasks}
\label{fig:correlation-with-generation}
\end{figure}

To get a more precise measurement of these relationships, we calculated the Pearson correlation between generation performance and each of correctness checking, execution prediction, and repair performance across problems. The resulting histogram showing the distribution of correlations for each task is shown in Fig. \ref{fig:correlation-with-generation}.

\subsection{Do models perceive their own samples differently?}
\begin{figure}[H]
    \centering
    \hspace*{\fill}
    \begin{subfigure}[b]{0.4\columnwidth}
        \centering
        \includegraphics[width=\textwidth]{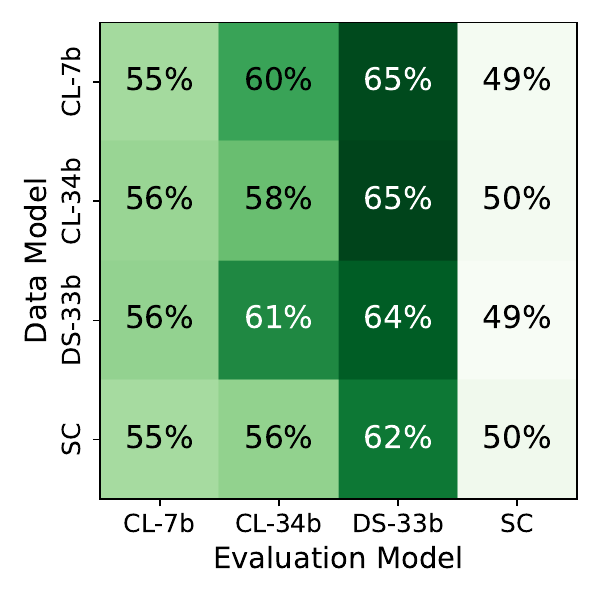}
        \subcaption{Correctness (ODEX)}
    \end{subfigure}
    \hspace*{\fill}
    \begin{subfigure}[b]{0.4\columnwidth}
        \centering
        \includegraphics[width=\textwidth]{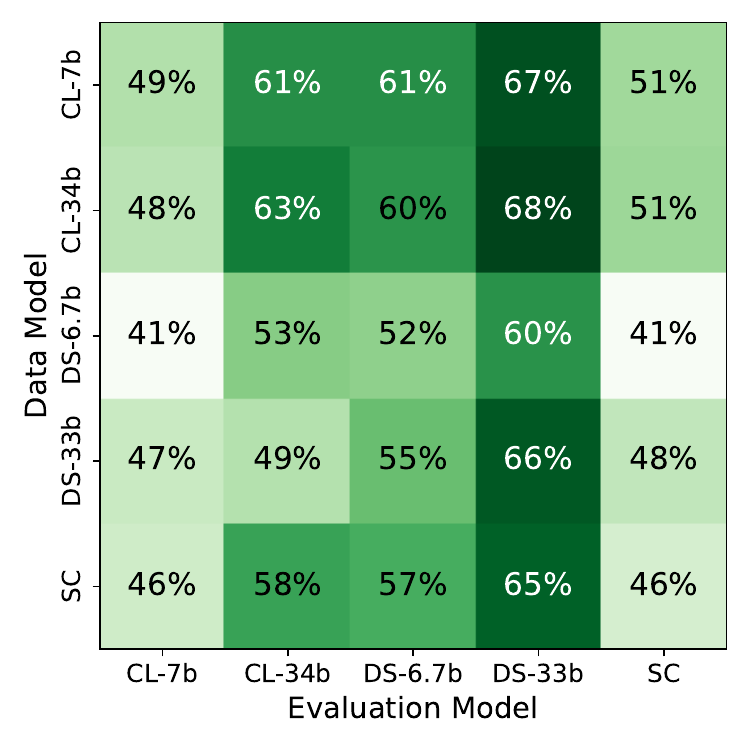}
        \subcaption{Execution (HumanEval)}
    \end{subfigure}
    \hspace*{\fill}
    \caption{Heatmap of accuracies for correctness checking and execution prediction. Models do not seem to perceive their own generations differently from those of other models.}
    \label{fig:self-understanding-heatmaps}
\end{figure}

For a given model, its counterfeit samples had a high enough log-likelihood to be generated by the model, so one may hypothesize that models might have a harder time than other models at distinguishing their own counterfeit samples. In Fig. \ref{fig:self-understanding-heatmaps}, we plot heatmaps showing the performance of various models on datasets generated by other models for the correctness checking task (left) and execution prediction task for HumanEval (right). For both tasks, the relative performance of different models is similar across datasets, we find no evidence that models falter more on their own samples. This suggests that counterfeit samples may be general: those from one model are generally difficult for other models to understand as well.

\subsection{Do stronger models generate harder counterfeit samples?}
\begin{figure}[ht!]
    \centering
    \begin{subfigure}[b]{0.49\columnwidth}
        \centering
        \includegraphics[width=\textwidth]{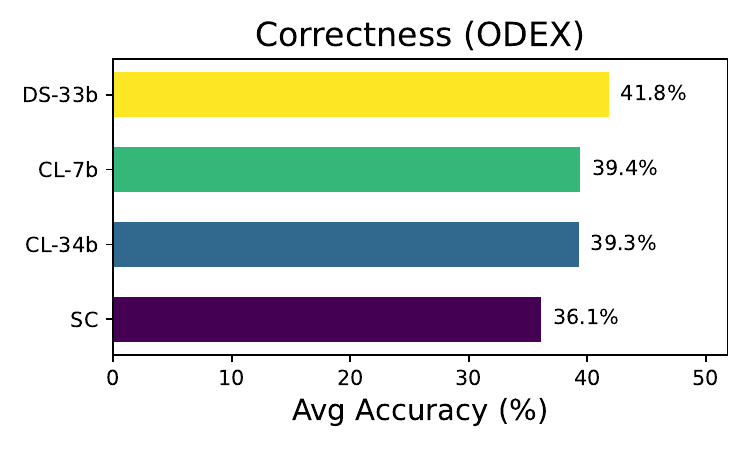}
        \subcaption{Correctness (ODEX)}
    \end{subfigure}
    \hfill
    \begin{subfigure}[b]{0.49\columnwidth}
        \centering
        \includegraphics[width=\textwidth]{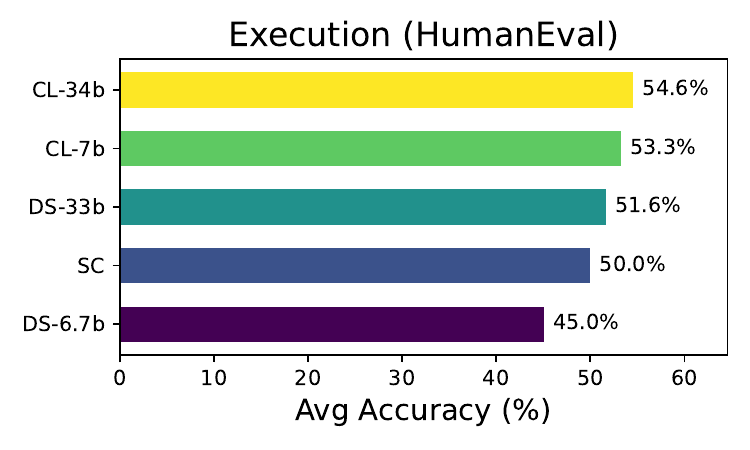}
        \subcaption{Execution (HumanEval)}
    \end{subfigure}
    \caption{Accuracy of correctness and execution tasks on counterfeit samples. Both strong and weak models are able to generate difficult counterfeit samples.}
    \label{fig:average-counterfeit-accuracy}
\end{figure}

One might also expect that counterfeit samples of stronger models are harder to verify than those of weaker models, as stronger models are less likely to generate obvious mistakes. In Fig. \ref{fig:average-counterfeit-accuracy}, we compare the average scores of two tasks on counterfeit samples for datasets generated by stronger (DS-33B, CL-34B) and weaker (CL-7B, SC) models. Note that this is the same as Fig. \ref{fig:self-understanding-heatmaps} with each row aggregated and filtered to only include counterfeit samples. Since there does not seem to be a significant difference between the difficulties, models of all strengths can be used to generate counterfeit samples that are challenging for models to understand.

\section{Qualitative Analysis} \label{sec:qual-analysis}
In general, most counterfeit programs fall into one of three broad categories: (1) error in algorithmic design or implementation, (2) incorrectly understanding or completely ignoring details in the specification, (3) failing to address corner cases in the input space. Examples of each are provided in Appendix~\ref{appendix:qual-sample-categories}.

\subsection{Failure Modes on Correctness Checking}
In Sec. \ref{sec:understanding-counterfeit}, we saw that GPT-4 is significantly more performant than other models at both correctness checking and execution prediction. Yet, we still find a significant fraction of failures. In this section, we take a closer look at the performance of GPT-3.5 and GPT-4 to understand where today's cutting-edge models still fall short. Through a manual inspection of examples, we uncover three main failure modes for verifying the correctness of counterfeit samples. Here, we present one example of each, deferring more to Appendix~\ref{appendix:qual-sample-mistakes}. For conciseness and readability, some of the problem statements have been altered from their original form.


\textbf{1) The model does not catch misunderstood or ignored details in the specification}: Sometimes, the verification model doesn't catch important specification details that are misunderstood or ignored by the counterfeit sample. 




For example, GPT-3.5 fails to find the reasoning error in the below counterfeit sample because it appears to have misunderstood the definition of ``corresponding closing bracket''. However, when asked to implement a function for the specification, it correctly understands that the corresponding closing bracket must come after the opening bracket, highlighting an inconsistency between generation and correctness checking.

\begin{lstlisting}[language=python, frame=single, ]
def correct_bracketing(brackets: str):
    """ brackets is a string of "<" and ">". return True if every opening bracket has a corresponding closing bracket.
    >>> correct_bracketing("<<><>>")
    True
    >>> correct_bracketing("><<>")
    False
    """
    opening = brackets.count("<")
    closing = brackets.count(">")
    return opening == closing
\end{lstlisting}

\begin{lstlisting}[style=reasoning, frame=single]
GPT-3.5: The code appears to be correct. It counts the number of opening brackets ("<") and closing brackets (">") in the input string and returns True if the counts are equal, indicating that every opening bracket has a corresponding closing bracket.
\end{lstlisting}

\textbf{2) The model fails to catch subtle implementation mistakes}: Often, counterfeit samples follow a generally correct algorithm design but contain a subtle implementation mistake that models do not catch. Below, \verb|if num = target| check should consider whether the remaining digits in \verb|square[index:]| sum to 0 for this implementation to be correct, but GPT-4 misses this subtle detail.

\begin{lstlisting}[language=python, frame=single]
"""Given a positive integer n, return the punishment number of n, defined as the sum of the squares of all integers i such that 1) 1 <= i <= n and 2) The decimal representation of i * i can be partitioned into contiguous substrings such that the sum of the integer values of these substrings equals i."""

def punishmentNumber(n: int) -> int:
    def check(num, target, index, square):
        if num == target:
            return True
        if index == len(square):
            return False
        cur = 0
        for i in range(index, len(square)):
            cur = cur * 10 + int(square[i])
            if cur > target:
                break
            if check(num + cur, target, i + 1, square):
                return True
        return False

    res = 0
    for i in range(1, n + 1):
        square = str(i * i)
        if check(0, i, 0, square):
            res += i * i
    return res

# GPT-4 CoT verdict: Correct
\end{lstlisting}

\textbf{3) The model forgives an identified error}: A third failure mode occurs when the model correctly reasons that the code is incorrect, but then makes up a justification for the error and judges the code as correct. In the counterfeit sample below, the code fails to consider capital consonants. While GPT-4 correctly identifies this issue, it proceeds to state that assuming that concern is not a real issue, the code is correct. However, the implementation fails one of the examples.

\begin{lstlisting}[language=python, frame=single]
def select_words(s, n):
    """Given a string s and a natural number n, return a list of all words from string s that contain exactly n consonants, in the order these words appear in the string s.
    Examples:
    select_words("Mary had a little lamb", 4) ==> ["little"]
    select_words("Mary had a little lamb", 3) ==> ["Mary", "lamb"] 
    """
    # convert string to list
    word_list = s.split()
    consonants = "bcdfghjklmnpqrstvwxyz"
    # filter out all words that contain n consonants
    return [word for word in word_list if sum(c in consonants for c in word) == n]
\end{lstlisting}

\begin{lstlisting}[style=reasoning, frame=single]
GPT-4: ... However, there is a potential issue with case sensitivity. The `consonants` string only contains lowercase consonants, and the program does not explicitly convert the input string `s` or the words to lowercase before checking if a character is a consonant. This means that if the input string contains uppercase letters, the program might not count them as consonants, leading to incorrect results. [...] the!* program correctly implements the functionality described in the docstring under the assumption that the input string contains only lowercase letters or that case sensitivity is not a concern*!...
\end{lstlisting}

\textbf{Execution Prediction}: To conclude this section, we highlight an error in execution prediction. In the example below, even though the statement \texttt{if ")" in brackets} is clearly true, GPT-4 is biased by the intended functionality of the program and does not follow its actual execution semantics. More examples can be found in Appendix \ref{appendix:gpt4-execution-errors}.

\begin{lstlisting}[language=Python, frame=single]
def correct_bracketing(brackets: str):
    if ")" in brackets:
        return False
    open_brackets = 0
    for bracket in brackets:
        if bracket == "(":
            open_brackets += 1
        else:
            open_brackets -= 1
            if open_brackets < 0:
                return False
    return open_brackets == 0
assert correct_bracketing('()') == ??
# Correct Answer: False
\end{lstlisting}
\begin{lstlisting}[style=reasoning, frame=single]
GPT-4 Execution: [...] 2. The function is called with the argument "()".
!* 3. The first if condition checks if ")" is in brackets. Since it is, but only after "(", this condition does not lead to a return of False at this point [...]
\end{lstlisting}

\section{Related Work}
\paragraph{Language models as judges and verifiers}
LLMs are increasingly explored for their potential in evaluation and verification tasks in various domains. In open-ended question answering, \citet{zheng2023judging} demonstrate that models like GPT-4 align with human preferences, indicating their potential as tools for assessing LLM-generated responses. 

In domains like mathematics and commonsense reasoning, various forms of automated LLM feedback have led to improved reasoning abilities \citep{cobbe2021training, zhou2023solving, weng2022large, lightman2023let, chen2023universal,wang2023math,shao2024deepseekmath}. 
Additionally, Studies like \citep{schneider2023towards, matelsky2023large} investigate LLMs as auto-graders or judges for educators.
On the other hand, \citet{valmeekam2023can, tyen2023llms, stechly2023gpt, chen2024tree} find that LLMs can struggle to find when they are wrong and that critiquing plans could harm performance.

\paragraph{Evaluation and verification for code synthesis}
The challenge of LLMs producing incorrect code in response to natural language prompts has led to a significant focus on automated evaluation and verification of generated code samples. Various studies have demonstrated that postprocessing the samples from LLMs can substantially enhance the accuracy of the system~\citep{chen2022codet, ridnik2024code, key2022speak, zhang2023algo, li2022competition, huang2023enhancing}.

Also, \citet{inala2022fault,zhang2023coder,ni2023lever} have employed a neural model to verify code samples, with the aim of ranking more accurate codes higher.

\paragraph{Code understanding in language models}
Many benchmarks evaluate aspects of code understanding and code intelligence such as code summarization \citep{iyer2016summarizing, hasan2021codesc}, commit message generation \citep{liu2020atom}, code comprehension \citep{singhal2024nofuneval}, clone detection \citep{lu2021codexglue}, code question answering \citep{sahu2022learning}, and code explaining \citep{muennighoff2023octopack}. Neural-based code execution has been studied in \citep{austin2021program, nye2021show, gu2024cruxeval, la2024code}, and code repair has been studied in \citep{madaan2023self, chen2024teaching, zhang2023self, olausson2024repair}, and \citet{liu2024codemind} examine a suite of code reasoning benchmarks.

A few controlled studies highlight the extent to which language models understand code. For example, code generation abilities have been shown to drop after syntactic changes like identifier swaps \citep{miceli2023larger} and semantic changes like 1-indexing \citep{wu2023reasoning}. \citet{dinh2024large} show that models fail at completing code with bugs. \citet{jin2023evidence} provide evidence that LMs can learn meaningful representations when trained on programs, \citet{zhang2023can} explore the behavior of transformers to simulate recursive functions, and \citet{min2023beyond} discover that code language models are inconsistent on various coding tasks.

\paragraph{Models understanding their own generations}
Some recent works investigate the extent to which models understand their generations. \citet{huang2023large, chen2024teaching, tyen2023llms, olausson2024repair} find that LLMs struggle to find their own reasoning errors, but are able to correct them with adequate external feedback. \citet{singhal2024nofuneval} discover that models are better at fixing buggy code than distinguishing between correct and buggy code. Relevant to our work, \citet{west2023generative} and \citet{oh2024generative} argue that generative capability may not be contingent on understanding capability in textual domains.
\section{Conclusion}
In this work, we bring attention to the \textit{counterfeit samples} of a code language model: incorrect programs that a model thinks are correct and can pass surface-level correctness checks. We observe that in a sense, these counterfeit samples are adversarial to the model: models often cannot assess their correctness, reason about their execution, and struggle to repair them. Compared to other models, GPT-4 may be different from other evaluated models in this regard, in that they are much less susceptible to the traps we observe on counterfeit samples from other models.

While we operate in the domain of code, where it is simple to precisely check a model's understanding, we suspect that the same phenomena occur more generally in language models, which is consistent with the findings from \citet{west2023generative}. Because models being able to understand their own counterfeit samples is a prerequisite to strong self-repair and self-verification schemes, we recommend that others be critical and careful in light of our findings.

\section{Limitations}
We identify a few limitations below:

\textbf{Vague labels for HumanEval samples}: We find that HumanEval specifications can often be vague with the inputs and outputs that are tested on. Therefore, some programs can be argued to be either correct or counterfeit. When manually inspecting programs and their scores, we find that base tests are too weak while EvalPlus tests are too strong. Therefore, for correctness, we use the criteria that the program must pass all base samples and at least 95\% of EvalPlus samples. However, this only affects a small fraction of samples and we do not believe changes any of our main claims (which are also supported by LeetCode and ODEX). 

\textbf{Filter for counterfeit samples}: In this work, we use a relatively liberal filter for counterfeit samples that consists of mostly basic syntax and/or correctness checks. While we believe our results would hold for slight alterations of our filter, we do not assess this.

\textbf{Nature of counterfeit samples}: The scope of this work is limited to counterfeit samples that are generated by sampling from a natural language description. It is unclear how these samples differ from human-written incorrect samples or samples constructed in a different way, for example by synthetically injecting bugs into correct samples as in HumanEvalFix \citep{muennighoff2023octopack}.

\textbf{Dataset and prompting variation}: While we make a best-effort attempt to use standardized prompts that lead to the best performance, evaluation has been found to be quite sensitive to the prompt and task format \citep{mizrahi2023state}. In addition, there is variation across the datasets generated by various models. We try to mitigate this by showcasing that our conclusions remain robust across a variety of datasets and models. 

\textbf{Other perspectives on code understanding}: Although the three tasks we evaluate capture important aspects of code understanding, our claims do not necessarily extrapolate to other aspects of code understanding such as code summarization, translation, or optimization. We believe that other dimensions of code understanding are equally important and encourage future evaluation beyond the tasks we present here.

\textbf{Limited results for GPT-3.5 and GPT-4}: All our counterfeit samples are generated from CodeLlama, DeepSeekInstruct, and StarCoder, so it is unknown whether the same insights apply to GPT-3.5 and GPT-4 counterfeits. In addition, due to budget constraints, we only evaluate these two models on a limited subset of our counterfeit datasets, decreasing the statistical significance of our results on these models. 
\section{Acknowledgements}
In alphabetical order, we thank Ansong Ni, Dylan Zhang, Federico Cassano, and Terry Yue Zhao for feedback during various stages of the project.

A. Gu is supported by the National Science Foundation (NSF) Graduate Research Fellowship under Grant No. 2141064. 
N. Jain is supported by NSF grants CCF:1900968, CCF:1908870, and by SKY Lab industrial sponsors and affiliates. 
T.X. Olausson is supported by the Defense Advanced Research Projects Agency (DARPA) under the ASKEM program, award HR00112220042. 
A. Solar-Lezama is supported by the National Science Foundation (NSF) and Intel Corporation through NSF Grant CCF:2217064.

\bibliography{custom}
\bibliographystyle{plainnat}

\appendix
\raggedbottom

\section{Detailed Experimental Setup} \label{appendix:detailed-experimental-setup}
\subsection{Correct and Counterfeit Data Generation for Correctness Checking}
In Table \ref{tab:dataset-sizes-correctness-checking}, we show statistics about the datasets used for correctness checking. Recall that each dataset consists of 5 correct and 5 counterfeit samples per problem. We also show the average pass@1 score of problems in the dataset at $T=0.6$. A few examples of correct and counterfeit samples are shown in Listings \ref{lst:dataset-humaneval-example}, \ref{lst:dataset-leetcode-example}, and \ref{lst:dataset-odex-example}.

\begin{table}[H]
    \centering
    \caption{Correctness Checking Dataset Sizes}
    \begin{tabular}{cccc}
        \toprule
        \textbf{Dataset} & \textbf{Model} & \textbf{Pass@1} & \textbf{Size} \\
	\midrule
	\multirow{5}{*}{HumanEval}
	& CL-34b & 42.0 & 850 \\
	& CL-7b & 36.9 & 870 \\
	& DS-I-33B & 45.1 & 830 \\
	& StarCoder & 32.9 & 660 \\
	& DS-I-6.7B & 56.7 & 810 \\
        \midrule
	\multirow{2}{*}{LeetCode}
	& DS-I-33B & 49.4 & 460 \\
	& DS-I-6.7B & 37.3 & 360 \\
	\midrule
	\multirow{5}{*}{ODEX}
	& CL-34B & 49.2 & 1070 \\
	& CL-7B & 52.2 & 1190 \\
	& DS-I-33B & 44.3 & 520 \\
	& StarCoder & 46.2 & 1060 \\
	& CL-13B & 50.1 & 1090 \\
    \bottomrule
    \end{tabular}
    \label{tab:dataset-sizes-correctness-checking}
\end{table}

\textbf{HumanEval}: HumanEval \citep{chen2021evaluating} is a dataset of 164 relatively simple natural language to programming problems in Python. We sample $200$ generations at $T=0.6$. We use both the original HumanEval tests and EvalPlus tests, which are more comprehensive \citep{liu2023your}. In order to filter out trivial solutions and keep the task interesting, we only consider a counterfeit sample to be a program with an EvalPlus score of over 10\% and manually inspect the resulting dataset. On manual inspection, we found that EvalPlus tests can sometimes be too strong and filter out correct solutions due to very subtle errors like precision and floating point issues, we consider a solution as correct if it passes all the base tests and at least 95\% of EvalPlus tests. Our manual inspection shows that this is a fairer criteria for assessing program correctness.

\textbf{LeetCode}: LeetCode is a dataset of 130 LeetCode problems used for programming interviews. We sample $200$ generations at $T=0.6$. Unlike in HumanEval, run-time and computational complexity is an important consideration for many LeetCode problems. However, determining whether a program can finish within the time limit can be difficult. Therefore, we consider counterfeit programs to be those that resulted in a ``Correct'' or ``Wrong Answer'' verdict, and remove programs that received a ``Runtime Error'' and ``Time Limit Exceeded''. 

\textbf{ODEX}: ODEX \citep{wang2022execution} is an open-domain, multilingual, execution-based natural language to code generation benchmark. We only use the English subset of 479 problems. Unlike HumanEval and LeetCode, ODEX contains problems using a wide variety of Python library functions such as \texttt{numpy}, \texttt{os}, and \texttt{pandas}. We sample $50$ generations at $T=0.6$. As ODEX does not come with cleanly separated test cases, we consider a program as counterfeit if it can be parsed successfully by \texttt{ast.parse} and have a length of under 500 characters. From manual inspection, some of the problems in ODEX can be quite vague, making it difficult to discern if a solution is correct or counterfeit without seeing the input-output format. To mitigate this, for ODEX only, we include \textit{both the generated program and the assertions that is checked}. This leads to a slightly easier setting than the previous two tasks, but we find that it still poses a significant challenge for models.

\begin{lstlisting}[language=python,caption={Example of counterfeit sample for HumanEval, generated by CL-34B},style=boxed,label={lst:dataset-humaneval-example}]
def sort_even(l: list):
    """This function takes a list l and returns a list l' such that
    l' is identical to l in the odd indicies, while its values at the even indicies are equal
    to the values of the even indicies of l, but sorted.
    >>> sort_even([1, 2, 3])
    [1, 2, 3]
    >>> sort_even([5, 6, 3, 4])
    [3, 6, 5, 4]
    """
    even_nums = l[::2]
    even_nums.sort()
    odd_nums = l[1::2]
    ans = []
    for i in range(len(even_nums)):
        ans.append(even_nums[i])
        ans.append(odd_nums[i])
    return ans
\end{lstlisting}

\begin{lstlisting}[language=python,caption={Example of correct sample for LeetCode, generated by DS-I-33B},style=boxed,label={lst:dataset-leetcode-example}]
"""
You are given a 0-indexed permutation of n integers nums.
A permutation is called semi-ordered if the first number equals 1 and the last number equals n. You can perform the below operation as many times as you want until you make nums a semi-ordered permutation:

Pick two adjacent elements in nums, then swap them.

Return the minimum number of operations to make nums a semi-ordered permutation.
A permutation is a sequence of integers from 1 to n of length n containing each number exactly once.
 
Example 1:

Input: nums = [2,1,4,3]
Output: 2
Explanation: We can make the permutation semi-ordered using these sequence of operations: 
1 - swap i = 0 and j = 1. The permutation becomes [1,2,4,3].
2 - swap i = 2 and j = 3. The permutation becomes [1,2,3,4].
It can be proved that there is no sequence of less than two operations that make nums a semi-ordered permutation. 

Example 2:

Input: nums = [2,4,1,3]
Output: 3
Explanation: We can make the permutation semi-ordered using these sequence of operations:
1 - swap i = 1 and j = 2. The permutation becomes [2,1,4,3].
2 - swap i = 0 and j = 1. The permutation becomes [1,2,4,3].
3 - swap i = 2 and j = 3. The permutation becomes [1,2,3,4].
It can be proved that there is no sequence of less than three operations that make nums a semi-ordered permutation.

Example 3:

Input: nums = [1,3,4,2,5]
Output: 0
Explanation: The permutation is already a semi-ordered permutation.

 
Constraints:

2 <= nums.length == n <= 50
1 <= nums[i] <= 50
nums is a permutation.
"""

def semiOrderedPermutation(nums: List[int]) -> int:
    n = len(nums)
    min_idx = nums.index(1)
    max_idx = nums.index(n)
    
    if min_idx < max_idx:
        return min_idx + n - 1 - max_idx
    else:
        return min_idx + n - 2 - max_idx
\end{lstlisting}

\begin{lstlisting}[language=python,caption={Example of counterfeit sample for ODEX, generated by DS-I-33B},style=boxed,label={lst:dataset-odex-example}]
def f_41861705(split_df, csv_df):
    """Create new DataFrame object by merging columns "key" of  dataframes `split_df` and `csv_df` and rename the columns from dataframes `split_df` and `csv_df` with suffix `_left` and `_right` respectively
    """
    return split_df.merge(csv_df, how='left', left_on='key', right_on='key', suffixes=('_left', '_right'))

import pandas as pd

def check(candidate):
    split_df = pd.DataFrame({'key': ['foo', 'bar'], 'value': [1, 2]})
    csv_df = pd.DataFrame({'key': ['foo', 'baz'], 'value': [3, 4]})
    result = pd.DataFrame({'key': ['foo'], 'value_left': [1],'value_right': [3]})
    assert all(candidate(csv_df, split_df) == result)

check(f_41861705)
\end{lstlisting}

\subsection{Data Generation for Execution Prediction}
We perform code execution experiments on HumanEval and LeetCode programs. The inputs and outputs for these datasets are primitive Python objects (mostly \texttt{int}, \texttt{str}, \texttt{bool}, \texttt{list}). While it is possible, we do not evaluate execution for ODEX because many of the programs involve file modifications and cannot easily be represented. For each dataset and data-generating model, we use the same set of programs used in the correctness checking experiment for consistency. As of today, we cannot expect a language model to follow the execution of arbitrary Python programs. Therefore, we ensure that the execution samples in our benchmark are reasonable by applying a filter following the setup in \citep{gu2024cruxeval}. One key difference from their work is that instead of using arbitrary programs, the programs we use here are seeded from a natural language specification and are semantically meaningful. This allows us to analyze how models behave differently when asked to reason about correct and counterfeit programs.

We create our dataset of samples to evaluate code execution as follows: first, we take the programs generated for the correctness checking dataset. The docstring containing the problem statement is stripped away to force the model to use the provided code. Second, we run the program on the tests provided in the original problem statement and examples, which are generally simple and concise to create a large set of model-generated programs, inputs, and outputs. Third, we apply a compile-time and runtime based filter using Python bytecode to remove programs that are too long, require complex arithmetic/floating point operations, and have too many steps in the execution. The final step is a manual inspection of programs, inputs, and outputs passing the filter to ensure that they seem reasonable. The resulting dataset sizes are shown in Table \ref{tab:dataset-sizes-execution}, and examples are shown in Listings \ref{lst:dataset-humaneval-execution-example}, \ref{lst:dataset-leetcode-execution-example}.

\begin{table}[H]
    \centering
    \caption{Execution Dataset Sizes}
    \begin{tabular}{ccc}
        \toprule
        \textbf{Dataset} & \textbf{Model} & \textbf{Dataset Size} \\
	\midrule
	\multirow{5}{*}{HumanEval}
	& CL-34B & 1406 \\
	& CL-7B & 1528 \\
	& DS-I-33B & 1964 \\
	& StarCoder & 1622 \\
	& DS-I-6.7B & 1917 \\
        \midrule
	\multirow{2}{*}{LeetCode}
	& DS-I-33B & 845 \\
	& DS-I-6.7B & 694 \\
    \bottomrule
    \end{tabular}
    \label{tab:dataset-sizes-execution}
\end{table}

\begin{lstlisting}[language=python,caption={Example of HumanEval execution prediction example, generated by StarCoder},style=boxed,label={lst:dataset-humaneval-execution-example}]
from typing import List


def string_xor(a: str, b: str) -> str:
    assert len(a) == len(b)
    res = ""
    for i in range(len(a)):
        if a[i] == b[i]:
            res += "0"
        else:
            res += "1"
    return res
assert string_xor('1', '1') == ??
# Answer: '0'
\end{lstlisting}

\begin{lstlisting}[language=python,caption={Example of LeetCode execution prediction example, generated by DS-I-6.7B},style=boxed,label={lst:dataset-leetcode-execution-example}]
def relocateMarbles(nums: List[int], moveFrom: List[int], moveTo: List[int]) -> List[int]:
    # Create a dictionary to store the number of marbles at each position
    marbles = {}
    for num in nums:
        marbles[num] = marbles.get(num, 0) + 1

    # Apply the moves
    for f, t in zip(moveFrom, moveTo):
        # Remove the marbles at the source position
        count = marbles.pop(f)
        # Add the marbles at the target position
        marbles[t] = marbles.get(t, 0) + count

    # Return the sorted keys of the dictionary
    return sorted(marbles.keys())
assert relocateMarbles(nums = [1, 6, 7, 8], moveFrom = [1, 7, 2], moveTo = [2, 9, 5]) == ??
# Answer: [5, 6, 8, 9]
\end{lstlisting}

\section{Models, Task Evaluation, and Prompts}\label{appendix:task-evaluation-prompts}
\subsection{Models}
We use DeepSeek \citep{guo2024deepseek}, CodeLlama \citep{roziere2023code}, and StarCoder \citep{li2023starcoder} models. The HuggingFace URLs are listed in Table \ref{tab:model-url-list}. Experiments were run on A100 (80 GB) and A6000 (40 GB) machines.
\begin{table}[h]
    \centering
    \small
    \caption{Model Links}
    \begin{tabular}{|ll|}
        \hline
        \textbf{Model Name} & \textbf{HuggingFace URL} \\
        \hline
        DeepSeek Instruct (6.7B) & \url{https://huggingface.co/deepseek-ai/deepseek-coder-6.7b-instruct} \\
        DeepSeek Instruct (33B) & \url{https://huggingface.co/deepseek-ai/deepseek-coder-33b-instruct} \\
        StarCoder (15.5B) & \url{https://huggingface.co/bigcode/starcoder} \\
        CodeLlama (7B) & \url{https://huggingface.co/codellama/CodeLlama-7b-hf} \\
        CodeLlama (13B) & \url{https://huggingface.co/codellama/CodeLlama-13b-hf} \\
        CodeLlama (34B) & \url{https://huggingface.co/codellama/CodeLlama-34b-hf} \\
        CodeLlama Instruct (34B) & \url{https://huggingface.co/codellama/CodeLlama-34b-Instruct-hf} \\
        \hline
    \end{tabular}
    \label{tab:model-url-list}
\end{table}

\subsection{Task Evaluation}
\label{sec:task-eval}
\textit{Correctness Checking}: For this task, we use an autoregressive-style CoT prompt from Listing \ref{lst:correctness-checking-prompt-cot}. We perform majority voting on the binary label (correct/incorrect) with $N=10$ samples and temperature $T=0.2$ and report accuracy on these labels. We do this because greedy decoding can be noisy for chain-of-thought prompting and majority voting has been shown to help \citep{wei2022chain, wang2022self}.

\begin{figure}[H]
    \centering
    \includegraphics[width=0.5\textwidth]{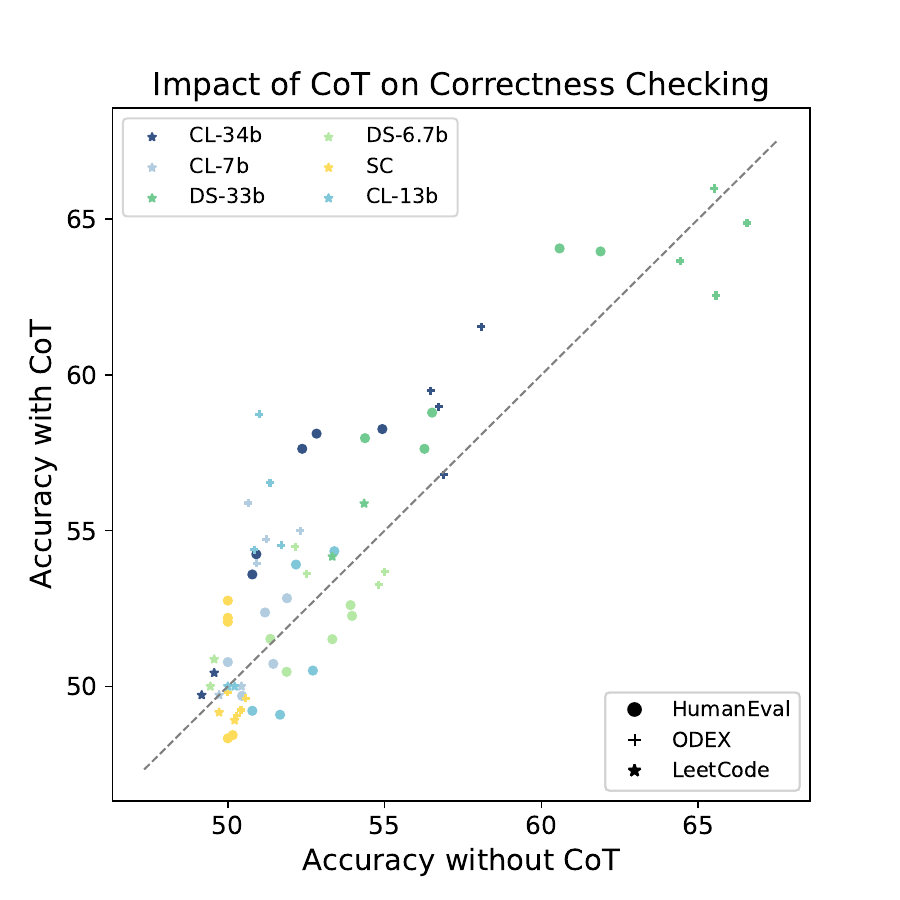}
    \caption{Models are slightly better when using CoT than without}
    \label{fig:correctness-checking-cot-impact}
\end{figure}

We also compared this with an autoregressive-style prompt without CoT, where the model is simply asked to predict Correct/Incorrect. In this case, we have the direct log-probabilities of each outcome $p_{\text{correct}}$ and $p_{\text{incorrect}} = 1 - p_{\text{correct}}$, so the predicted label is taken to be $p_{\text{correct}} \ge 0.5$. In Fig. \ref{fig:correctness-checking-cot-impact}, we observe that for a majority of settings and samples, CoT helps the accuracy of this task, motivating our use of CoT.

\textit{Execution Prediction}: For this task, we use the same prompt format as in \citep{gu2024cruxeval} with modified few-shot examples to better resemble our dataset format. We tested both CoT and non-CoT prompts, discovering that CoT did not help models other than GPT-3.5 and GPT-4. This is relatively consistent with the  results from \citet{gu2024cruxeval}\footnote{See their \href{https://crux-eval.github.io/leaderboard.html}{leaderboard}}, which only saw a 1.2\% improvement for Code Llama 34B and no improvement for Code Llama 13B. Therefore, we use CoT for GPT models, and non-CoT prompts for the others. The accuracy is calculated using pass@1 with $N=10, T=0.2$. 

\textit{Repair}: For this task, we base our prompt format on those employed in prior work by \citet{olausson2024repair}.
This prompt format is reminiscent of Chain-of-Thought in that it instructs the model to generate a textual explanation of what is wrong with the code, before generating the fixed version of the program.
Note that in our version of this prompt format, the model is not given any details as to what test test the program failed, and so has to relate the program to the natural language specification to debug it.
Unlike the other tasks, the prompt format we use for repair is zero-shot. 
Preliminary experiments indicated that this led to better results, particularly for smaller models which showed a tendency to debug the example program instead of the target. For the experiments with DeepSeek-based models, we replaced the HTML-style tags with Markdown-style tags (e.g., [PYTHON] $\to$ \`{}\`{}\`{}python).
Since repair requires generating a rather long answer, with both a textual explanation and a fixed version of the program, variance can be higher than in the other settings we consider.
To reduce this variance, we generate a large amount ($R=50$) of repair candidates for each counterfeit sample, using a temperature of $T=0.6$. We then average over all $5 \cdot 50 = 250$ samples to compute the mean success rate for each task, and also show a 95\% confidence interval on the mean.\footnote{Recall that our curated datasets contain 5 counterfeit samples per problem.}
Note that due to this increased computational burden, we do not carry out repair experiments for the full Cartesian product of models considered before, instead focusing on those open-source models that performed best on each dataset.

\subsection{Prompts}
\label{sec:prompts}
In this section, we list the HumanEval prompts. The prompts for other tasks can be found in our codebase \footnote{\url{https://github.com/update-after-deanonymization}}. Listings \ref{lst:correctness-checking-prompt-no-cot}, \ref{lst:correctness-checking-prompt-cot} show the correctness checking prompt without and with CoT, and Listings \ref{lst:execution-prompt-no-cot}, \ref{lst:execution-prompt-cot} show the execution prediction prompts. We give credit to \citet{gu2024cruxeval} and \citet{olausson2024repair} for their execution prediction and repair prompts.

\begin{lstlisting}[caption={Prompt for correctness checking (HumanEval)},label={lst:correctness-checking-prompt-no-cot}, style=boxed]
You will be given a Python coding problem with its specification and input/output examples in docstrings. 
Your goal is to determine whether the program exactly matches the specification. 
A correct program must be correct for all inputs, including hidden test cases not listed in the docstring.
In [ANSWER] and [/ANSWER] tags, write "Correct" if the program is correct, and "Incorrect" otherwise.

[PYTHON]
from typing import List


def is_at_least_zero(numbers: List[int]) -> bool:
    """ For a given list of numbers, checks if their sum is at least 0.
    >>> is_at_least_zero([1, -2, 3])
    True
    >>> is_at_least_zero([-1, -2, 2])
    False
    """
    return sum(numbers) >= 0
[/PYTHON]
[ANSWER]
Correct
[/ANSWER]

[PYTHON]
from typing import List


def identical_length(s: List[str]) -> bool:
    """ Check if in given list of strings, there exist two different
    strings with the same length.
    >>> identical_length(["abc", "def", "ghi"])
    True
    >>> identical_length(["abc", "gh"])
    False
    """
    for i in range(len(s)):
        for j in range(len(s)):
            if len(s[i]) == len(s[j]):
                return True
    return False
[/PYTHON]
[ANSWER]
Incorrect
[/ANSWER]
\end{lstlisting}

\begin{lstlisting}[caption={Prompt for correctness checking, CoT (HumanEval)},label={lst:correctness-checking-prompt-cot}, style=boxed]
You will be given a Python coding problem with its specification and input/output examples in docstrings. 
Your goal is to judge whether the program exactly matches the behavior specified in the docstring.
A correct program must be correct for all inputs, including hidden test cases not listed in the docstring.
First, think step by step about the program in [THOUGHT] and [/THOUGHT] tags.
In [ANSWER] and [/ANSWER] tags, write "Correct" if the program is correct, and "Incorrect" otherwise.

[PYTHON]
from typing import List


def is_at_least_zero(numbers: List[int]) -> bool:
    """ For a given list of numbers, checks if their sum is at least 0.
    >>> is_at_least_zero([1, -2, 3])
    True
    >>> is_at_least_zero([-1, -2, 2])
    False
    """
    return sum(numbers) >= 0
[/PYTHON]
[THOUGHT]
The function calculates the sum of a list of integers and checks if the sum is at least 0. The implementation uses Python's built-in sum function to calculate the total sum of the list and compares it with 0 using the >= operator. This approach should work correctly for any list of integers, including empty lists, where the sum would be 0, thus meeting the specification for all possible inputs.
[/THOUGHT]
[ANSWER]
Correct
[/ANSWER]

[PYTHON]
from typing import List


def identical_length(s: List[str]) -> bool:
    """ Check if in given list of strings, there exist two different
    strings with the same length.
    >>> identical_length(["abc", "def", "ghi"])
    True
    >>> identical_length(["abc", "gh"])
    False
    """
    for i in range(len(s)):
        for j in range(len(s)):
            if len(s[i]) == len(s[j]):
                return True
    return False
[/PYTHON]
[THOUGHT]
The program checks if any two strings in the list have the same length. However, it also compares each string with itself due to the loops' range, which means it will always find two strings (the same string compared with itself) with identical length, returning True incorrectly for any non-empty list. The correct approach should exclude the case where i equals j.
[/THOUGHT]
[ANSWER]
Incorrect
[/ANSWER]
\end{lstlisting}

\begin{lstlisting}[caption={Prompt for execution prediction (HumanEval)},label={lst:execution-prompt-no-cot}, style=boxed]
You are given a Python function and an assertion containing an input to the function.
Complete the assertion with a literal (no unsimplified expressions, no function calls) containing the output when executing the provided code on the given input.
Even if the function is incorrect or incomplete, give the output when executing the Python code as provided.
Assume all required imports have been included.
Do NOT output any extra information. Provide the full assertion with the correct output in [ANSWER] and [/ANSWER] tags, following the examples.

[PYTHON]
def add_one(number : int) -> int:
    return number + 2
assert add_one(17) == ??
[/PYTHON]
[ANSWER]
assert add_one(17) == 19
[/ANSWER]

[PYTHON]
def add_character_a(string : str) -> str:
    return string + "a"
assert add_character_a("x9j") == ??
[/PYTHON]
[ANSWER]
assert add_character_a("x9j") == "x9ja"
[/ANSWER]

[PYTHON]
{solution}
assert {input} == ??
[/PYTHON]
[ANSWER]
\end{lstlisting}

\begin{lstlisting}[caption={Prompt for execution prediction, CoT (HumanEval)},label={lst:execution-prompt-cot}, style=boxed]
You are given a Python function and an assertion containing an input to the function.
Complete the assertion with a literal (no unsimplified expressions, no function calls) containing the output when executing the provided code on the given input.
Even if the function is incorrect or incomplete, give the output when executing the Python code as provided.
Assume all required imports have been included. Think through the execution of the program in [THOUGHT] and [/THOUGHT] tags.
Provide the full assertion with the correct output in [ANSWER] and [/ANSWER] tags, following the examples.

[PYTHON]
def performOperation(s : str) -> str:
    s = s + s
    return "b" + s + "a"
assert performOperation("hi") == ??
[/PYTHON]
[THOUGHT]
Let's execute the code step by step:

1. The function performOperation is defined, which takes a single argument s.
2. The function is called with the argument "hi", so within the function, s is initially "hi".
3. Inside the function, s is concatenated with itself, so s becomes "hihi".
4. The function then returns a new string that starts with "b", followed by the value of s (which is now "hihi"), and ends with "a".
5. The return value of the function is therefore "bhihia".
[/THOUGHT]
[ANSWER]
assert performOperation("hi") == "bhihia"
[/ANSWER]

[PYTHON]
{solution}
assert {input} == ??
[/PYTHON]
[THOUGHT]
\end{lstlisting}

\begin{lstlisting}[caption={Prompt for (self-)repair (HumanEval)},label={lst:repair-prompt}, style=boxed]
=== system prompt ===
You are a helpful programming assistant and an expert Python programmer.
You are helping a user write a program.
The user has been given a function signature, along with a doc-string explaining its specification, and has then written an attempted implementation of the function.
Unfortunately, their code has some bugs and is not passing all of the hidden unit tests.

You will help the user by first giving a concise textual explanation of what is wrong with the code.
After you have pointed out what is wrong with the code, you will then generate a fixed version of the program.
Put your fixed program within code delimiters, for example:
[PYTHON]
# YOUR CODE HERE
[/PYTHON]
Do not change the function signature or doc-string in any way: they must be exactly as given by the user.

=== user prompt ===
### INCORRECT CODE
[PYTHON]
{code}
[/PYTHON]
The program does not pass all of the hidden test cases. Please fix it.
\end{lstlisting}

\section{Accuracy Results for All Tasks} \label{appendix:all-accuracy-results}
\subsection{Correctness Checking}
Fig. \ref{fig:correctness-checking-results-full} shows the full set of correctness checking results across all models for each of the three datasets.

\begin{figure}[H]
    \centering
    \begin{subfigure}[b]{0.65\textwidth}
        \includegraphics[width=\linewidth]{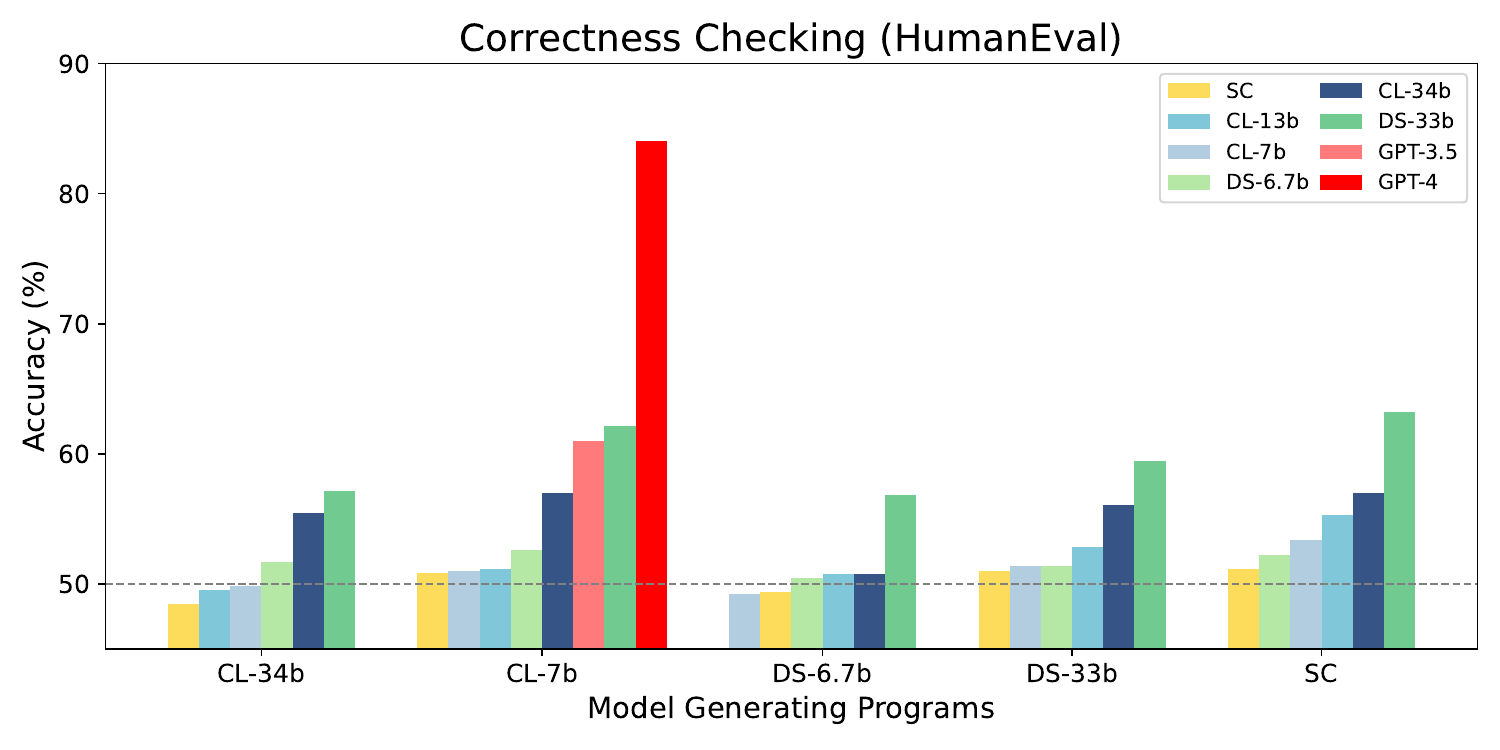}
        \subcaption{}
    \end{subfigure}
    \begin{subfigure}[b]{0.33\textwidth}
        \includegraphics[width=\linewidth]{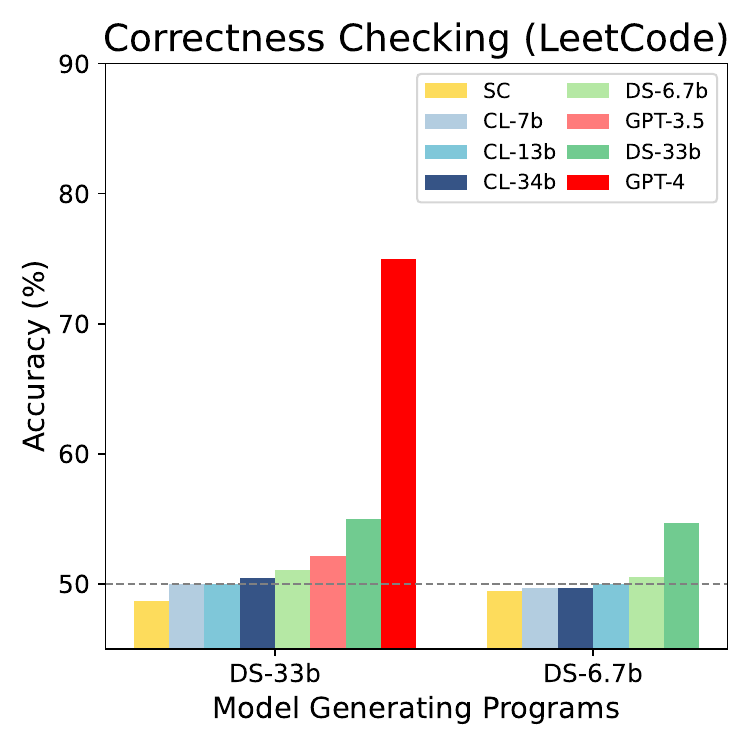}
        \subcaption{}
    \end{subfigure}
\end{figure}
\begin{figure}[H]
    \ContinuedFloat
    \centering
    \begin{subfigure}[b]{0.65\textwidth}
        \centering
        \includegraphics[width=\linewidth]{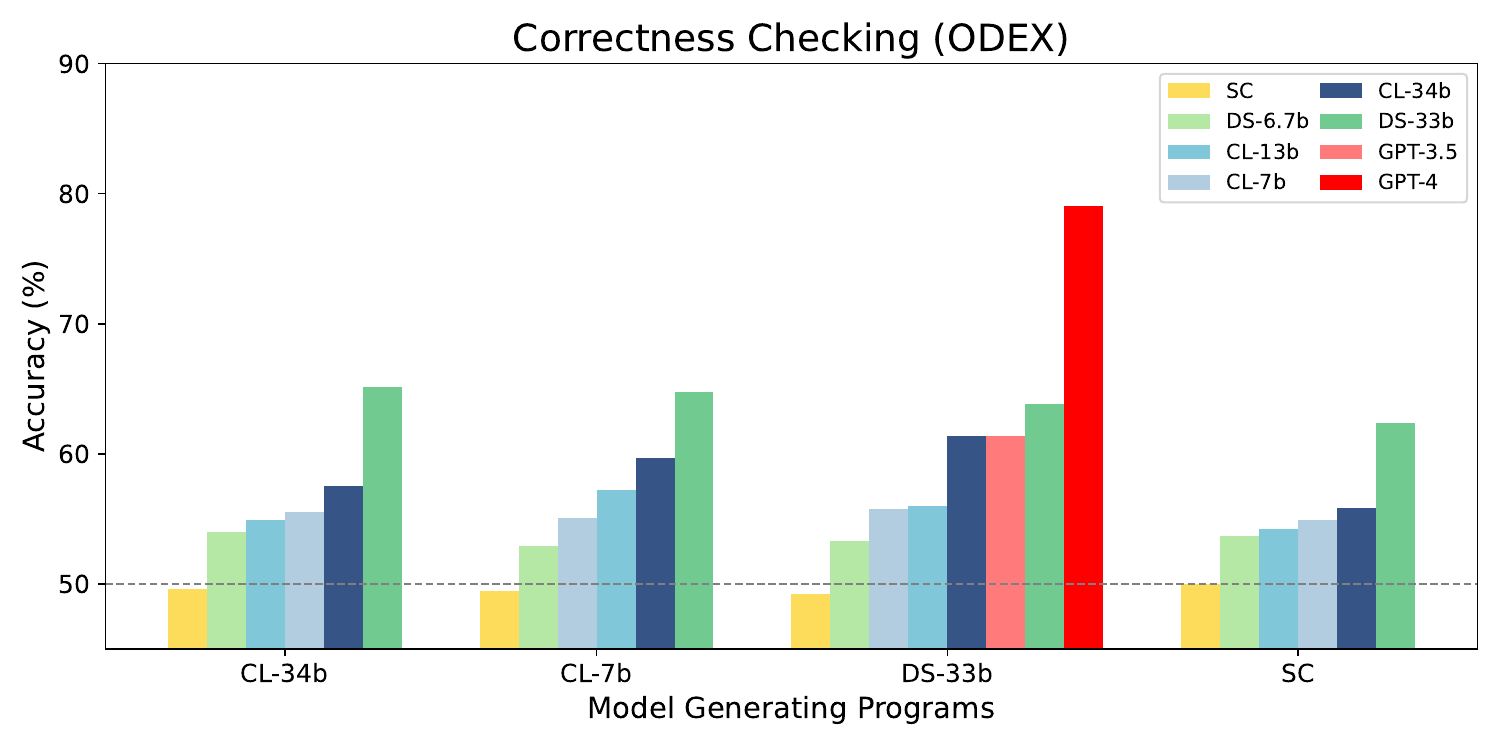}
        \subcaption{}
    \end{subfigure}
\caption{Correctness checking results across all models and datasets}
\label{fig:correctness-checking-results-full}
\end{figure}

\begin{figure}[H]
    \centering
    \begin{subfigure}[b]{0.32\columnwidth}
        \centering
        \includegraphics[width=\textwidth]{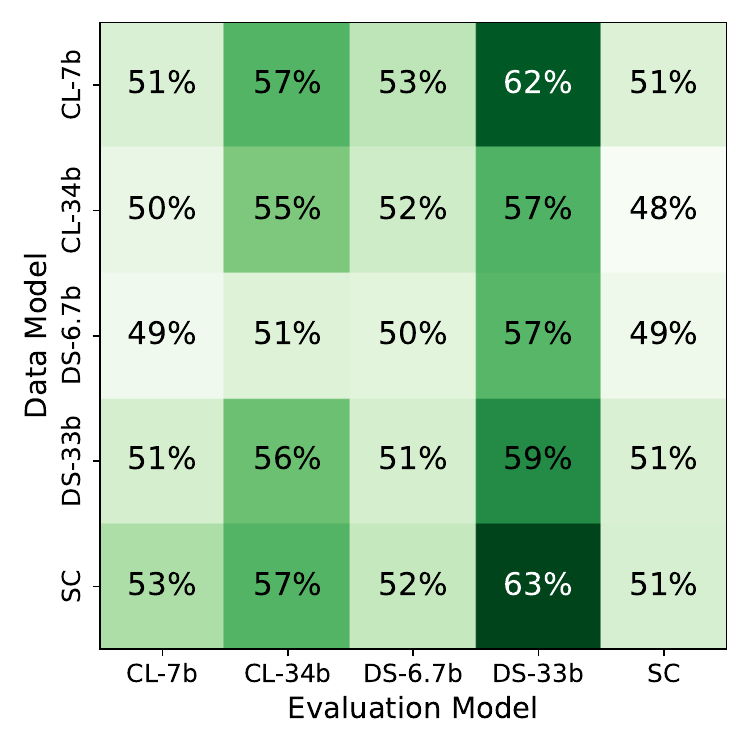}
        \subcaption{Correctness (HumanEval)}
    \end{subfigure}
    \hfill
    \begin{subfigure}[b]{0.32\columnwidth}
        \centering
        \includegraphics[width=\textwidth]{figures_v2/main/accuracy_heatmaps/odex_verification.pdf}
        \subcaption{Correctness (ODEX)}
    \end{subfigure}
    \hfill
    \begin{subfigure}[b]{0.32\columnwidth}
        \centering
        \includegraphics[width=\textwidth]{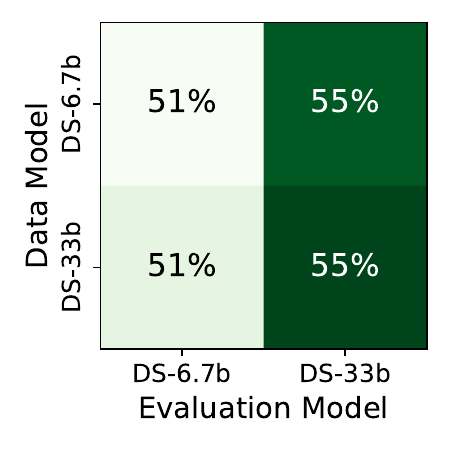}
        \subcaption{Correctness (LeetCode)}
    \end{subfigure}
    \caption{Heatmap of accuracies for correctness checking.}
    \label{fig:correctness-checking-heatmaps}
\end{figure}

\pagebreak These results are shown in heatmap form in Fig. \ref{fig:correctness-checking-heatmaps}.

In Fig. \ref{fig:confusion-matrix-all}, we show the confusion matrices of predicted and correct labels, which highlights the prediction biases of various models.

\begin{figure}[H]
    \centering
    \begin{subfigure}[b]{\textwidth}
        \centering
        \includegraphics[width=\linewidth]{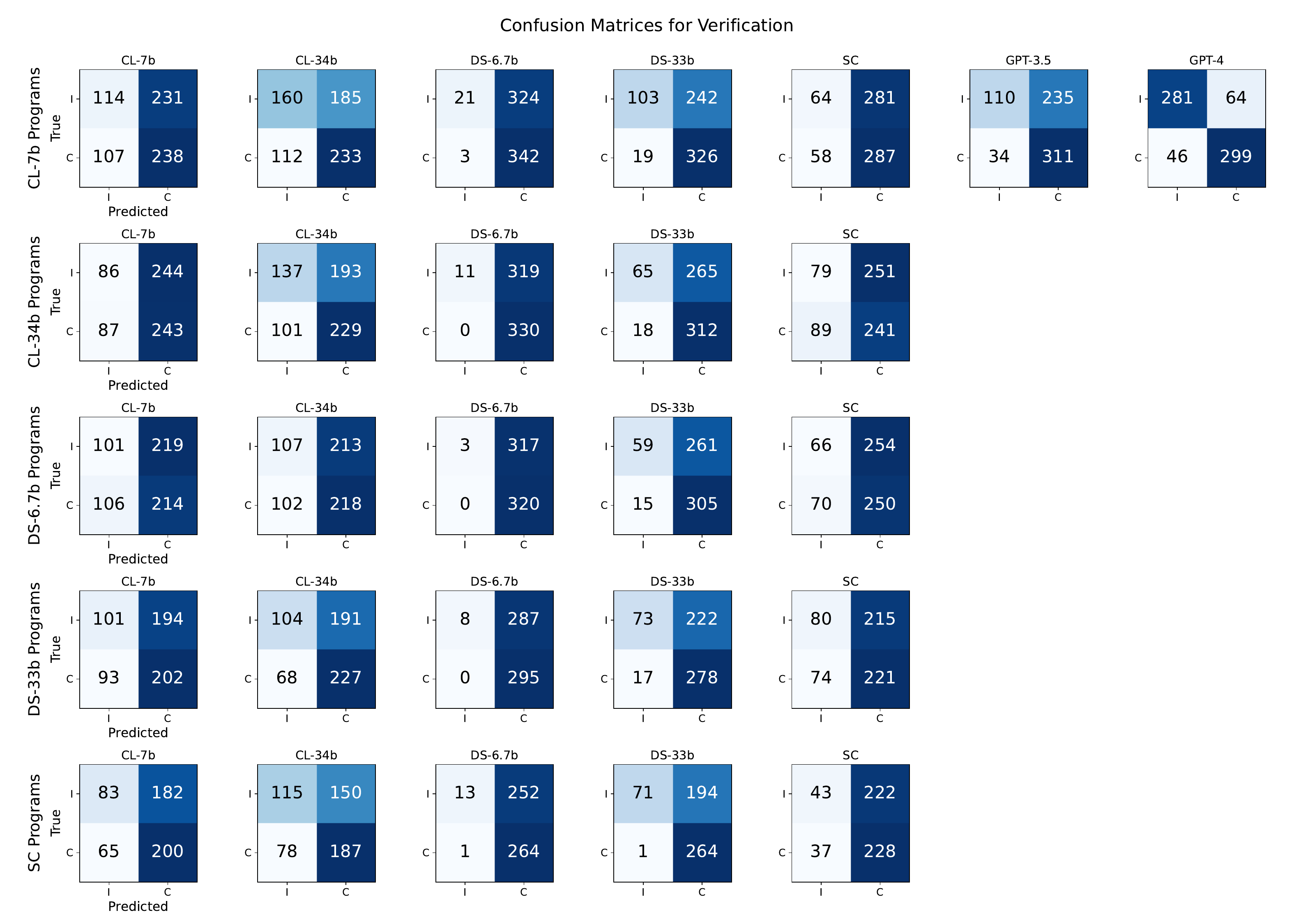}
        \subcaption{HumanEval}
    \end{subfigure}
\end{figure}
\begin{figure}[H]
    \ContinuedFloat
    \centering
    \begin{subfigure}[b]{\textwidth}
        \centering
        \includegraphics[width=\linewidth]{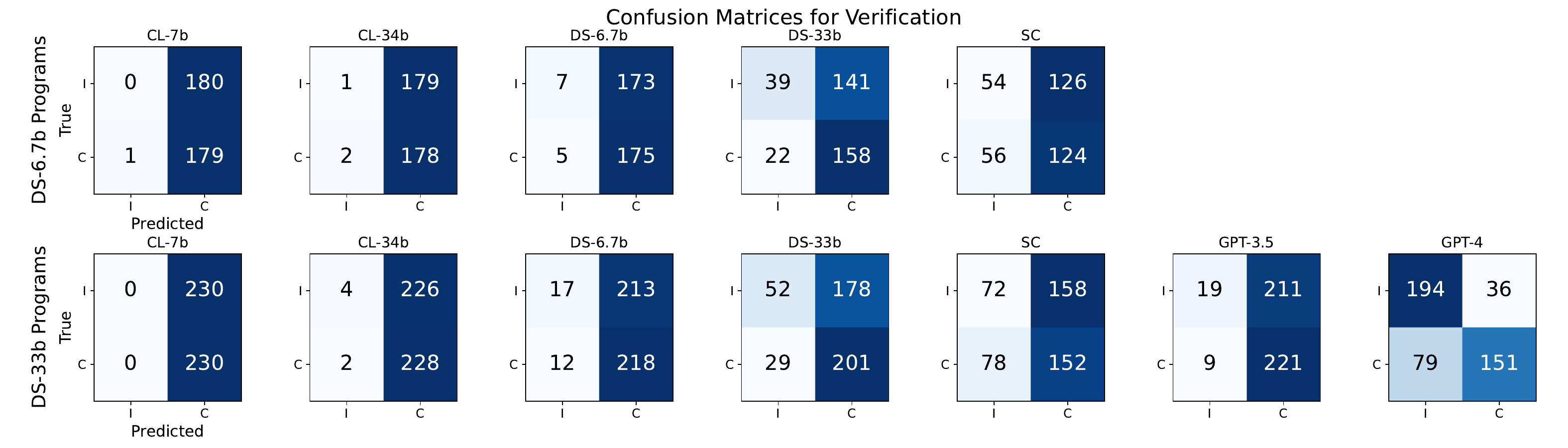}
        \subcaption{LeetCode}
    \end{subfigure}
    \begin{subfigure}[b]{0.9\textwidth}
        \centering
        \includegraphics[width=\linewidth]{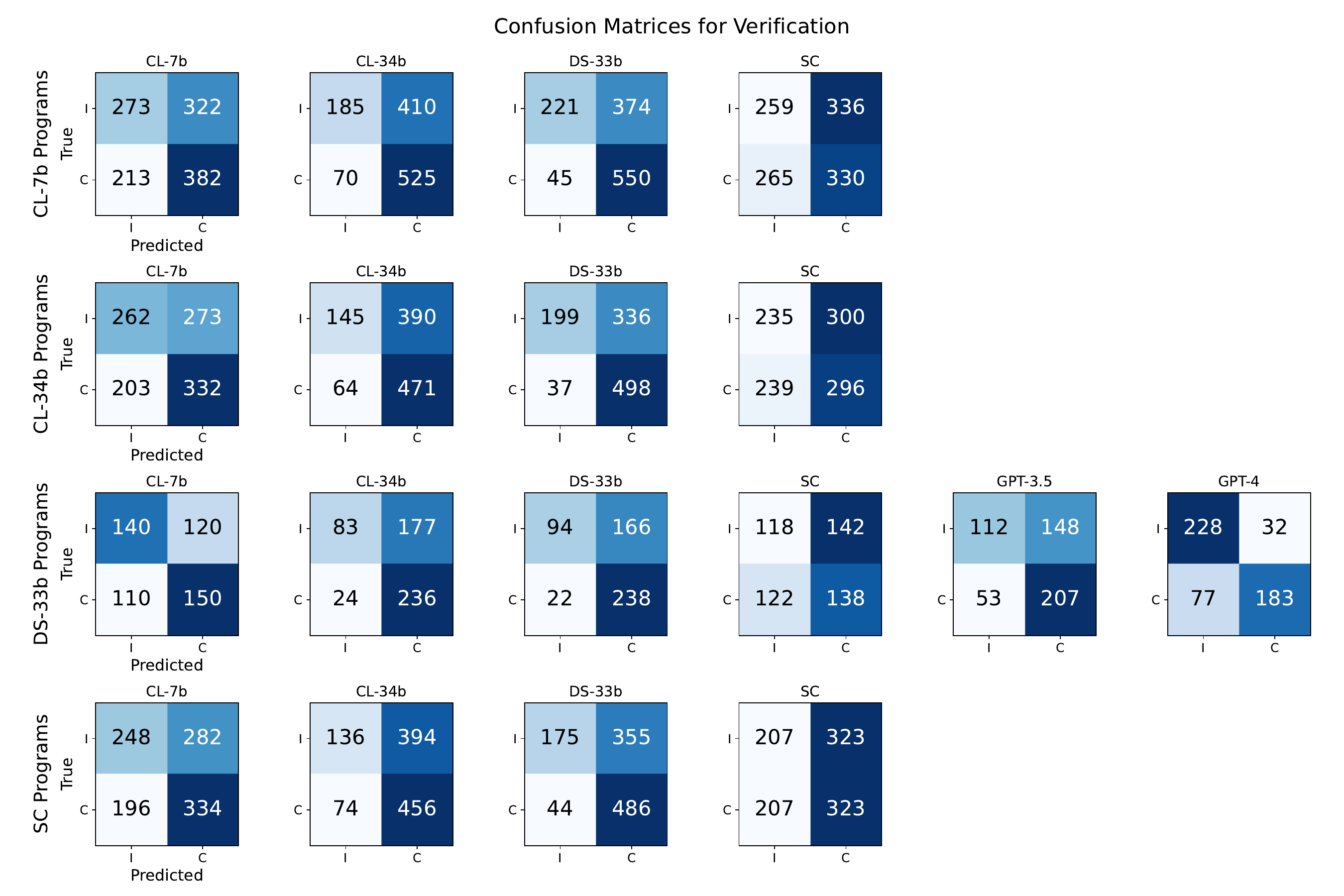}
        \subcaption{ODEX}
    \end{subfigure}
    \caption{Confusion matrices of predictions vs. labels}
    \label{fig:confusion-matrix-all}
\end{figure}

\begin{figure}[ht!]
    \centering
    \begin{subfigure}[b]{0.56\columnwidth}
        \centering
        \includegraphics[width=\textwidth]{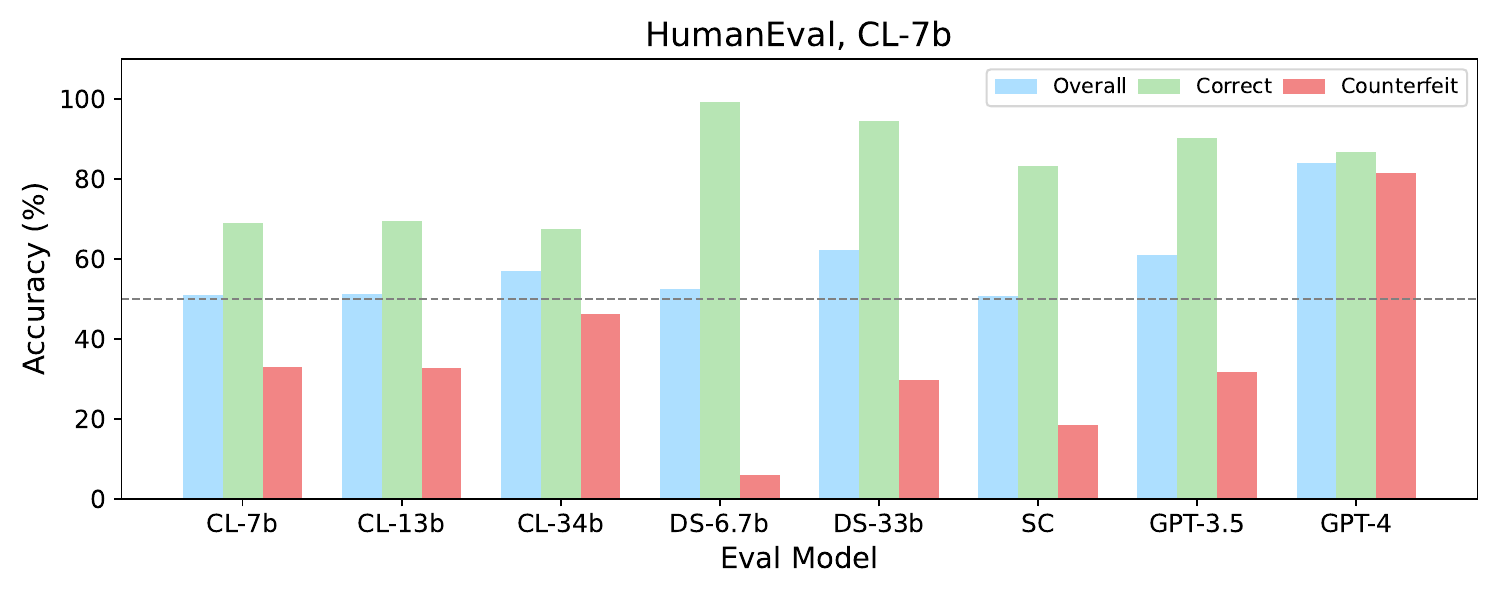}
        \subcaption{HumanEval (CL-7B)}
    \end{subfigure}
    \hfill
    \begin{subfigure}[b]{0.42\columnwidth}
        \centering
        \includegraphics[width=\textwidth]{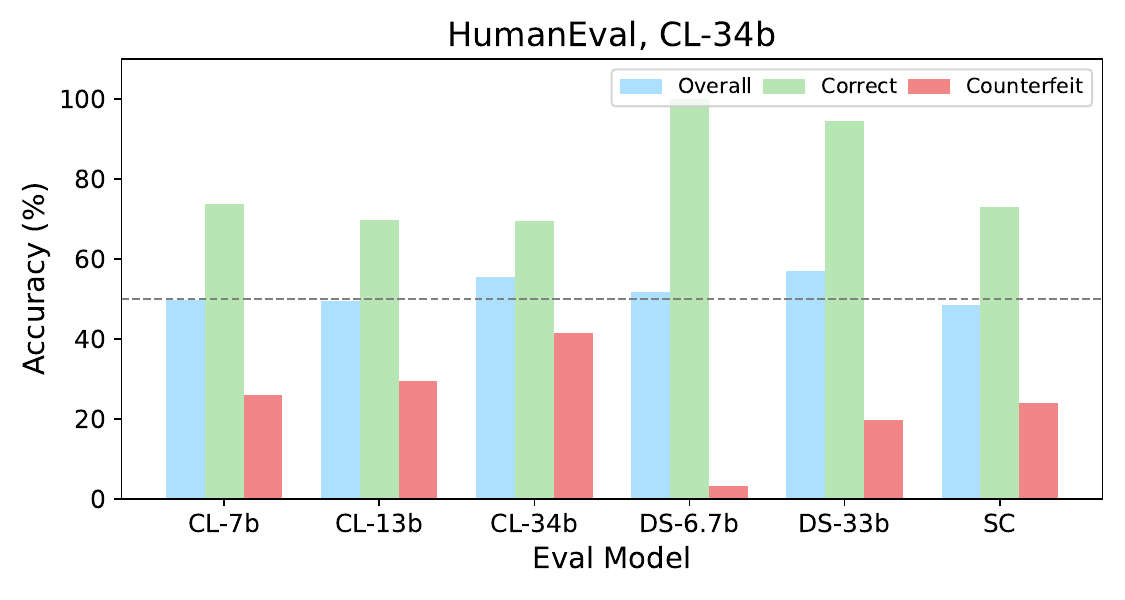}
        \subcaption{HumanEval (CL-34B)}
    \end{subfigure}
    \hfill
    \begin{subfigure}[b]{0.42\columnwidth}
        \centering
        \includegraphics[width=\textwidth]{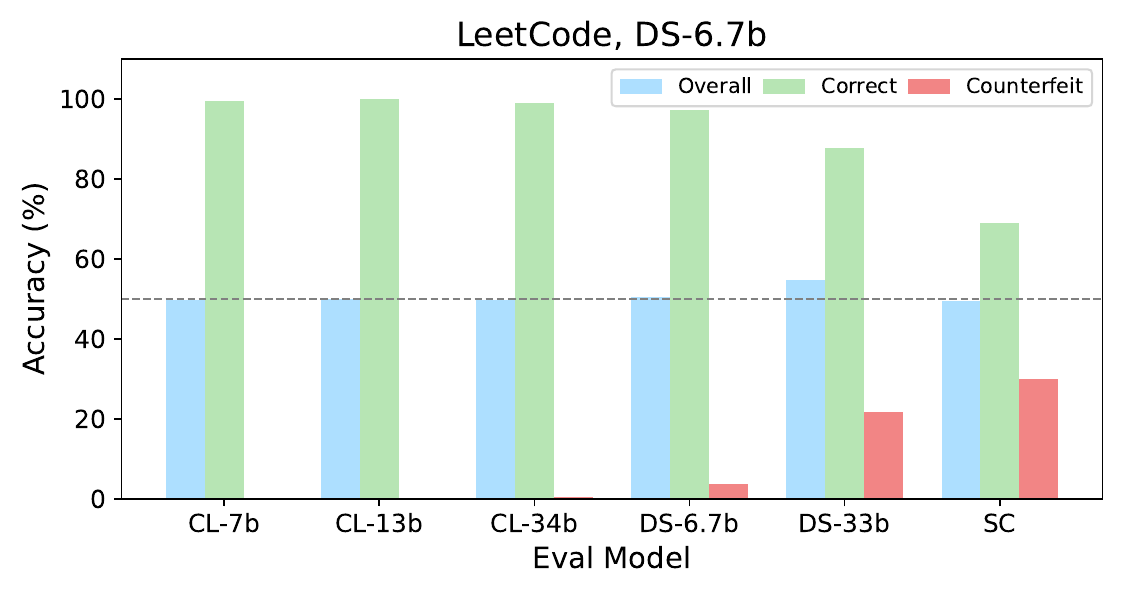}
        \subcaption{LeetCode (DS-6.7B)}
    \end{subfigure}
    \hfill
    \begin{subfigure}[b]{0.56\columnwidth}
        \centering
        \includegraphics[width=\textwidth]{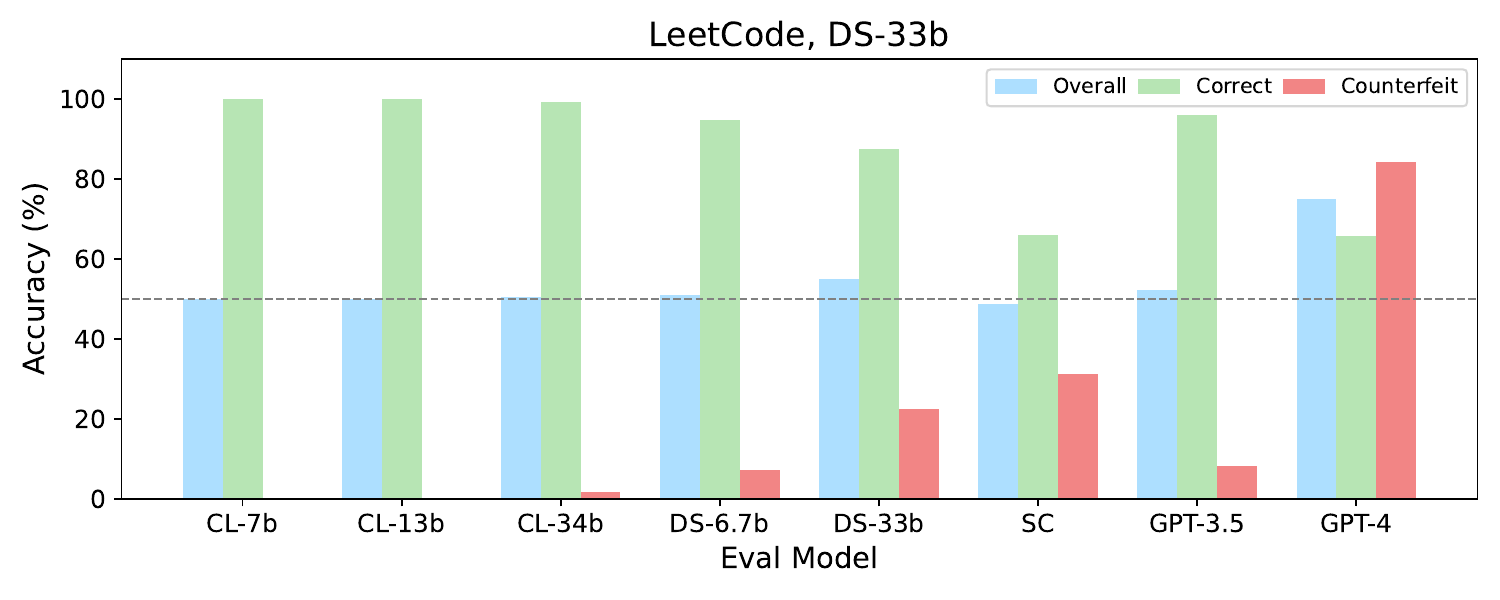}
        \subcaption{LeetCode (DS-33B)}
    \end{subfigure}
    \hfill
    \begin{subfigure}[b]{0.42\columnwidth}
        \centering
        \includegraphics[width=\textwidth]{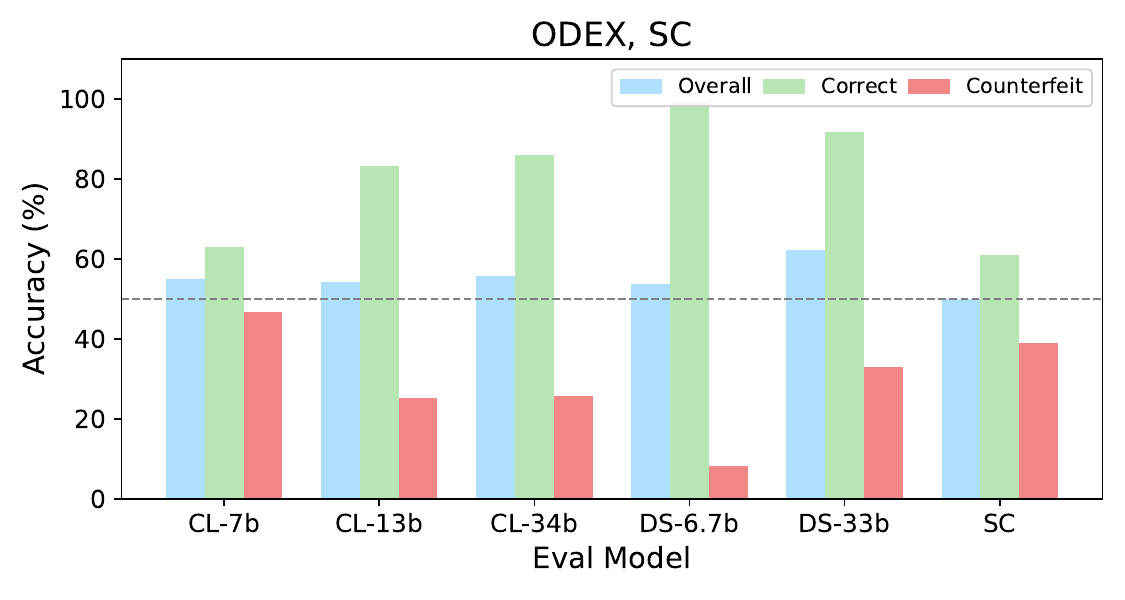}
        \subcaption{ODEX (StarCoder)}
    \end{subfigure}
    \hfill
    \begin{subfigure}[b]{0.55\columnwidth}
        \centering
        \includegraphics[width=\textwidth]{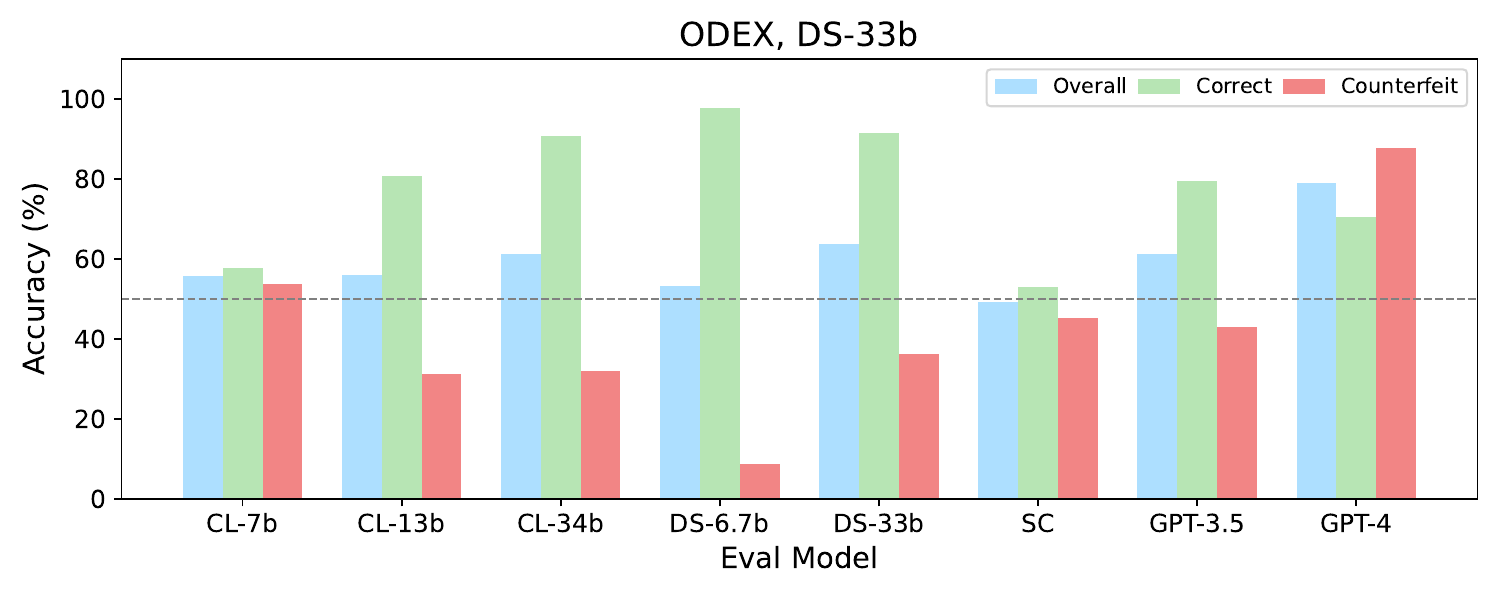}
        \subcaption{ODEX (DS-33B)}
    \end{subfigure}
    \hfill
    \caption{Models are much better at assessing the correctness of correct samples than counterfeit samples.}
    \label{fig:correctness-checking-counterfeit}
\end{figure}

\pagebreak
In Fig. \ref{fig:correctness-checking-counterfeit}, we also show a few more plots highlighting that models often mispredict counterfeit samples as correct.

\clearpage
\subsection{Execution Prediction}
\begin{figure}[H]
    \centering
    \begin{subfigure}[b]{0.65\textwidth}
        \includegraphics[width=\linewidth]{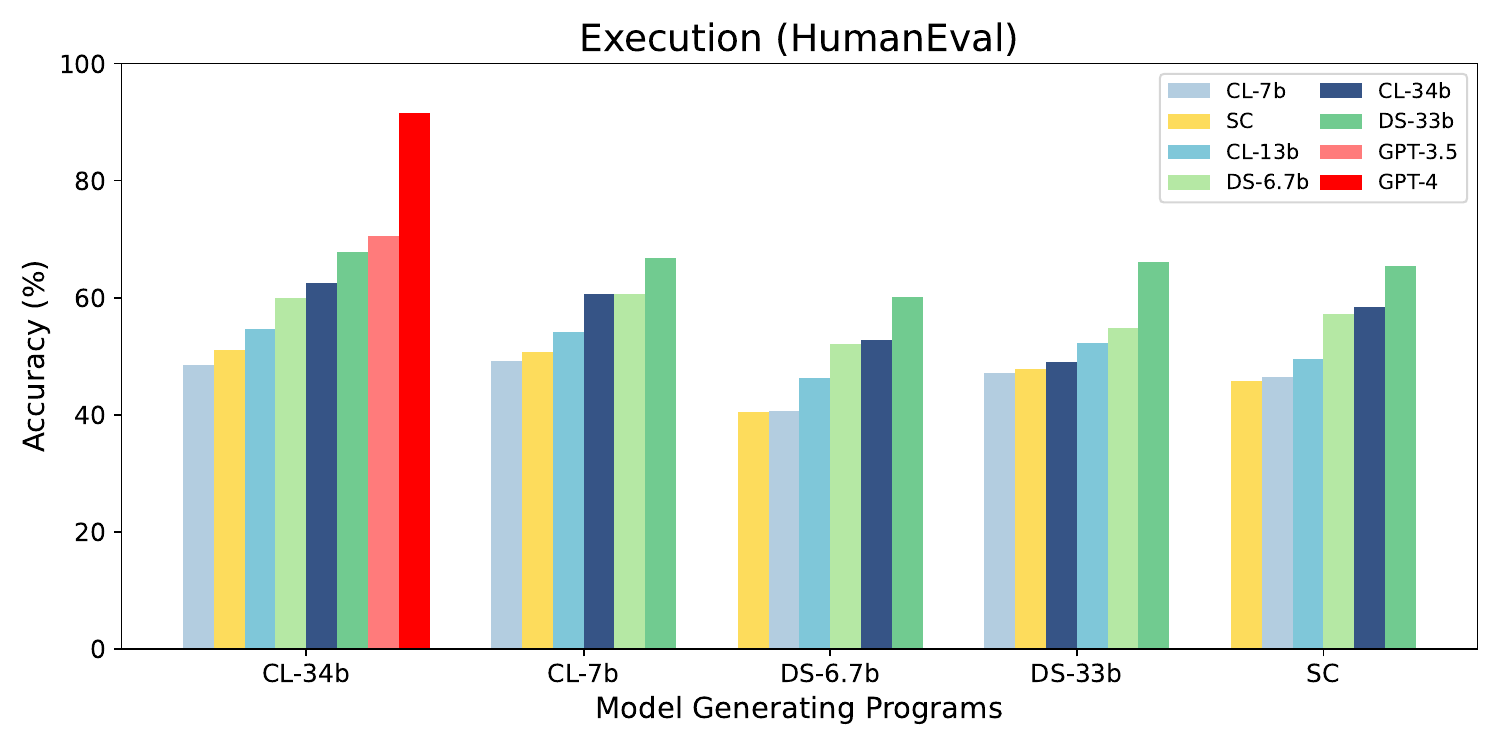}
        \subcaption{}
    \end{subfigure}
    \begin{subfigure}[b]{0.33\textwidth}
        \includegraphics[width=\linewidth]{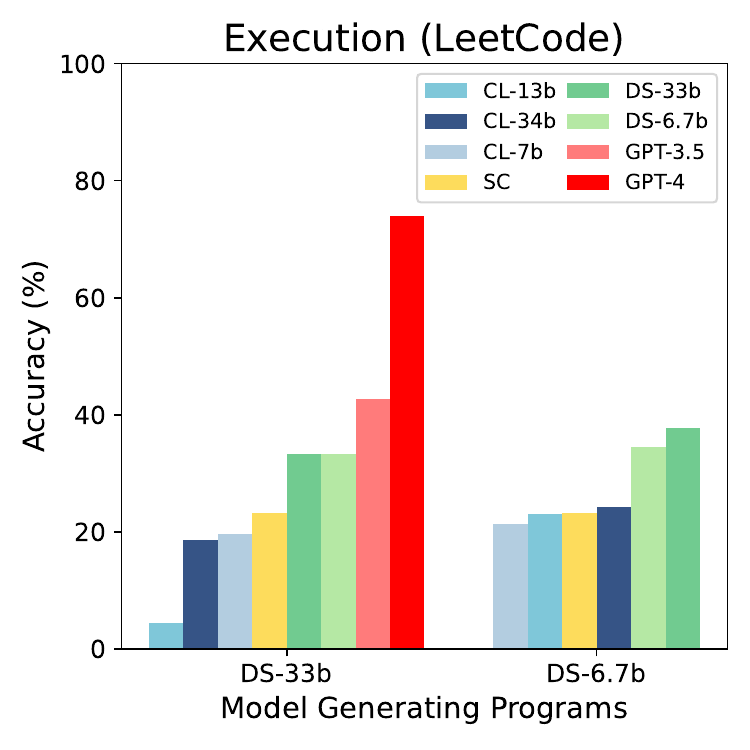}
        \subcaption{}
    \end{subfigure}
\caption{Execution prediction results across all models and datasets}
\label{fig:execution-prediction-results-full}
\end{figure}

Fig. \ref{fig:execution-prediction-results-full} shows the full set of correctness checking results across all models for each of the three datasets.

\begin{figure}[H]
    \centering
    \begin{subfigure}[b]{0.35\columnwidth}
        \centering
        \includegraphics[width=\textwidth]{figures_v2/main/accuracy_heatmaps/humaneval_execution.pdf}
        \subcaption{Execution (HumanEval)}
    \end{subfigure}
    \begin{subfigure}[b]{0.35\columnwidth}
        \centering
        \includegraphics[width=\textwidth]{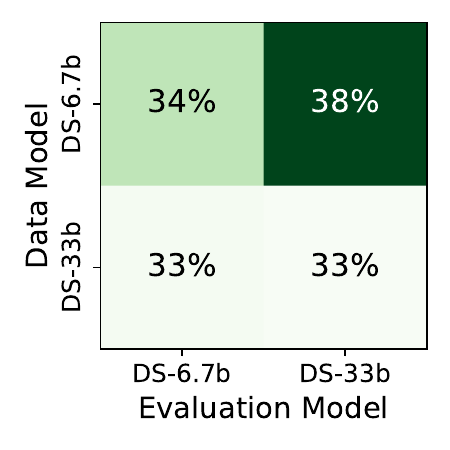}
        \subcaption{Execution (LeetCode)}
    \end{subfigure}
    \caption{Heatmap of accuracies for correctness checking.}
    \label{fig:execution-prediction-heatmaps}
\end{figure}

These results are shown in heatmap form in Fig. \ref{fig:execution-prediction-heatmaps}.

\begin{figure}[ht!]
    \centering
    \begin{subfigure}[b]{0.42\columnwidth}
        \centering
        \includegraphics[width=\textwidth]{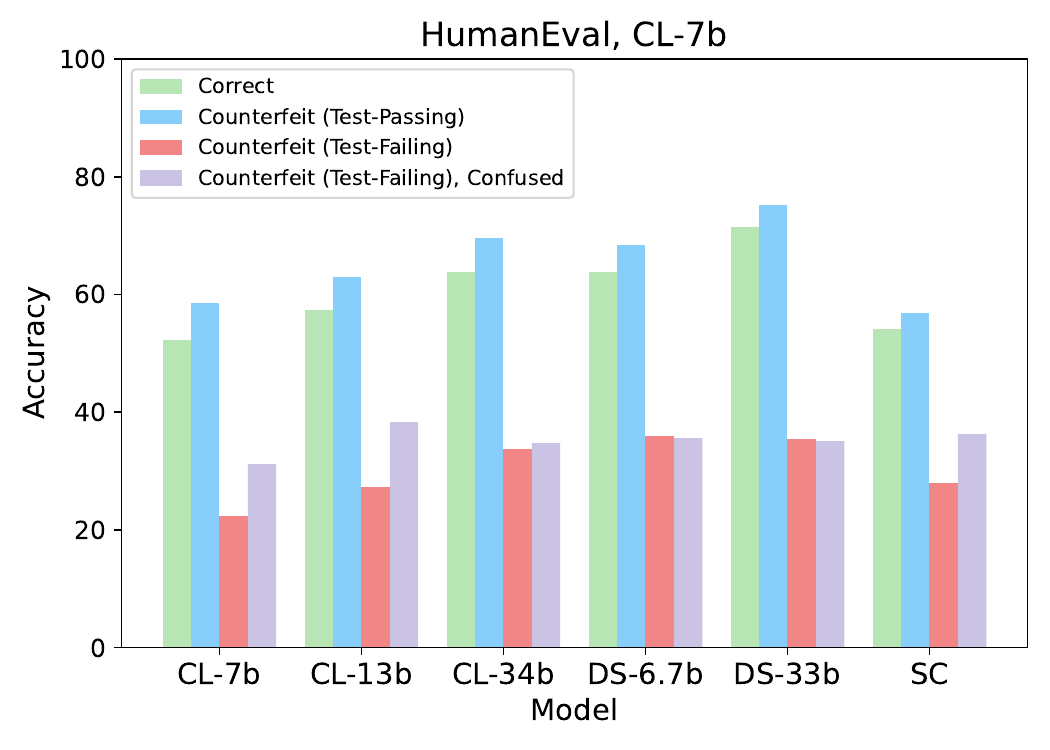}
        \subcaption{HumanEval (CL-7B)}
    \end{subfigure}
    \hfill
    \begin{subfigure}[b]{0.57\columnwidth}
        \centering
        \includegraphics[width=\textwidth]{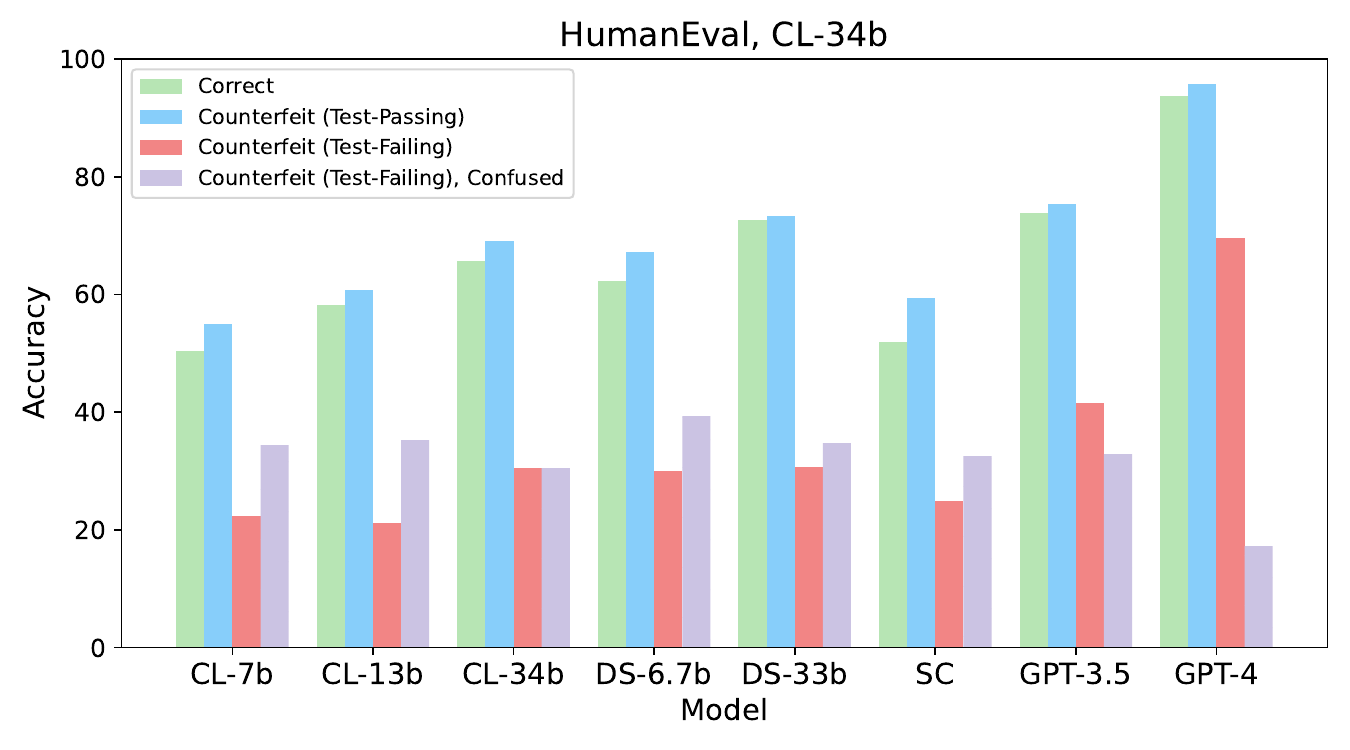}
        \subcaption{HumanEval (CL-34B)}
    \end{subfigure}
    \hfill
    \begin{subfigure}[b]{0.49\columnwidth}
        \centering
        \includegraphics[width=\textwidth]{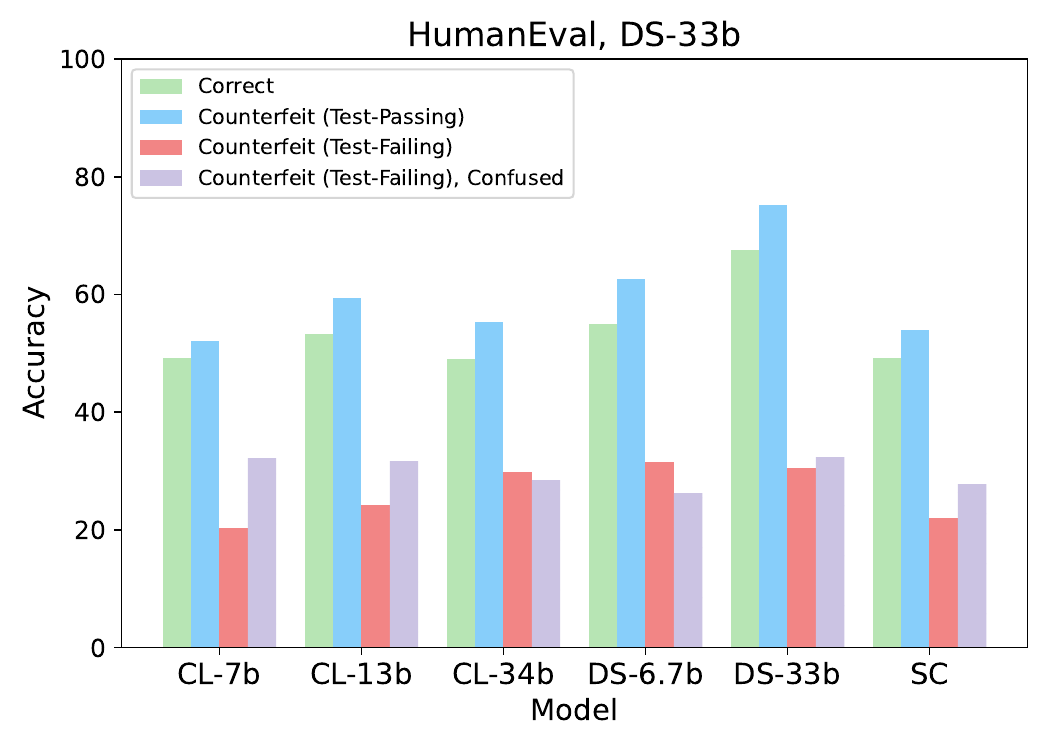}
        \subcaption{HumanEval (DS-33B)}
    \end{subfigure}
    \hfill
    \begin{subfigure}[b]{0.49\columnwidth}
        \centering
        \includegraphics[width=\textwidth]{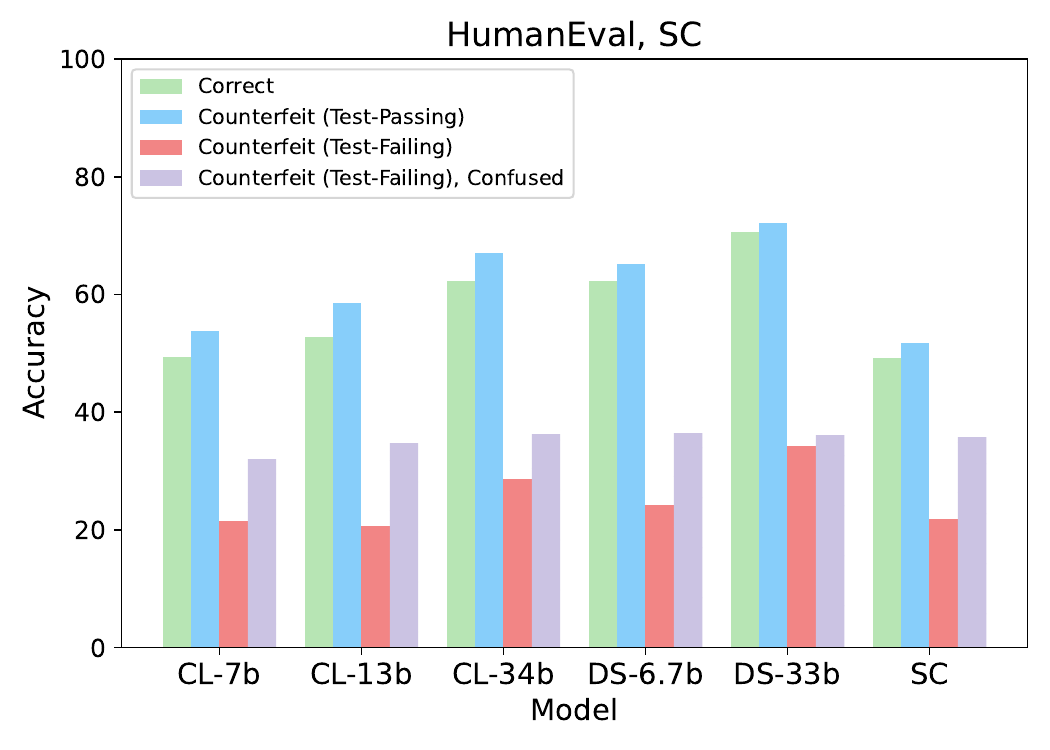}
        \subcaption{HumanEval (StarCoder)}
    \end{subfigure}
    \hfill
    \begin{subfigure}[b]{0.42\columnwidth}
        \centering
        \includegraphics[width=\textwidth]{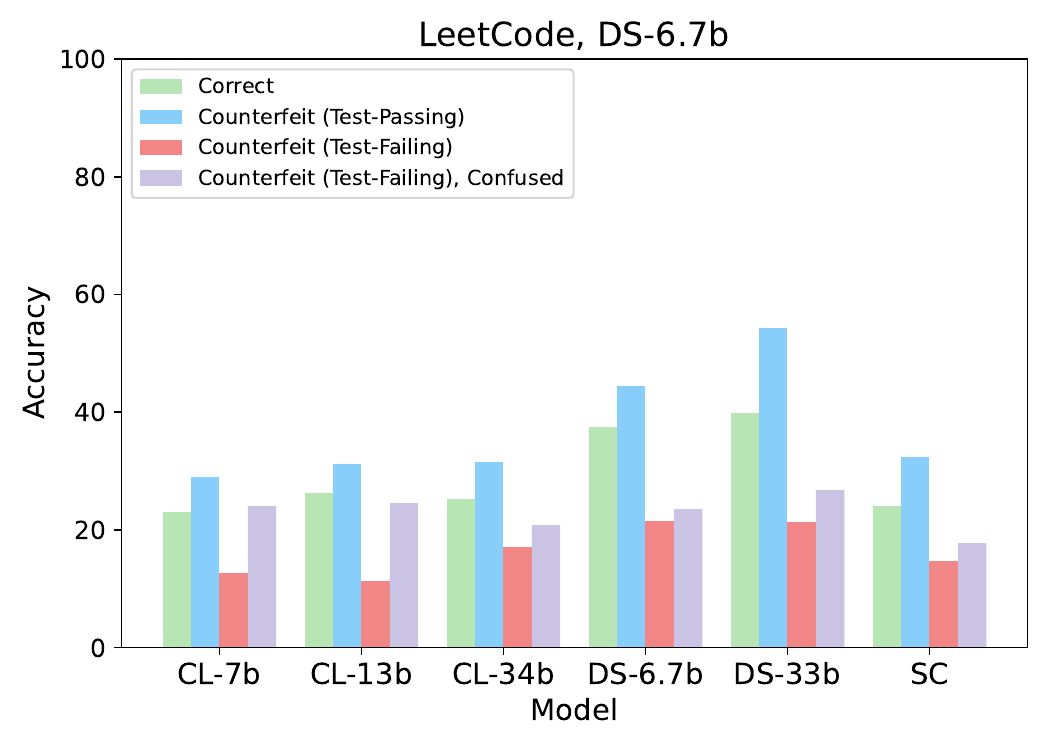}
        \subcaption{LeetCode (DS-6.7B)}
    \end{subfigure}
    \hfill
    \begin{subfigure}[b]{0.55\columnwidth}
        \centering
        \includegraphics[width=\textwidth]{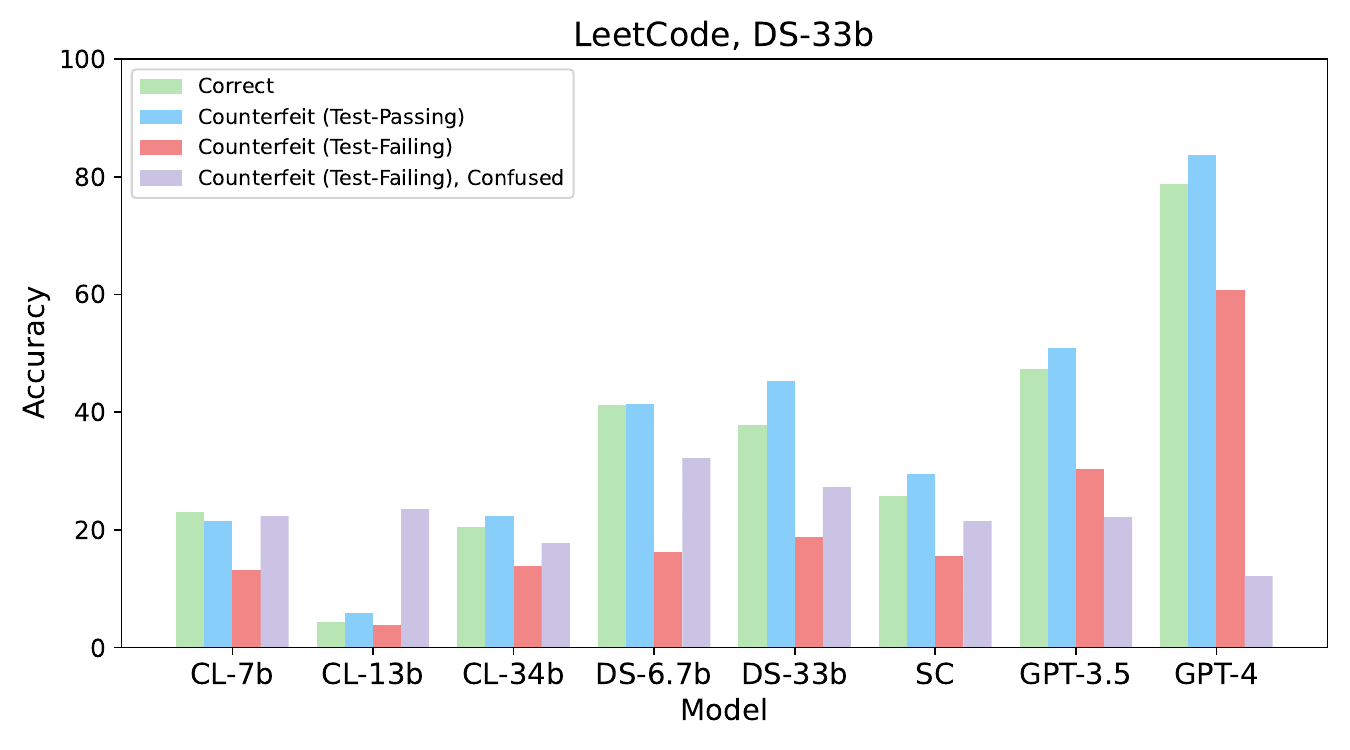}
        \subcaption{LeetCode (DS-33B)}
    \end{subfigure}
    \hfill
    \caption{Models are much better at executing correct samples than counterfeit samples, and even often execute counterfeit samples as if they were correct.}
    \label{fig:execution-counterfeit}
\end{figure}

\pagebreak In Fig. \ref{fig:execution-counterfeit}, we show that on other datasets and models, models are generally better at executing correct samples than counterfeit samples with outputs that don't match those of the correct samples, and that models often predict the output of the correct sample when asked to execute these counterfeit samples.

\clearpage
\subsection{Repair}
\label{appendix:repair}
\begin{figure}[H]
    \centering
    \begin{subfigure}[b]{0.33\textwidth}
        \includegraphics[width=\linewidth]{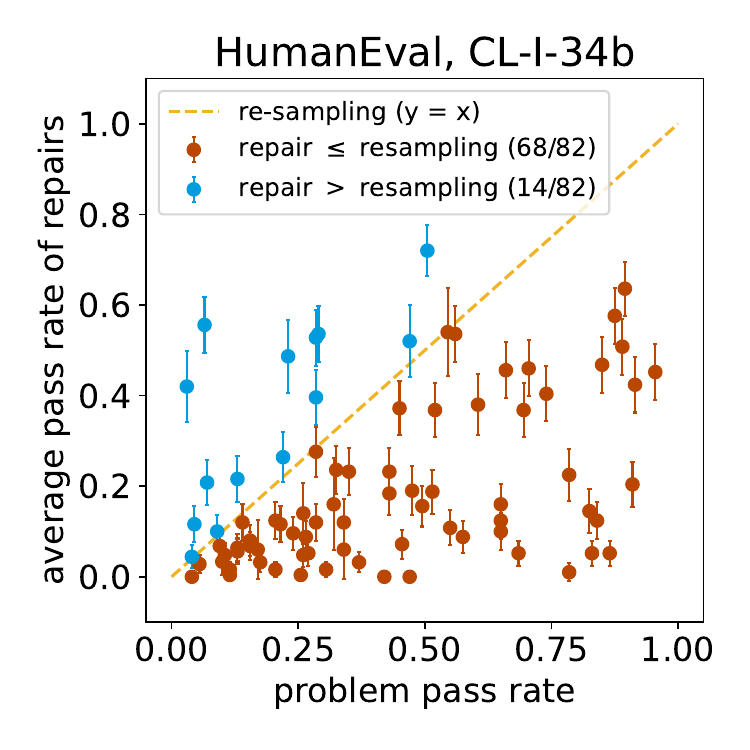}
        \subcaption{CL-I-34b}
    \end{subfigure}
    \begin{subfigure}[b]{0.33\textwidth}
        \includegraphics[width=\linewidth]{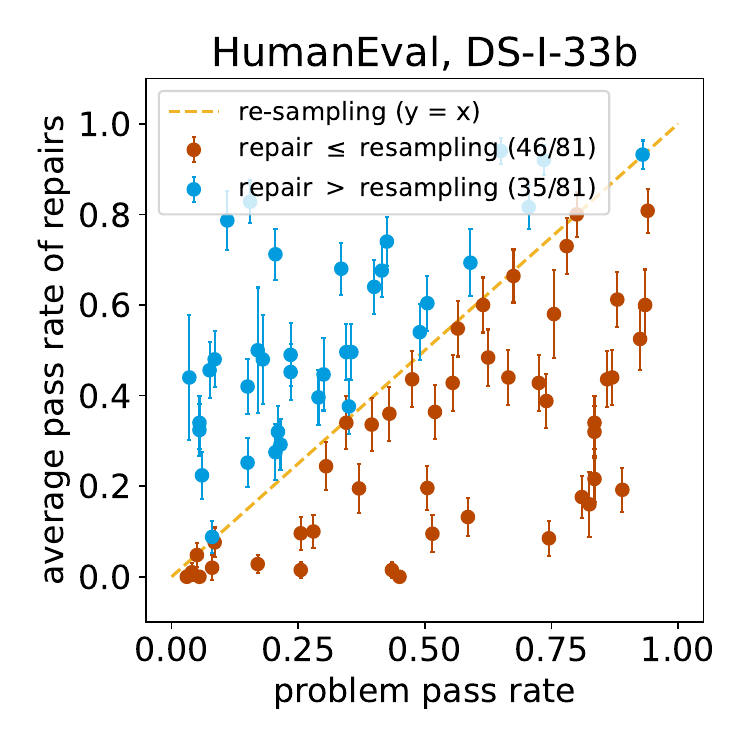}
        \subcaption{DS-I-33b}
        \label{fig:repair-hull-humaneval-ds33}
    \end{subfigure}
\caption{Repair success vs. baseline pass@1 on HumanEval.} 
\label{fig:repair-full-humaneval}
\end{figure}

\begin{figure}[H]
    \centering
    \begin{subfigure}[b]{0.33\textwidth}
        \includegraphics[width=\linewidth]{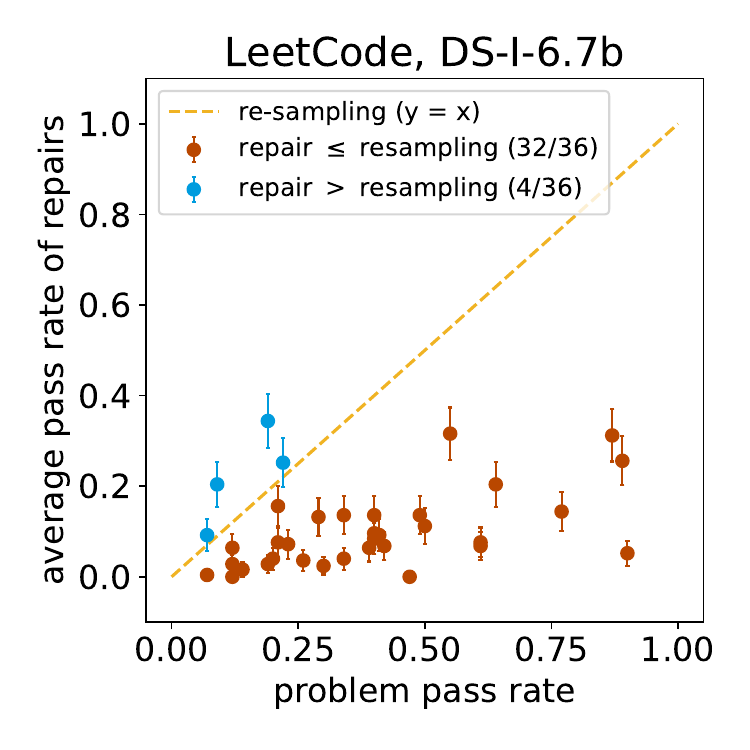}
        \subcaption{DS-I-6.7b}
    \end{subfigure}
    \begin{subfigure}[b]{0.33\textwidth}
        \includegraphics[width=\linewidth]{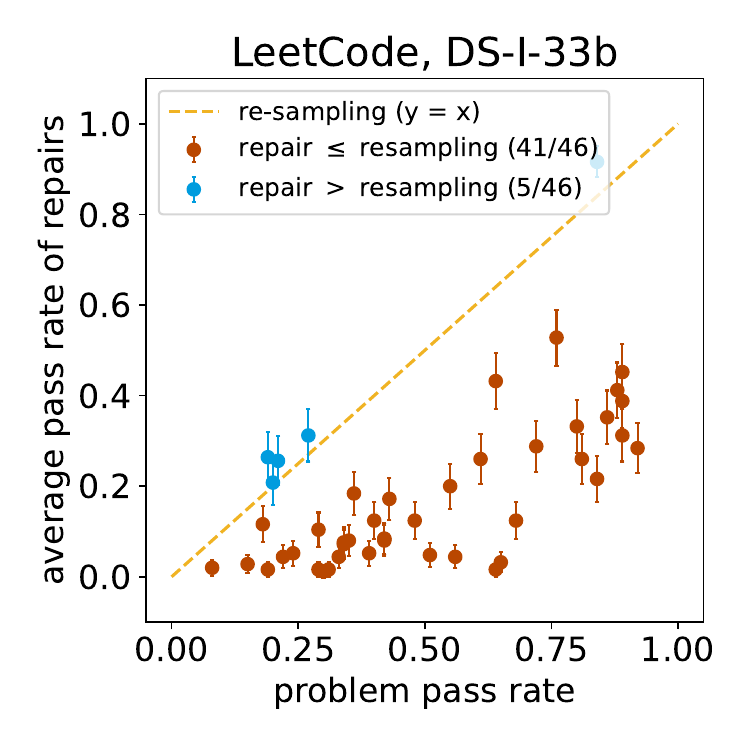}
        \subcaption{DS-I-33b}
    \end{subfigure}
\caption{Repair success vs. baseline pass@1 on LeetCode.} 
\label{fig:repair-full-leetcode}
\end{figure}

\begin{figure}[H]
    \centering
    \begin{subfigure}[b]{0.33\textwidth}
        \includegraphics[width=\linewidth]{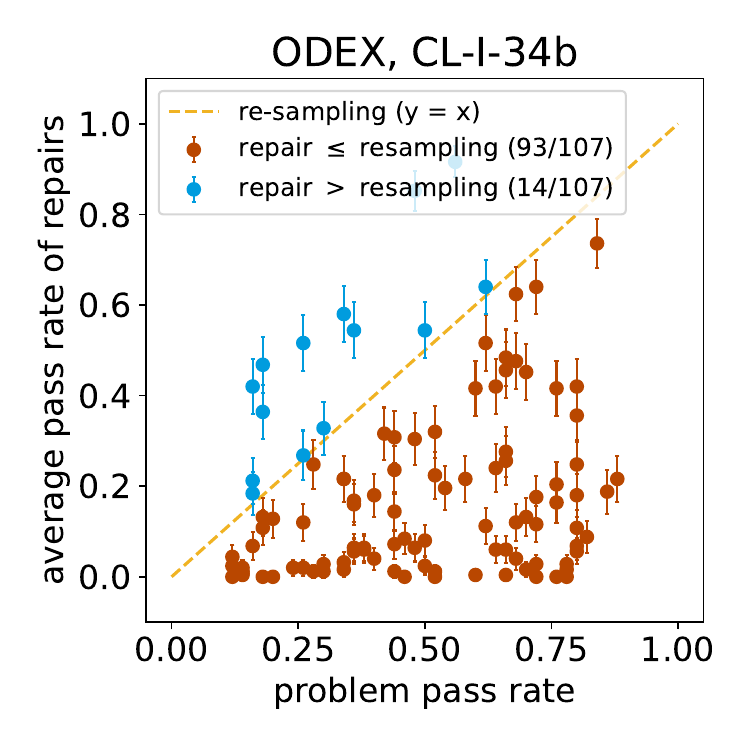}
        \subcaption{CL-I-34b}
    \end{subfigure}
    \begin{subfigure}[b]{0.33\textwidth}
        \includegraphics[width=\linewidth]{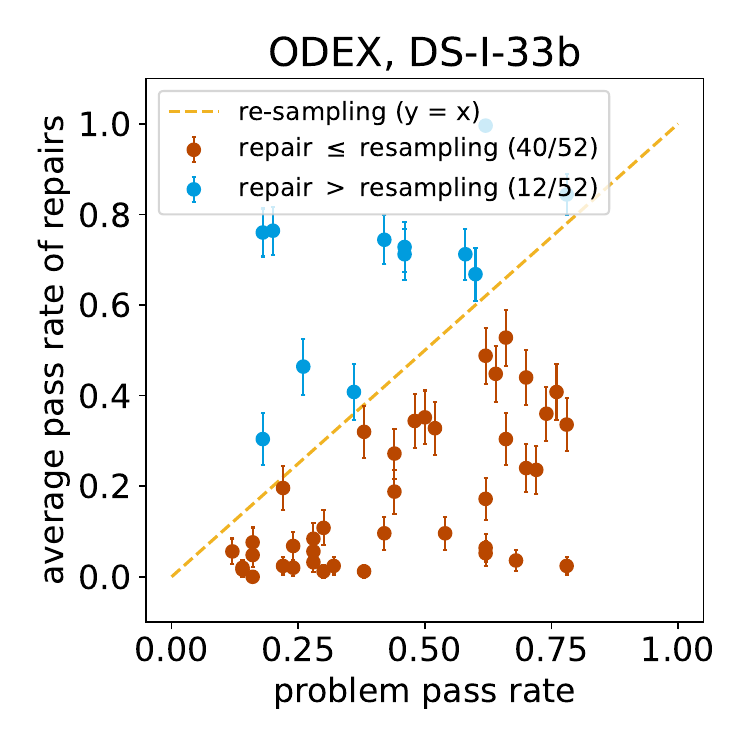}
        \subcaption{DS-I-33b}
    \end{subfigure}
\caption{Repair success vs. baseline pass@1 on ODEX.} 
\label{fig:repair-full-odex}
\end{figure}

Figures~\ref{fig:repair-full-humaneval}-\ref{fig:repair-full-odex} show the full set of scatterplots for the repair experiments in Sec.~\ref{sec:repair}.
In these plots, \emph{the same model is used for both repair and the initial code generation}, so that the resampling strategy can be represented by the line $y=x$ (simplifying exposition).
Note that even in the most successful setting, DS-I-33b on HumanEval, the number of problems for which repair is more successful than simply resampling is still in the minority (35/81).

\begin{figure}[H]
    \centering
    \begin{subfigure}[b]{0.25\textwidth}
        \includegraphics[width=\linewidth]{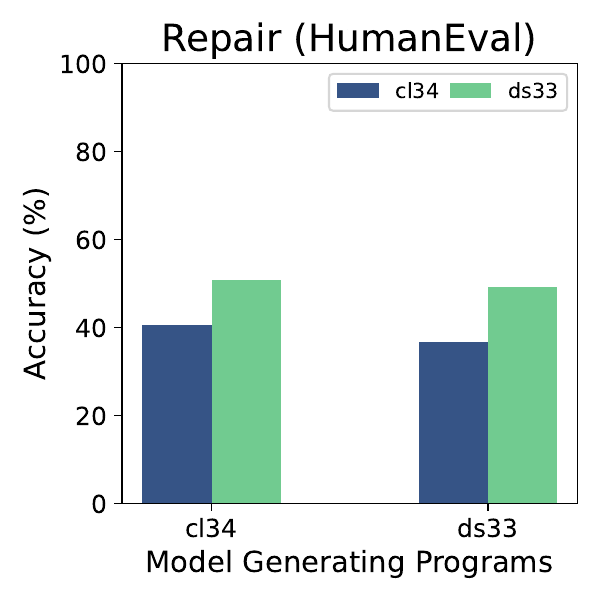}
        \subcaption{HumanEval}
    \end{subfigure}
    \begin{subfigure}[b]{0.25\textwidth}
        \includegraphics[width=\linewidth]{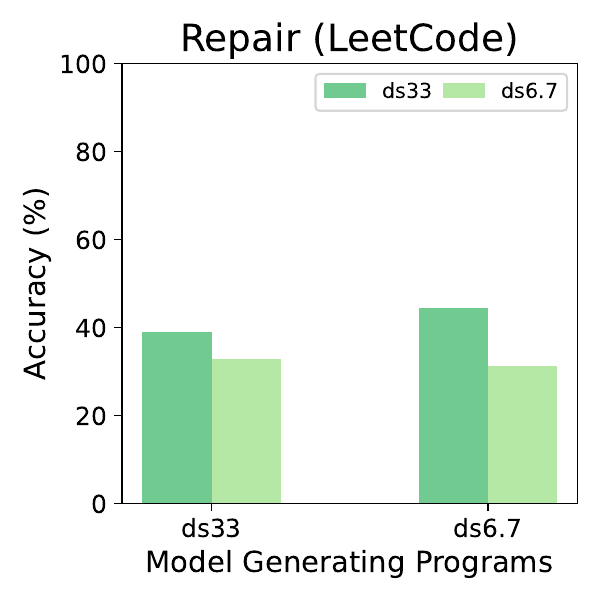}
        \subcaption{LeetCode}
    \end{subfigure}
    \begin{subfigure}[b]{0.25\textwidth}
        \includegraphics[width=\linewidth]{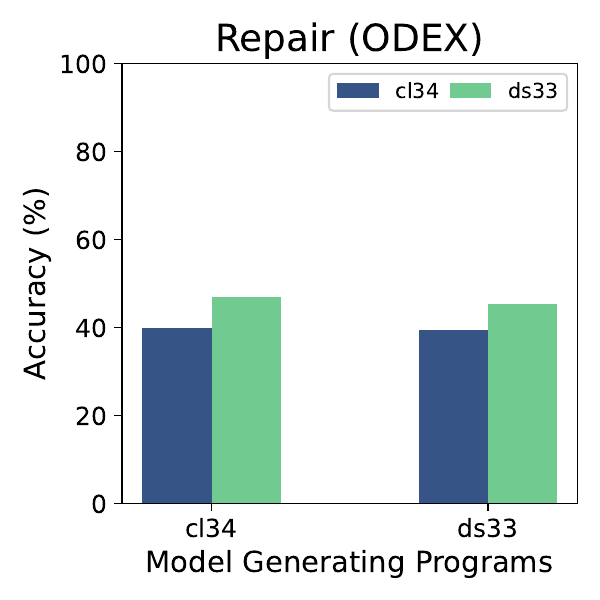}
        \subcaption{ODEX}
    \end{subfigure}
\caption{Average repair accuracy across all models and datasets.}
\label{fig:repair-bars}
\end{figure}

\begin{figure}[H]
    \centering
    \begin{subfigure}[b]{0.25\columnwidth}
        \centering
        \includegraphics[width=\textwidth]{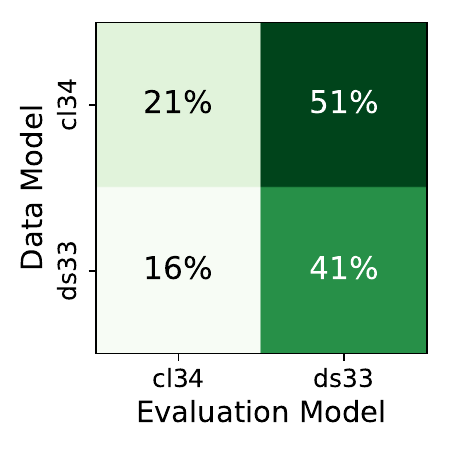}
        \subcaption{HumanEval}
    \end{subfigure}
    \begin{subfigure}[b]{0.25\columnwidth}
        \centering
        \includegraphics[width=\textwidth]{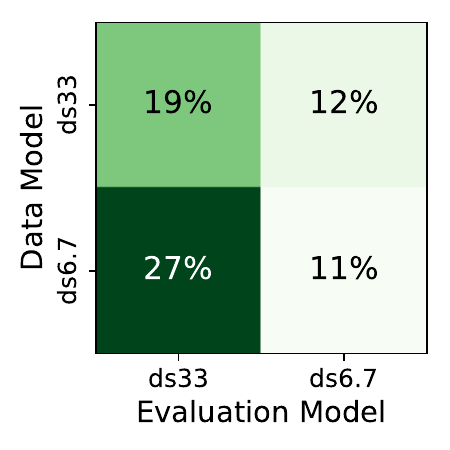}
        \subcaption{LeetCode}
    \end{subfigure}
    \begin{subfigure}[b]{0.25\columnwidth}
        \centering
        \includegraphics[width=\textwidth]{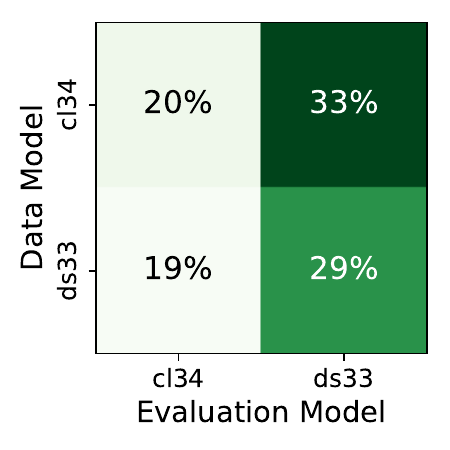}
        \subcaption{ODEX}
    \end{subfigure}
    \caption{Heatmaps of average repair accuracy across all models and datasets.}
    \label{fig:repair-heatmaps}
\end{figure}

Figures~\ref{fig:repair-bars}-\ref{fig:repair-heatmaps} also shows the absolute mean success rate of repair across tasks for each model and dataset, similarly to how was done in the previous sections.
Note that these absolute numbers should not be paid to much attention to, since repair must always be compared to the accuracy of the simple resampling strategy; however, they do show that models do not appear to be better at repairing their own counterfeit samples than those generated by other models.


\section{Correlations by Problem Difficulty}
\subsection{Problem Difficulty vs. Correctness Checking} \label{appendix:difficulty-correctness-full}
In Fig. \ref{fig:correctness-checking-difficulty-all}, we show the accuracy of HumanEval (top), LeetCode (middle), and ODEX (bottom) across different models. We see an absence of correlation across the board.

\begin{figure}[H]
    \centering
    \begin{subfigure}[b]{\textwidth}
        \centering
        \includegraphics[width=\linewidth]{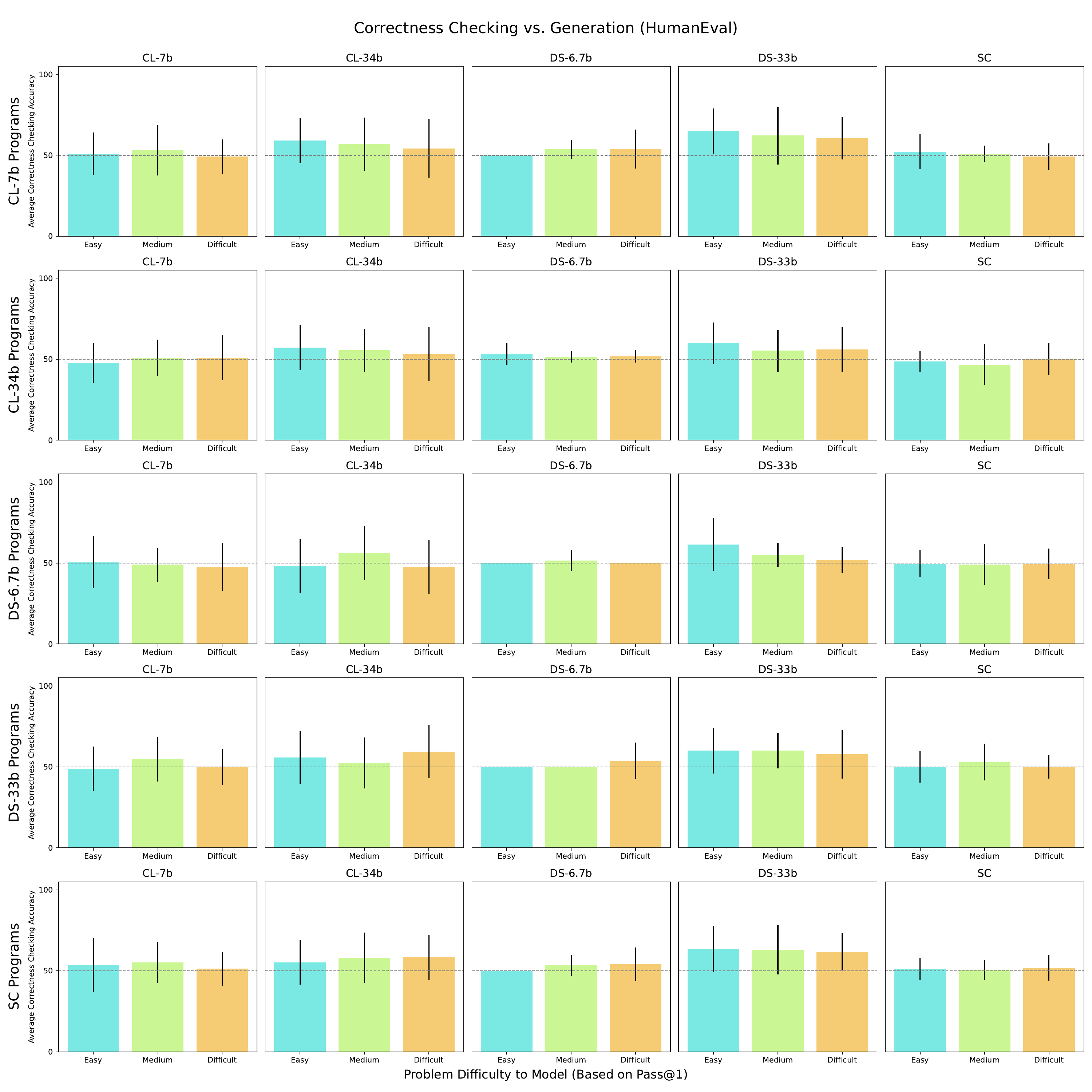}
    \end{subfigure}
\end{figure}
\begin{figure}[H] 
    \ContinuedFloat
    \centering
    \begin{subfigure}[b]{0.5\textwidth}
        \centering
        \includegraphics[width=\linewidth]{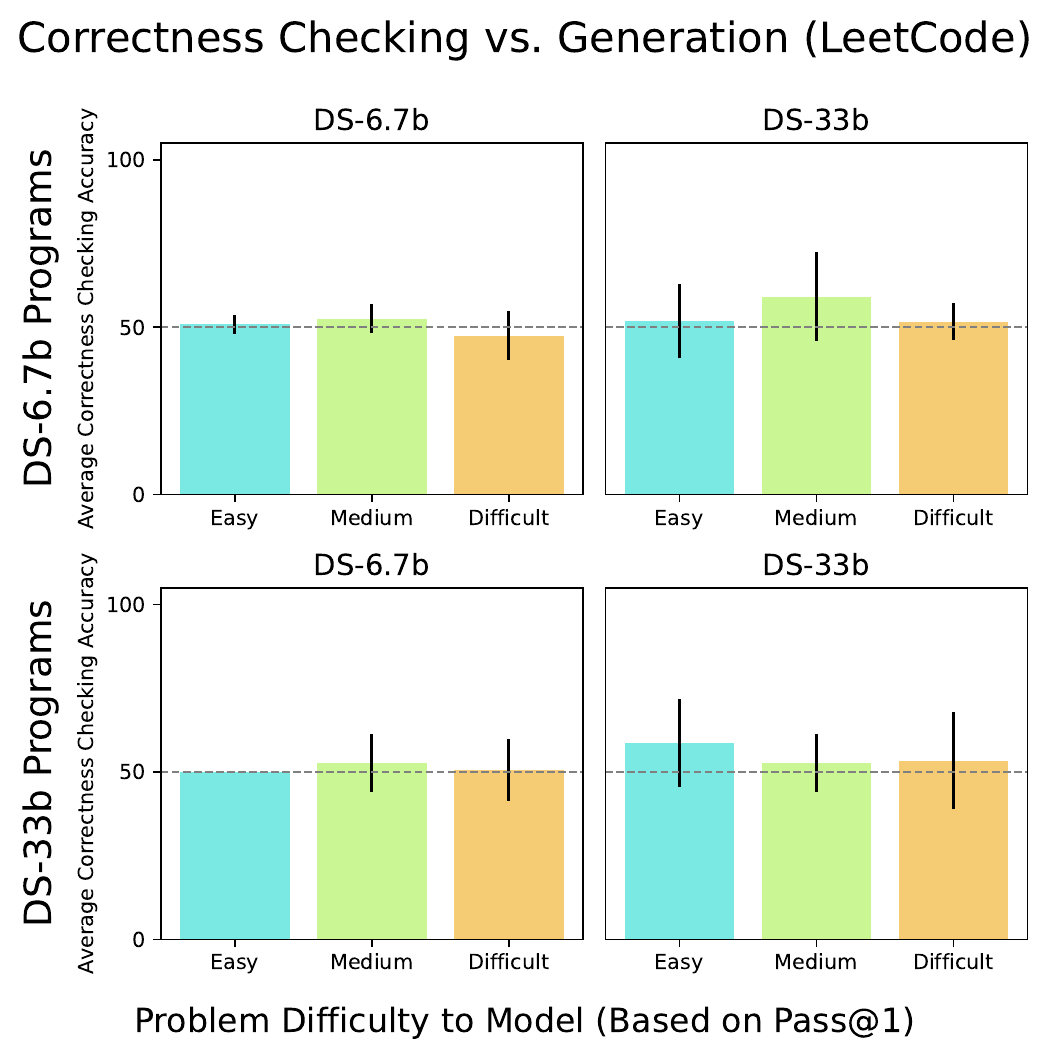}
    \end{subfigure}
    \begin{subfigure}[b]{0.9\textwidth}
        \centering
        \includegraphics[width=\linewidth]{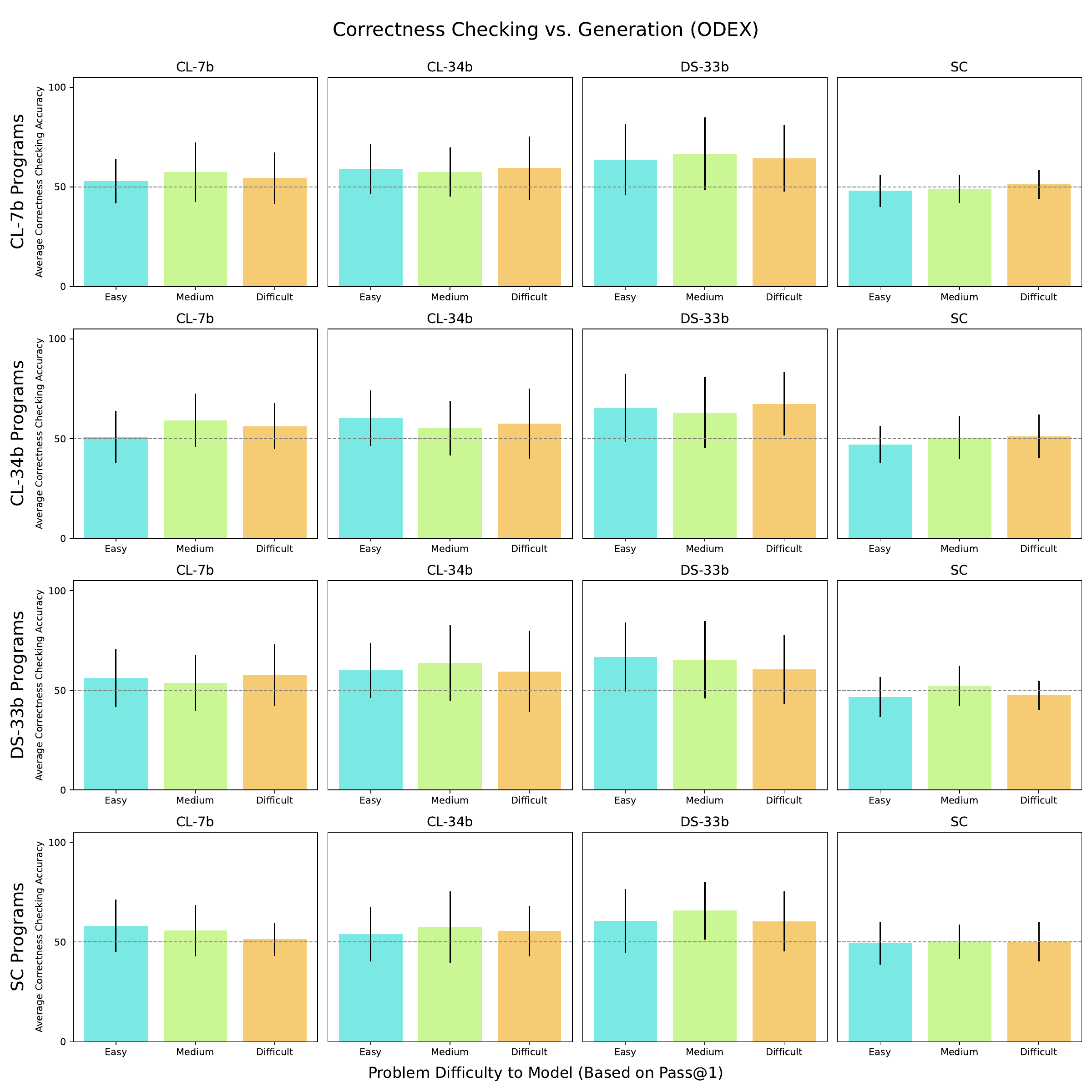}
    \end{subfigure}
    \caption{Accuracies for correctness checking task bucketed by difficulty.}
    \label{fig:correctness-checking-difficulty-all}
\end{figure}

\subsection{Problem Difficulty vs. Execution} \label{appendix:difficulty-execution-full}

In Fig. \ref{fig:execution-difficulty-all}, we show the accuracy of HumanEval (a), LeetCode (b), and ODEX (c) across different models. We see a slight correlation, where programs for more difficult problems are harder to execute.

\begin{figure}[H]
    \centering
    \begin{subfigure}[b]{\textwidth}
        \centering
        \includegraphics[width=\linewidth]{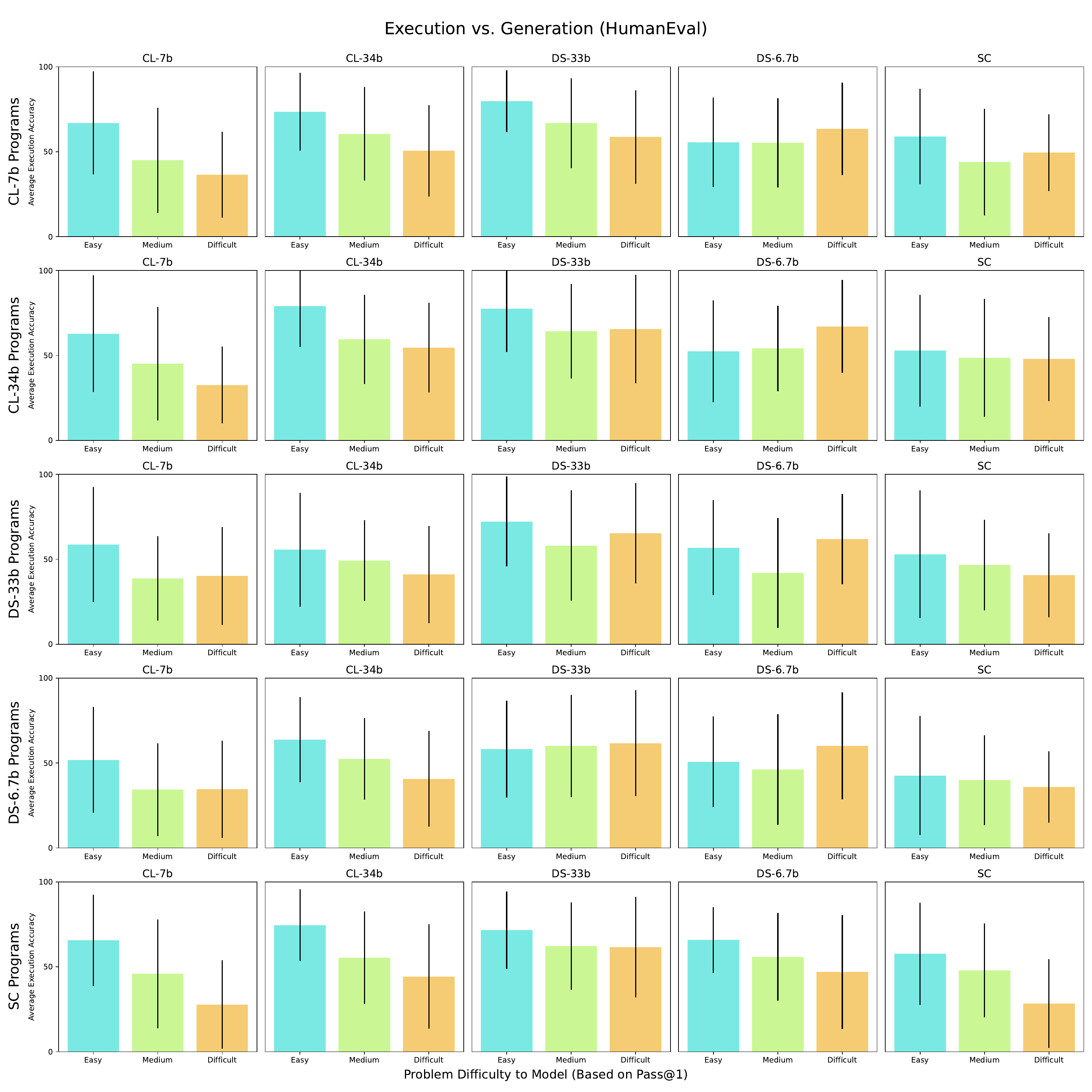}
        \subcaption{HumanEval}
    \end{subfigure}
\end{figure}
\begin{figure}[H]
    \ContinuedFloat
    \centering
    \begin{subfigure}[b]{0.5\textwidth}
        \centering
        \includegraphics[width=\linewidth]{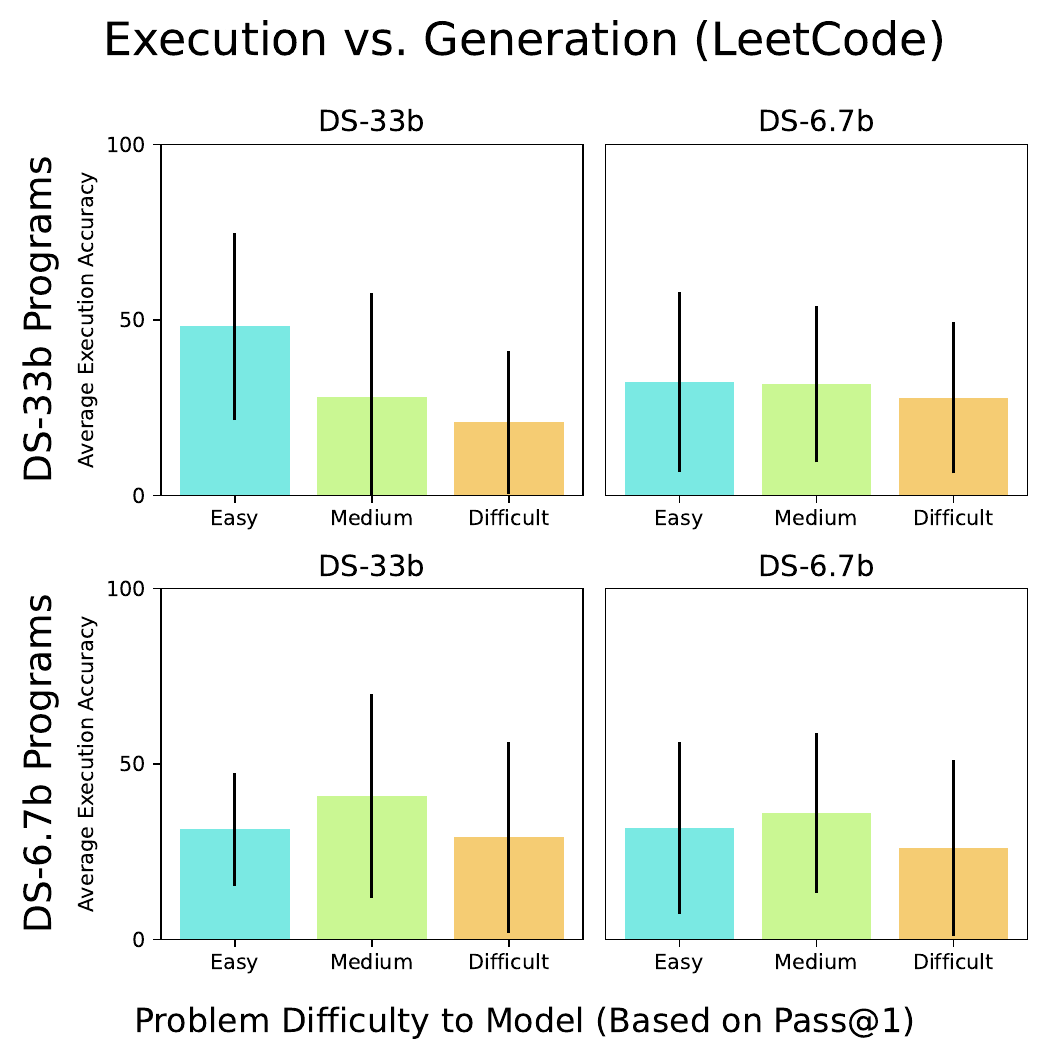}
        \subcaption{LeetCode}
    \end{subfigure}
    \caption{Accuracies for execution task bucketed by difficulty.}
    \label{fig:execution-difficulty-all}
\end{figure}

\clearpage
\subsection{HumanEval Pass Rate vs. Correctness Prediction}
\begin{figure}[H]
    \centering
    \includegraphics[width=\linewidth]{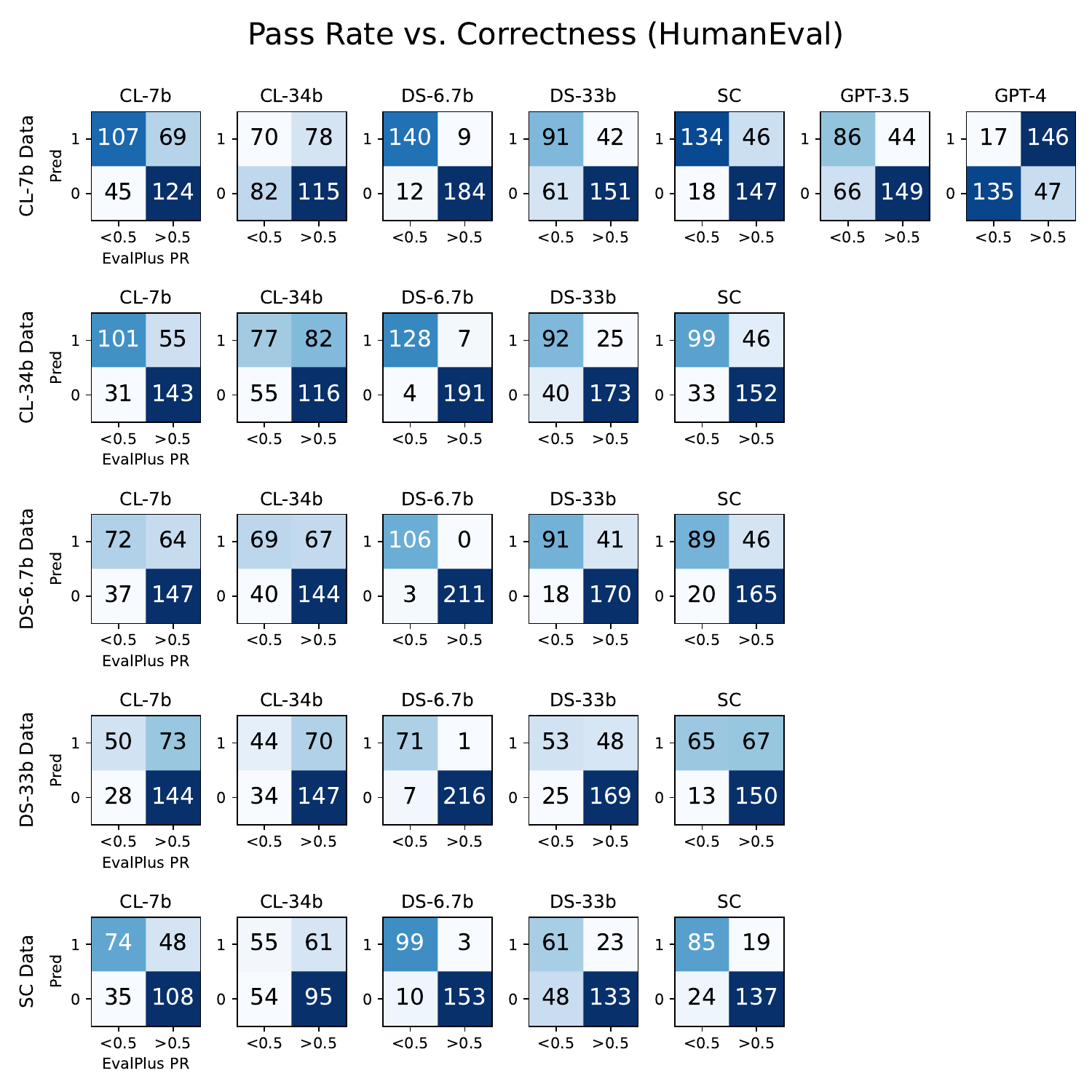}
    \caption{Models other than GPT-4 show a lack of correlation between a problem's pass rate and its correctness prediction.}
    \label{fig:pass-rate-cm}
\end{figure}

In Fig. \ref{fig:pass-rate-cm}, we investigate the correlation between a program's pass rate on HumanEval (using EvalPlus tests) and its prediction. Since a problem's pass rate is indicative of how close it is to correct, we might expect that programs with a higher pass rate have a higher chance of being predicted as correct. For most models, this does not seem ot be the case, though we do see this trend for GPT-4.

\section{Examples of Counterfeits} \label{appendix:qual-sample-categories}
\subsection{Counterfeits with Algorithmic Errors}

\begin{lstlisting}[language=python, frame=single]
def multiply(a, b):
    """Complete the function that takes two integers and returns 
    the product of their unit digits.
    Assume the input is always valid.
    Examples:
    multiply(148, 412) should return 16.
    multiply(19, 28) should return 72.
    multiply(2020, 1851) should return 0.
    multiply(14,-15) should return 20.
    """
    return a % 10 * b % 10
\end{lstlisting}
\begin{lstlisting}[language=python, frame=single]
def sort_numbers(numbers: str) -> str:
    """ Input is a space-delimited string of numberals from 'zero' to 'nine'.
    Valid choices are 'zero', 'one', 'two', 'three', 'four', 'five', 'six', 'seven', 'eight' and 'nine'.
    Return the string with numbers sorted from smallest to largest
    >>> sort_numbers('three one five')
    'one three five'
    """
    return ' '.join(sorted(numbers.split()))
\end{lstlisting}
\begin{lstlisting}[language=python, frame=single]
def eat(number, need, remaining):
    """
    You're a hungry rabbit, and you already have eaten a certain number of carrots,
    but now you need to eat more carrots to complete the day's meals.
    you should return an array of [ total number of eaten carrots after your meals,
                                    the number of carrots left after your meals ]
    if there are not enough remaining carrots, you will eat all remaining carrots, but will still be hungry.
    
    Example:
    * eat(5, 6, 10) -> [11, 4]
    * eat(4, 8, 9) -> [12, 1]
    * eat(1, 10, 10) -> [11, 0]
    * eat(2, 11, 5) -> [7, 0]
    
    Variables:
    @number : integer
        the number of carrots that you have eaten.
    @need : integer
        the number of carrots that you need to eat.
    @remaining : integer
        the number of remaining carrots thet exist in stock
    
    Constrain:
    * 0 <= number <= 1000
    * 0 <= need <= 1000
    * 0 <= remaining <= 1000

    Have fun :)
    """
    return [number + need, remaining - need] if remaining - need >= 0 else [remaining, 0]
\end{lstlisting}

\subsection{Counterfeits ignoring aspects of the specification}
\begin{lstlisting}[language=python, frame=single]
def specialFilter(nums):
    """Write a function that takes an array of numbers as input and returns 
    the number of elements in the array that are greater than 10 and both 
    first and last digits of a number are odd (1, 3, 5, 7, 9).
    For example:
    specialFilter([15, -73, 14, -15]) => 1 
    specialFilter([33, -2, -3, 45, 21, 109]) => 2
    """
    # Define a function that checks if the first and last digit of a number is odd
    def isSpecial(n):
        if n < 0: n = -n  # make sure n is positive
        return n % 10 in [1, 3, 5, 7, 9] and n//10 % 10 in [1, 3, 5, 7, 9]

    # Use a list comprehension to filter the numbers and count the results
    return sum(1 for num in nums if num > 10 and isSpecial(num))
\end{lstlisting}
\begin{lstlisting}[language=python, frame=single]
def common(l1: list, l2: list):
    """Return sorted unique common elements for two lists.
    >>> common([1, 4, 3, 34, 653, 2, 5], [5, 7, 1, 5, 9, 653, 121])
    [1, 5, 653]
    >>> common([5, 3, 2, 8], [3, 2])
    [2, 3]

    """
    return list(set(l1) & set(l2))
\end{lstlisting}
\begin{lstlisting}[language=python, frame=single]
def find_max(words):
    """Write a function that accepts a list of strings.
    The list contains different words. Return the word with maximum number
    of unique characters. If multiple strings have maximum number of unique
    characters, return the one which comes first in lexicographical order.

    find_max(["name", "of", "string"]) == "string"
    find_max(["name", "enam", "game"]) == "enam"
    find_max(["aaaaaaa", "bb" ,"cc"]) == ""aaaaaaa"
    """
    max_word = ""
    max_len = 0
    for word in words:
        word_len = len(set(word))
        if word_len > max_len:
            max_len = word_len
            max_word = word
    return max_word
\end{lstlisting}

\begin{lstlisting}[language=python, frame=single]
def double_the_difference(lst):
    '''
    Given a list of numbers, return the sum of squares of the numbers
    in the list that are odd. Ignore numbers that are negative or not integers.
    
    double_the_difference([1, 3, 2, 0]) == 1 + 9 + 0 + 0 = 10
    double_the_difference([-1, -2, 0]) == 0
    double_the_difference([9, -2]) == 81
    double_the_difference([0]) == 0  
    
    If the input list is empty, return 0.
    '''
    return sum(elem**2 for elem in lst if elem%2!=0)
\end{lstlisting}

\subsection{Counterfeits that fail to handle important edge cases}
\begin{lstlisting}[language=python, frame=single]
def intersperse(numbers: List[int], delimeter: int) -> List[int]:
    """ Insert a number 'delimeter' between every two consecutive elements of input list `numbers'
    >>> intersperse([], 4)
    []
    >>> intersperse([1, 2, 3], 4)
    [1, 4, 2, 4, 3]
    """
    result = []
    for number in numbers[:-1]:
        result.extend([number, delimeter])
    result.append(numbers[-1])
    return result
\end{lstlisting}

\begin{lstlisting}[language=python, frame=single]
def next_smallest(lst):
    """
    Return the 2nd smallest element of a list of integers
    Return None if there is no such element.
    next_smallest([2, 1, 3]) == 2
    next_smallest([1, 1]) == None
    """
    if len(lst) < 2:
        return None
    lst_copy = lst.copy()
    lst_copy.sort()
    return lst_copy[1]
\end{lstlisting}

\begin{lstlisting}[language=python, frame=single]
def sum_to_n(n: int):
    """sum_to_n is a function that sums numbers from 1 to n.
    >>> sum_to_n(30)
    465
    >>> sum_to_n(100)
    5050
    >>> sum_to_n(5)
    15
    >>> sum_to_n(10)
    55
    >>> sum_to_n(1)
    1
    """
    if n == 1:
        return 1
    else:
        return n + sum_to_n(n - 1)
\end{lstlisting}
\begin{lstlisting}[language=python, frame=single]
def get_odd_collatz(n):
    """
    Given a positive integer n, return a sorted list that has the odd numbers in collatz sequence.

    The Collatz conjecture is a conjecture in mathematics that concerns a sequence defined
    as follows: start with any positive integer n. Then each term is obtained from the 
    previous term as follows: if the previous term is even, the next term is one half of 
    the previous term. If the previous term is odd, the next term is 3 times the previous
    term plus 1. The conjecture is that no matter what value of n, the sequence will always reach 1.

    Note: 
        1. Collatz(1) is [1].
        2. returned list sorted in increasing order.

    For example:
    get_odd_collatz(5) returns [1, 5] # The collatz sequence for 5 is [5, 16, 8, 4, 2, 1], so the odd numbers are only 1, and 5.
    """
    collatz_sequence = [n]

    while n != 1:
        if n % 2 == 0:
            n = n / 2
        else:
            n = 3 * n + 1
        collatz_sequence.append(n)

    odd_numbers = [i for i in collatz_sequence if i % 2 != 0]
    return sorted(odd_numbers)
\end{lstlisting}

\section{Examples of Counterfeits Misclassifications} \label{appendix:qual-sample-mistakes}
Here, we present some examples of counterfeit programs that are misclassified.

\subsection{Counterfeit samples that are mistaken as correct}
\noindent In the following example, the code does not follow the instruction to \textit{``Ignore numbers that are negative or not integers''}.
\begin{lstlisting}[language=python, frame=single, belowskip=20pt]
def double_the_difference(lst):
    '''
    Given a list of numbers, return the sum of squares of the numbers
    in the list that are odd. Ignore numbers that are negative or not integers.
    
    double_the_difference([1, 3, 2, 0]) == 1 + 9 + 0 + 0 = 10
    double_the_difference([-1, -2, 0]) == 0
    double_the_difference([9, -2]) == 81
    double_the_difference([0]) == 0  
    
    If the input list is empty, return 0.
    '''
    # Your code here
    # if lst == []:
    #     return 0
    # else:
    #     return sum([int(x)**2 for x in lst if int(x) % 2 == 1])
    return sum(x**2 for x in lst if x % 2 != 0)
\end{lstlisting}
\begin{lstlisting}[frame=single]
Model: The program defines a function `double_the_difference` that takes a list of numbers as input and returns the sum of squares of the numbers in the list that are odd. It ignores numbers that are negative or not integers. The provided examples in the docstring are correct and match the expected behavior. The program correctly handles the case when the input list is empty by returning 0. The list comprehension filters out negative numbers and non-integers before squaring and summing the odd numbers.
\end{lstlisting}

\noindent In this example, the model excludes the starting \verb|"0b"| despite the examples in the specification including them.
\begin{lstlisting}[language=python, frame=single, belowskip=20pt]
def rounded_avg(n, m):
    """You are given two positive integers n and m, and your task is to compute the
    average of the integers from n through m (including n and m). 
    Round the answer to the nearest integer and convert that to binary.
    If n is greater than m, return -1.
    Example:
    rounded_avg(1, 5) => "0b11"
    rounded_avg(7, 5) => -1
    rounded_avg(10, 20) => "0b1111"
    rounded_avg(20, 33) => "0b11010"
    """
    if n > m:
        return -1

    return bin(round((m+n)/2))[2:]
\end{lstlisting}

\begin{lstlisting}[frame=single]
Model: The program takes two positive integers n and m, computes the average of integers from n through m (inclusive), rounds the answer to the nearest integer, and converts that to binary. If n is greater than m, it returns -1. The rounding is done using the `round` function which rounds to the nearest even number in case of a tie.

For example:
- rounded_avg(1, 5) => average of (1+2+3+4+5)/2 = 3, binary representation is "0b11"
- rounded_avg(7, 5) => n > m, so return -1
- rounded_avg(10, 20) => average of (10+11+12+13+14+15+16+17+18+19+20)/2 = 15.5, rounded to 16, binary representation is "0b1111"
- rounded_avg(20, 33) => average of (20+21+22+23+24+25+26+27+28+29+30+31+32+33)/2 = 26.5, rounded to 26, binary representation is "0b11010"

The program correctly implements the specified logic.
\end{lstlisting}

\noindent In the following example, the \verb|max_heap| will never get initialized because \verb|i| starts at value \verb|1|. Then the remainder of the algorithm will never run.
\begin{lstlisting}[language=python, frame=single]
"""
Problem: You are given a 0-indexed array nums of n integers and an integer target.
You are initially positioned at index 0. In one step, you can jump from index i to any index j such that:

0 <= i < j < n
-target <= nums[j] - nums[i] <= target

Return the maximum number of jumps you can make to reach index n - 1.
If there is no way to reach index n - 1, return -1.
 
Example 1:

Input: nums = [1,3,6,4,1,2], target = 2
Output: 3
Explanation: To go from index 0 to index n - 1 with the maximum number of jumps, you can perform the following jumping sequence:
- Jump from index 0 to index 1. 
- Jump from index 1 to index 3.
- Jump from index 3 to index 5.
It can be proven that there is no other jumping sequence that goes from 0 to n - 1 with more than 3 jumps. Hence, the answer is 3. 
Example 2:

Input: nums = [1,3,6,4,1,2], target = 3
Output: 5
Explanation: To go from index 0 to index n - 1 with the maximum number of jumps, you can perform the following jumping sequence:
- Jump from index 0 to index 1.
- Jump from index 1 to index 2.
- Jump from index 2 to index 3.
- Jump from index 3 to index 4.
- Jump from index 4 to index 5.
It can be proven that there is no other jumping sequence that goes from 0 to n - 1 with more than 5 jumps. Hence, the answer is 5. 
Example 3:

Input: nums = [1,3,6,4,1,2], target = 0
Output: -1
Explanation: It can be proven that there is no jumping sequence that goes from 0 to n - 1. Hence, the answer is -1. 

 
Constraints:

2 <= nums.length == n <= 1000
-10^9 <= nums[i] <= 10^9
0 <= target <= 2 * 10^9
"""

def maximumJumps(nums: List[int], target: int) -> int:
    n = len(nums)
    dp = [0] + [-1] * (n - 1)
    max_heap = []

    for i in range(1, n):
        while max_heap and max_heap[0][1] < i - target:
            heapq.heappop(max_heap)

        while max_heap and abs(nums[i] - nums[max_heap[0][1]]) <= target:
            dp[i] = max(dp[i], max_heap[0][0] + 1)

        if dp[i] >= 0:
            heapq.heappush(max_heap, (-dp[i], i))

    return dp[-1]
\end{lstlisting}
\begin{lstlisting}[frame=single]
Model: The given program defines a function `maximumJumps` that takes a list of integers `nums` and an integer `target` as input. It uses dynamic programming to keep track of the maximum number of jumps that can be made to reach each index. It also uses a max heap to efficiently find the maximum number of jumps that can be made to reach the current index based on the constraints given. The function iterates through the array and updates the dynamic programming array and the max heap accordingly. Finally, it returns the maximum number of jumps needed to reach the last index.
\end{lstlisting}

\noindent In this next example, when checking whether another team is stronger than the current \verb|team|, it should instead check that \verb|grid[opponent][team] == 1|.
\begin{lstlisting}[language=python, frame=single]
"""
Problem: There are n teams numbered from 0 to n - 1 in a tournament.
Given a 0-indexed 2D boolean matrix grid of size n * n. For all i, j that 0 <= i, j <= n - 1 and i != j 
team i is stronger than team j if grid[i][j] == 1, otherwise, team j is stronger than team i.
Team a will be the champion of the tournament if there is no team b that is stronger than team a.
Return the team that will be the champion of the tournament.
 
Example 1:

Input: grid = [[0,1],[0,0]]
Output: 0
Explanation: There are two teams in this tournament.
grid[0][1] == 1 means that team 0 is stronger than team 1. So team 0 will be the champion.

Example 2:

Input: grid = [[0,0,1],[1,0,1],[0,0,0]]
Output: 1
Explanation: There are three teams in this tournament.
grid[1][0] == 1 means that team 1 is stronger than team 0.
grid[1][2] == 1 means that team 1 is stronger than team 2.
So team 1 will be the champion.

 
Constraints:

n == grid.length
n == grid[i].length
2 <= n <= 100
grid[i][j] is either 0 or 1.
For all i grid[i][i] is 0.
For all i, j that i != j, grid[i][j] != grid[j][i].
The input is generated such that if team a is stronger than team b and team b is stronger than team c, 
then team a is stronger than team c.
"""

def findChampion(grid: List[List[int]]) -> int:
    n = len(grid)
    champion = -1
    for team in range(n):
        strong = True
        for opponent in range(n):
            if team != opponent and grid[team][opponent] == 1:
                strong = False
                break
        if strong:
            champion = team
            break
    return champion
\end{lstlisting}
\begin{lstlisting}[frame=single]
Model: The given program takes a 2D boolean matrix `grid` representing the strengths between teams in a tournament. It iterates through each team and checks if there is no other team that is stronger than the current team. If such a team is found, it updates the champion and breaks out of the loop. The program then returns the champion team number.

For each team, it checks against all other teams to determine if it is stronger than all of them. If it finds a team that is stronger, it marks the current team as not strong enough to be the champion. If no such team is found, the current team is marked as the champion.

The program correctly implements the logic of finding the champion team based on the given rules of team strengths in the tournament.
\end{lstlisting}

\noindent Following are several counterfeit samples that are misclassified in verification because of forgiving an identified error:

\begin{lstlisting}[language=python, frame=single]
def f_30551576(s):
    """Get all urls within text `s`
    """
    return re.findall('(?P<url>https?://[^\s]+)', s)
\end{lstlisting}
\begin{lstlisting}[frame=single]
Model: [...] However, it's important to consider edge cases that are not covered by the provided test cases. For example:
- URLs that are immediately followed by punctuation (e.g., a comma, period, or semicolon) which are common in natural text, might be included in the URL by this regex pattern.
- The regex does not account for URLs embedded within parentheses or other characters that are not considered whitespace but should not be part of the URL.

Despite these considerations, based on the information provided and the test cases, the program seems to meet the requirements specified in the docstring for the given examples. It's important to note that for more comprehensive validation, additional test cases covering a wider range of scenarios, including those mentioned above, would be beneficial.
\end{lstlisting}

\begin{lstlisting}[language=python, frame=single]
def f_22187233():
    """delete all occurrences of character 'i' in string 'it is icy'
    """
    return 't s  cy'
\end{lstlisting}

\begin{lstlisting}[frame=single]
Model: The program defines a function `f_22187233` with the purpose of deleting all occurrences of the character 'i' in the string 'it is icy' as per the docstring. The function returns a hardcoded string 't s  cy', which is the expected result after removing all 'i's from the given string.[...] However, it's important to note that the program's approach is not dynamic and does not actually perform any operation to remove 'i's from a given string. It simply returns a hardcoded string that matches the expected output for the specific input 'it is icy'. This means the function is not versatile and will not work correctly for any other input string or requirement to remove different characters. But, based on the very specific requirement outlined in the docstring, the program does meet the criteria for correctness.
\end{lstlisting}

\begin{lstlisting}[language=python, frame=single]
def reverse_delete(s,c):
    """Task
    We are given two strings s and c, you have to deleted all the characters in s that are equal to any character in c
    then check if the result string is palindrome.
    A string is called palindrome if it reads the same backward as forward.
    You should return a tuple containing the result string and True/False for the check.
    Example
    For s = "abcde", c = "ae", the result should be ('bcd',False)
    For s = "abcdef", c = "b"  the result should be ('acdef',False)
    For s = "abcdedcba", c = "ab", the result should be ('cdedc',True)
    """
    # take a copy of the string
    s1 = s
    # loop over the string
    for i in range(len(s)):
        # check if the character is in the list
        if s[i] in c:
            # remove the character
            s = s.replace(s[i], "")
    # return the string and check if it is a palindrome
    return s, s == s[::-1]
\end{lstlisting}

\begin{lstlisting}[frame=single, breakindent=0pt]
[...] However, there's a potential issue with the approach of removing characters by iterating and using `replace` within the same loop: if a character appears multiple times in `s` and is also in `c`, each occurrence of that character will be removed in a single iteration, which is the intended behavior. 

[...] In conclusion, the program appears to correctly implement the specified functionality, despite the initial misinterpretation of the potential issue with modifying `s` during iteration. The key point is that `replace` creates a new string each time, avoiding the problem of modifying the iterable during iteration.
\end{lstlisting}

\begin{lstlisting}[language=python, frame=single]
def largest_divisor(n: int) -> int:
    """ For a given number n, find the largest number that divides n evenly, smaller than n
    >>> largest_divisor(15)
    5
    """
    # your code here
    for i in range(n - 1, 1, -1):
        if n % i == 0:
            return i
\end{lstlisting}

\begin{lstlisting}[frame=single, breakindent=0pt]
Model: ...The program does not explicitly handle the case where `n` is less than or equal to 1. However, given 
the problem statement, it is reasonable to assume that `n` will be a positive integer greater than 1. 
For `n = 1` or `n = 0`, the function would not enter the loop and would not return a value, which 
implicitly aligns with the expectation that there are no divisors smaller than `n` in these cases 
(since `n` itself is not greater than 1).

Considering the above, the program correctly implements the functionality described in the docstring 
for all positive integers greater than 1. It efficiently finds and returns the largest divisor of `n` 
that is smaller than `n` itself by iterating backward from `n-1` and checking for divisibility.
\end{lstlisting}



\subsection{GPT-4 Errors in Execution Prediction} \label{appendix:gpt4-execution-errors}
In this section, we show a few examples where GPT-4 makes a mistake in execution prediction and arrives at the output had the program been correct. In the following listing, the model correctly identifies the operation to perform at each step, but incorrectly says that appending 6, 8, and 3 to \texttt{[5, 8, 6]} gives \texttt{[5, 8, 6, 8, 3]}. It is unclear if this is due to the model being biased by the correct output.
\newline
\begin{lstlisting}[language=Python, frame=single]
def intersperse(numbers: List[int], delimeter: int) -> List[int]:
    result = []
    for index, value in enumerate(numbers[1:], 1):
        result.append(numbers[index - 1])
        result.append(delimeter)
        result.append(value)
    return result
assert intersperse([5, 6, 3, 2], 8) == [5, 8, 6, 6, 8, 3, 3, 8, 2]
# (Incorrect) output of correct solution: [5, 8, 6, 8, 3, 8, 2]
\end{lstlisting}

\begin{lstlisting}[frame=single, belowskip=20pt]
Let's break down the execution of the intersperse function:

1. The function is called with the list [5, 6, 3, 2] and the delimiter 8.
2. An empty list named result is initialized.
3. The for loop iterates over the numbers list starting from the second element (due to numbers[1:]), with index starting from 1. This means the loop will iterate over elements 6, 3, and 2 with their respective indices 1, 2, and 3.
4. In the first iteration (index=1, value=6), the function appends the element at index 0 (5), then the delimiter (8), and then the current value (6) to the result list. So, result becomes [5, 8, 6].
5. In the second iteration (index=2, value=3), !* it appends the element at index 1 (6), then the delimiter (8), and then the current value (3) to the result list. Result now becomes [5, 8, 6, 8, 3]. *!
6. In the third iteration (index=3, value=2), it appends the element at index 2 (3), then the delimiter (8), and then the current value (2) to the result list. Result now becomes [5, 8, 6, 8, 3, 8, 2].
7. The function returns the result list, which is [5, 8, 6, 8, 3, 8, 2].
\end{lstlisting}

\noindent In the next listing, GPT-4 does everything correctly. It correctly says that zip function will stop at the shortest list but somehow arrives at the output of the correct solution.

\begin{lstlisting}[language=Python, frame=single]
def sort_even(l: list):
    even = l[::2]
    even.sort()
    odd = l[1::2]
    return [x for t in zip(even, odd) for x in t]
assert sort_even([1, 2, 3]) == [1, 2]
# (Incorrect) output of correct solution: [1, 2, 3]
\end{lstlisting}

\begin{lstlisting}[frame=single, belowskip=20pt]
Let's execute the code step by step:

1. The function sort_even is defined, which takes a single argument l (a list).
2. The function is called with the argument [1, 2, 3], so within the function, l is initially [1, 2, 3].
3. The variable even is assigned the value of l[::2], which means it takes every second element from l, starting from the first element (index 0). Therefore, even becomes [1, 3].
4. The even list is then sorted, but since it's already in ascending order ([1, 3]), it remains unchanged.
5. The variable odd is assigned the value of l[1::2], which means it takes every second element from l, starting from the second element (index 1). Therefore, odd becomes [2].
6. The function then returns a list that is created by zipping the even and odd lists together and flattening the resulting tuples into a single list. !* Since the even list has more elements than the odd list, the zip function will stop at the shortest list (odd), resulting in the final list being [1, 2, 3].
\end{lstlisting}

\noindent In the next listing, the error is in step 7, highlighted below. The model correctly identifies that the inner while loop will divide 27 by 3 until it is no longer divisible by 3, but incorrect concludes that this will end in $n=3$ instead of $n=1$.

\begin{lstlisting}[language=Python, frame=single]
def largest_prime_factor(n: int):
    # We don't use 1 as a factor, so we start with 2.
    i = 2
    # Check each potential factor from 2 to sqrt(n).
    while i * i <= n:
        if n % i == 0:
            # Found a factor! Check if it's prime.
            while n % i == 0:
                n //= i
            return largest_prime_factor(n)
        i += 1
    # All the while loop's conditions were false, so n is prime!
    return n
assert largest_prime_factor(27) == 1
# (Incorrect) output of correct solution: 3
\end{lstlisting}

\begin{lstlisting}[frame=single, belowskip=20pt]
Let's execute the code step by step for `largest_prime_factor(27)`:

1. The function `largest_prime_factor` is defined to find the largest prime factor of a given number `n`.
2. The function is called with `n = 27`.
3. The variable `i` is initialized to 2, which is the smallest prime number.
4. The while loop condition `i * i <= n` is checked. For `i = 2`, `2 * 2 = 4`, which is less than `27`, so the loop continues.
5. Inside the loop, `n % i == 0` is checked to see if `i` is a factor of `n`. For `i = 2`, `27 % 2 != 0`, so `i` is incremented to `3`.
6. With `i = 3`, `3 * 3 = 9`, which is less than `27`, so the loop continues.
!* 7. Now, `27 % 3 == 0`, so `3` is a factor of `27`. The inner while loop (`while n % i == 0`) will divide `n` by `3` until it is no longer divisible by `3`. After this loop, `n` becomes `27 / 3 / 3 = 3`. *!
8. Since `n` was modified, the function calls itself recursively with the new value of `n`, which is now `3`.
9. In the recursive call with `n = 3`, the loop starts again with `i = 2`. `2 * 2 = 4`, which is greater than `3`, so the loop ends.
10. Since `n` is not modified in this recursive call, the function returns `n`, which is `3`.
\end{lstlisting}

\noindent In the final listing, GPT-4 takes a shortcut based on the comments rather than following the execution of the code.

\begin{lstlisting}[language=Python, frame=single]
def search(lst):
    # Your code here
    # sort the list and check if the first number is 1
    lst.sort()
    if lst[0] != 1:
        return -1
    # find the index of 1
    index = lst.index(1)
    # iterate over the list starting from the index of 1
    for i in range(index, len(lst)):
        # if the frequency of the number is equal to the value of the number
        if lst.count(lst[i]) >= lst[i]:
            # return the number
            return lst[i]
    # if no number is found return -1
    return -1
assert search([4, 1, 4, 1, 4, 4]) == 1
# (Incorrect) output of correct solution: 4
\end{lstlisting}

\begin{lstlisting}[frame=single]
Let's execute the code step by step:

1. The function `search` is defined, which takes a list `lst` as an argument.
2. The function is called with the argument `[4, 1, 4, 1, 4, 4]`.
3. Inside the function, the list is sorted, so `lst` becomes `[1, 1, 4, 4, 4, 4]`.
4. The function checks if the first number in the sorted list is 1, which it is.
5. The function then finds the index of 1, which is 0 (but since the list is sorted and starts with 1, this step is somewhat redundant in this context).
!* 6. The function iterates over the list starting from the index of 1. It checks if the frequency of the number is equal to or greater than the value of the number.
7. When it gets to the number 4, it checks if the frequency of 4 is equal to or greater than its value. Since there are four 4s in the list, and the value of 4 is equal to its frequency, the condition is met. *!
8. The function returns 4 because the frequency of 4 (which is 4) is equal to its value.
\end{lstlisting}

\end{document}